\documentclass{JHEP3}
 
\usepackage{amsmath,amsfonts,a4,graphicx}

\numberwithin{equation}{section}

\def\hR{\textbf R} 
\def\hA{\textbf A} 
\def\2F1{ {{}_{2}\textrm{F}_{1}}}
\def\sg{\hat\sigma}
\def\Pinf{\hat{\mathcal{P}}^\infty}
\def\PP{\tilde P}
\def\ZZ{{\mathbb Z}}

\def\P{{\tilde P}}
\def\Pinf{\hat{\mathcal P}^\infty}
\def\D{\textrm{D}}
\def\nA{\mu}

\newcommand{\be}{\begin{equation}}
\newcommand{\ee}{\end{equation}}
\newcommand{\bea}{\begin{eqnarray}}
\newcommand{\eea}{\end{eqnarray}}
\def\p{\textbf{p}}

\def\a{\alpha} 
\def\al{{\alpha'}}

\def\td{\tilde }

\def\SL(2,Z){SL(2,\mathbb{Z})}

\def\non{\nonumber }

\def\ha{ {1\over 2}}

\def\calF{{\cal F}}
\def\calD{{\cal D}}
\def\calP{{\cal P}}
\def\calT{{\cal T}}
\def\calE{{\cal E}}
\def\calR{{\cal R}}

\def\nn{\nonumber}
\def\half{{\scriptstyle {1 \over 2}}}

\def\sevenh{{\scriptstyle {7 \over 2}}}
\def\ze{{\zeta_1,\zeta_2,\zeta_3,\zeta_4}}

\preprint{DAMTP-2007-96;
SPhT-T-07-126;
UB-ECM-PF 07/29}

\title{Low energy expansion of the four-particle genus-one amplitude in type II superstring
 theory}

\author{Michael B. Green\\
 Department of Applied Mathematics and
Theoretical Physics\\
Wilberforce Road, Cambridge CB3 0WA, UK\\
\email{\tt M.B.Green@damtp.cam.ac.uk}}
\author{Jorge G. Russo\\
Instituci\' o Catalana de Recerca i Estudis Avan\c{c}ats (ICREA),\\
University de Barcelona,  Facultat de Fisica\\
 Av. Diagonal, 647,  Barcelona 08028 SPAIN\\
\email{\tt jrusso@ub.edu}}
\author{Pierre Vanhove\\
Service de Physique Th{\'e}orique,\\
CEA/DSM/PhT, CEA/Saclay, Orme des Merisiers, CEA/Saclay\\
91191 Gif-sur-Yvette Cedex, France\\
and\\
Niels Bohr Institute, University of Copenhagen,\\
  Blegdamsvej 17, DK--2100 Copenhagen \O, Denmark\\
\email{pierre.vanhove@cea.fr}}

\abstract{A diagrammatic expansion of coefficients in the low-momentum
expansion of the genus-one four-particle amplitude in type II superstring theory is
developed.  This is applied to determine coefficients up to order $s^6\,
\hR^4$
(where $s$ is a Mandelstam invariant and $\hR$ the linearized
super-curvature), and partial results are
obtained beyond that order.
  This involves integrating powers of the scalar propagator on a
  toroidal world-sheet, as well as integrating over the modulus of the
  torus. At any given order in $s$ the coefficients of these terms
  are given by rational numbers multiplying multiple zeta values (or Euler--Zagier sums) that, up to
  the order studied here, reduce to  products of Riemann zeta values.  We are careful to
disentangle the analytic pieces from logarithmic threshold terms, which involves a discussion of
the conditions imposed by unitarity.
We further consider the compactification of the amplitude on a circle of radius $r$, which results in
a plethora of terms that are power-behaved in $r$.  These coefficients provide boundary `data' that
must be matched by any non-perturbative expression for the low-energy expansion of the
four-graviton amplitude.

\medskip

The paper includes an appendix by  Don Zagier.

 }

\date{\Date}


\newpage
\begin{document}

\tableofcontents

\section{Introduction}

The low-momentum expansion (or $\alpha'$ expansion)
of string theory scattering amplitudes produces an
infinite sequence of stringy corrections to Einstein supergravity.
Although the coefficients in the momentum expansion of tree-level amplitudes
are easily obtained to all orders, it is far more difficult to obtain
the coefficients in the expansion of higher-genus contributions.
Knowledge of such
terms might be of use in constraining non-perturbative extensions of
string amplitudes.  However, determining these coefficients is
technically challenging, as we will see.

We will here be interested in the low-momentum expansion of the
genus-one scattering  amplitude for four particles in the massless
supergravity multiplet.
Each external particle is labeled by its momentum $p_r$ ($r=1,2,3,4$), where $p_r^2 =0$, and its
superhelicity $\zeta_r$, which takes 256 values (the dimensionality of the maximal supergravity
multiplet).
The genus-one amplitude has the form
 \cite{Green:1981yb},
\be
\label{eLoopa}
\hA_\ze^{genus-1}= I\,\hR_\ze^4\,,
\ee
where $I$ is the integral of a
modular function,
\be
\label{eLoop}
I(s,t,u)= \int_\calF {d^2\tau\over \tau_2^2} F(s,t,u;\tau)\,,
\ee
where  $F(s,t,u;\tau)$ is defined in eq.~(\ref{eF}) and
$s$, $t$, $u$ are Mandelstam invariants\footnote{
The Mandelstam invariants are defined by $s=-(p_1+p_2)^2$, $t=-(p_1+p_4)^2$ and
$u=-(p_1+p_3)^2$ and satisfy the mass-shell constraint $s+t+u=0$.}
and  $\tau=\tau_1+i\tau_2$ and $d^2\tau \equiv d\tau_1d\tau_2 =
d\tau d\bar\tau/2$.  The integral is over a fundamental domain of $\SL(2,Z)$, defined by
\be\label{fundom}
{\cal F}=\{|\tau_1|\leq \half, |\tau|^2 \geq 1\}\,.
\ee
The kinematical factor in (\ref{eLoopa})
is given by (see  (7.4.57) of \cite{Green:1987sp})
  \begin{equation}
  \hR_\ze^4(p_{1},p_{2},p_{3},p_{4})= \zeta_{1}^{AA'} \zeta_{2}^{BB'}
  \zeta_{3}^{CC'}\zeta_{4}^{DD'}\, K_{ABCD}\,\tilde K_{A'B'C'D'}\,,
  \label{kinfact}
 \end{equation}
where the indices $A, B$ on the polarization tensors $\zeta_r^{AB}$  run over both
 vector and spinor values (for example, the graviton polarization is $\zeta^{\mu\nu}$,
 where $\mu,\nu = 0,1,\dots,9$) and the tensor $K\, \tilde K$ is defined in
 \cite{Green:1987sp}.   In the case of external gravitons
  $\hR$ reduces to the linearized Weyl curvature,
$R_{\mu\rho\nu\sigma}
 = -4p_{[\mu}\, \zeta_{\rho][\sigma}\,p_{\nu]}$, and the kinematic factor is $R^4$, which denotes
 the product of four Weyl curvatures
 contracted into each other by a well-known  sixteen-index  $t_{8}t_{8}$ tensor defined in
 appendix~9.A of \cite{Green:1987mn}.
The integrand in (\ref{eLoopa}) is given by the expectation value of the product of the
vertex operators for the four external states integrated over their insertion
positions on the toroidal world-sheet of complex structure $\tau$.

The expression (\ref{eLoopa}) is free from divergences. We
can anticipate it has a low-momentum expansion that is the sum of an
analytic part and a non-analytic part
 associated with threshold singularities,
\bea
\label{loopexp}
I(s,t,u) &=&  I_{an}(s,t,u) + I_{nonan}(s,t,u)\, .
\eea
As shown in \cite{gv:stringloop},
the analytic part can be expressed in a power series in the Mandelstam invariants,
\be
I_{an}(s,t,u) = \sum_{p=0}^\infty\sum_{q=0}^\infty \hat\sigma_2^p\hat\sigma_3^q\,
J^{(p,q)}\,,
\label{iform}
 \ee
where
\be
\sg_2= \left(\frac{\al}{4}\right)^2\, (s^2+t^2+u^2)\, \qquad
\sg_3= \left(\frac{\al}{4}\right)^3\, (s^3+t^3+u^3) = 3
 \left(\frac{\al}{4}\right)^3\, stu\,,
\label{sigtwothree}
\ee
and $J^{(p,q)}$ are constant coefficients that are to be determined.
 The series (\ref{iform}) is the most general power series in Mandelstam invariants
that is symmetric in $s$, $t$ and $u$, subject to $s+t+u=0$, as
follows by making use of the identity,
 \be
 \label{sigmadef}
 \sg_n =\left( {\a'\over 4} \right)^n \big( s^n+t^n+u^n\big)= n \sum_{2p+3q=n}\,
 {(p+q-1)!\over p!q!}\, \left(\sg_{2}\over 2\right)^p \left(\sg_{3}\over3\right)^q\,.
 \ee
Each term in the series is a symmetric monomial in the Mandelstam invariants
of order $r=2p+3q$.
 For $r <6$ there is a single kinematic structure (a single term for a given
 value of $r$),   but there is a two-fold
 degeneracy for $r=6$,
 associated with $J^{(3,0)}$ and $J^{(0,2)}$, and thereafter the degeneracy of the
 kinematic factors increases sporadically.

The nonanalytic terms have branch cuts with a structure that is determined by unitarity.  These
singularities arise as infrared effects of internal massless states (i.e., supergravity states)
in the loop.   The discontinuities across these branch cuts are given by products of string tree-level
amplitudes integrated over the phase space of the (massless) states.  The lowest order term is the
one-loop contribution of maximal supergravity, while higher-order nonanalytic terms arise as stringy effects
from higher-order terms in the expansion of the tree amplitudes. As
will become apparent later,  the resulting structure of the nonanalytic
contribution to $I(s,t,u)$ is given by a series of terms,
\be
I_{nonan}(s,t,u) = I_{SUGRA} +  I_{nonan\ (4)}
+I_{nonan\, (6)} +I_{nonan\, (7)}+\cdots\, ,
\label{nongen}
\ee
 where $I_{SUGRA}$ is the integral that arises in one-loop maximal
 supergravity.  This has a complicated set of threshold singularities,
  that have discontinuities
 in a single Mandelstam variable, as well as
  terms that have multiple discontinuities (as will be reviewed
 in detail in section~\ref{sec:threshold}).
 The nature of the singularities of $I_{nonan}$ depends on the space-time dimension.  In ten dimensions
the higher-order
 threshold terms, $I_{nonan\, (r)}$,  have logarithmic branch points,
schematically  of the form $s^k\, \ln (-\al s/\nA_r)$,
 where $\nA_r$ is a constant scale. After compactification on a circle to nine
 non-compact dimensions they are half-integer powers,
schematically of the form $(-\alpha' s)^{k+1/2}$
 with integer $k$.

One motivation for evaluating the constants $J^{(p,q)}$ is to constrain the complete,
non-perturbative, expression for the
low-energy expansion of the four-particle amplitude.  For example in the IIB case
the exact $\SL(2,Z)$-invariant amplitude must have analytic pieces of the form
\be
\sum_{p,q=0}^\infty\calE_{(p,q)}(\Omega, \bar\Omega)\, e^{(2p+3q-1)\phi/2 }\,
\hat\sigma_2^p\hat \sigma_3^q\, \hR^4\, ,
\label{ampgen}
\ee
where $\Omega = \Omega_1 +i\Omega_2 =
C^{(0)} + i e^{-\phi}$ is the complex coupling constant that is given in
terms of the Ramond--Ramond pseudoscalar, $C^{(0)}$, and the dilaton $\phi$.
The explicit power of the string coupling constant $g_{s}= e^\phi$  disappears in the Einstein frame and
 $\calE_{(p,q)}(\Omega, \bar\Omega)$ is a modular function that contains
 both perturbative terms (that are power-behaved in $\Omega_2^{-1}=g_{s}$
 for small $g_{s}$) and non-perturbative (instantonic) contributions (that behave like $e^{-2\pi n/g_s}$).
The genus-one terms arise in (\ref{ampgen}) from the piece
of $\calE_{(p,q)}$ proportional to $e^{(1-2p-3q)\phi/2}\, J^{(p,q)}$.
  This amplitude may be
interpreted in terms of a
local effective action which can be schematically written in the form
\be
{\alpha'}^{k-1}  \,\int d^{10}x\, e^{(2p+3q-1)\phi/2 }\,
\sqrt{-G}\,\calE_{(p,q)}(\Omega, \bar\Omega)\, D^{2r}\hR^4\, ,
\label{interact}
\ee
In this expression the derivatives
 are contracted into each other and act on the curvature tensors in $\hR^4$
 in a manner that is defined by the functions of the Mandelstam invariants given in (\ref{ampgen}).
 The functions $\calE_{(p,q)}$ are known for $(p,q)=(0,0)$, $(1,0)$, $(0,1)$,  and
there are various conjectures at higher orders \cite{grv:twoloop}.
 In order to test conjectured forms of $\calE_{(p,q)}$ it is helpful to have as much
information from string perturbation theory as possible.

The aim of the
present paper is to develop the expansion of the genus-one amplitude more systematically in order
to explicitly evaluate the coefficients of higher
momentum terms. Since the type
  IIA and IIB theories have identical
  massless four-particle scattering amplitudes up to at least genus three, the
  calculations in this paper apply to both types of theory.
In section~\ref{sec:expand} we will begin by reviewing  the procedure of
 \cite{gv:stringloop}.
This involves integrating powers of propagators on the
world-sheet torus with fixed modulus $\tau = \tau_1 +i\tau_2$, followed by an integral over
$\tau$.

The ten-dimensional theory is considered in  section~\ref{sec:amplitude},
where the analytic  terms in the expansion of the amplitude up to order $s^6\, \hR^4$
are evaluated.
In order to analyze these terms we have to separate the nonanalytic threshold
singularities in a well-defined manner.
Such terms arise from the degeneration limit of the torus, in which $\tau_2\to \infty$, so we will divide the integral over $\calF$
into two domains: (i) $\calF_L$, where $\tau_2\le L$; (ii) $\calR_L$, where $\tau_2\ge L$,
with $L \gg 1$.  Integration over the truncated fundamental domain $\calF_L$ is carried out
in section~\ref{sec:amplitude}
by making use of a  theorem of harmonic analysis \cite{Terras}, and gives rise to the analytic
terms together with an $L$-dependent piece.

The integral over $\calR_L$, considered in
section~\ref{sec:threshold}, generates the nonanalytic threshold behaviour as well as
canceling the $L$-dependence of the $\calF_L$ integral.
The lowest-order threshold behaviour is the same as in the
supergravity field theory one-loop amplitude and has the schematic form
${\alpha'}\,s\log(-{\alpha'}\,s/\mu_1)+{\alpha'}\,t\log(-{\alpha'}\,t/\mu_1)+{\alpha'}\,u
\log(-{\alpha'}\,u/\mu_1)$
where $s$, $t$ and $u$ are Mandelstam invariants and $\mu_1$ is a constant scale.
The precise form of this threshold term is considerably more complicated,
as will be seen in detail in
section~\ref{sec:masslessbox}, but it still possesses the property that
the scale of the logarithms cancels (using $s+t+u=0$), so there is no
ambiguity associated with the normalization of the logarithms.
However, an important new feature at order ${\alpha'}^4\, s^4\, \hR^4$
is the occurrence of the second nonanalytic massless threshold
singularity, $I_{nonan\, (4)}$, that arises
via unitarity from the presence of the tree-level $\hR^4$ interaction, which will be discussed
in section~\ref{sec:twounitarity}.
This is again a symmetric function of the Mandelstam variables and has the schematic form
$\al^4\, s^4 \log (-{\alpha'}\,s/\nA_4) + \dots$, where
 $\nA_4$ is another normalization constant.  Changing the value
 of $\nA_4$ changes the definition of the coefficient of the analytic terms of the form $s^4+\dots$,
 so the precise value of $\nA_4$
has to be determined.
We will also determine the terms at
 order $s^5 \,\hR^4$ and $s^6\, \hR^4$,
 where the next threshold singularity arises.  We will also discuss partial
 results for terms at order $s^7\, \hR^4$, and $s^8\, \hR^4$.
These  ten-dimensional  results are summarized in section~\ref{sec:discussion}.
 
In section~\ref{sec:ninedim}
we will study the compactification of the loop amplitude on a circle
 of dimensionless radius $r$ to nine dimensions.
The threshold singularities are now
half-integral powers of $s,t$ and $u$,
and can be uniquely disentangled from the contributions analytic
in $s$, $t$ and $u$, that give rise to the local effective action.
The expression for the compactified amplitude depends on $r$ as well as on
the nine-dimensional Mandelstam invariants.
Since T-duality equates the IIA theory at radius $r$ to the IIB theory
at radius $1/r$ and the four-particle genus-one amplitude of the IIA and
 IIB theories are equal,
the compactified genus-one amplitude is invariant under $r\to 1/r$.
The coefficient of a generic term of order $s^k\, \hR^4$
in the analytic part of the compactified genus-one amplitude
contains powers ranging from $r^{2k-1}$ to $r^{1-2k}$
 plus exponential terms $O(\exp(-r))$.  The
term linear in $r$ survives the ten-dimensional limit $r\to \infty$.
The infinite series of terms
 proportional to  $r^{2k-1}\,s^k$ for $k>1$, diverge in the ten-dimensional $r\to \infty$ limit.
  However,
 these can be resummed and thereby convert the nine-dimensional normal thresholds, which contain
 half-integer powers of the Mandelstam invariants,  into the ten-dimensional thresholds  containing
  factors of $\log (-{\alpha'}\,s)$, $\log (-{\alpha'}\,t)$ and $\log (-{\alpha'}\,u)$  \cite{grv:bigpaper,gkv:twoloop}.
 In addition there are terms of the form $s^k\, \log(r^2)$, which are analytic in $s$
 but not in $r$.
  The many coefficients of the low-energy expansion
 of the nine-dimensional and ten-dimensional  genus-one amplitude that we determine
have the form of rational numbers multiplying products of Riemann zeta values.
These nine-dimensional results are summarized in section~\ref{sec:nineD}.

\section{The general structure of the genus-one four-particle amplitude}\label{sec:expand}


Here we  review the general structure of the genus-one
amplitude in ten-dimensional Minkowski space and its $\alpha'$ expansion.
The dynamical factor $F$  in (\ref{eLoop}) is given by an integral over the positions
$\nu^{(i)} = \nu^{(i)}_1 + i \nu^{(i)}_2$ of the four vertex operators on the torus,
\bea
\label{eF}
F(s,t,u;\tau)& = & \prod_{i=1}^3 \int_{\cal T}
{d^2\nu^{(i)}\over \tau_2}
\left(\chi_{12}\chi_{34}\right)^{\alpha' s}
\left(\chi_{14}\chi_{23} \right)^{\alpha' t}
\left(\chi_{13}\chi_{24} \right)^{\alpha'u}
\nonumber\\
 &=& \int_{\cal T}
\prod_{i=1}^3 {d^2\nu^{(i)}\over \tau_2} \ e^\calD = \int_{\calT}
\prod_{i=1}^3 {d^2\nu^{(i)}\over \tau_2} \ \exp(\alpha's\Delta_s
+\alpha' t \Delta_t + \alpha' u\Delta_u),
\eea
where  $d^2\nu^{(i)} \equiv d\nu^{(i)}_1 d\nu^{(i)}_2$,
$\nu^{(4)}=\tau$,  and
\be\label{ppdef}
\calD =\alpha's\Delta_s
+\alpha' t \Delta_t + \alpha' u\Delta_u,
\ee
with
 \be
\label{deldefs}
\Delta_s =  \log (\chi_{12}\chi_{34}), \qquad
\Delta_t =  \log(\chi_{14}\chi_{23}),\qquad  \Delta_u = \log
(\chi_{13}\chi_{24})
\ee
and $\log\chi_{ij}\equiv \log \calP(\nu^{(i)}-\nu^{(j)}|\tau)$ where
\be
\label{propmom}
\calP(\nu|\tau)
=-{1\over4}\left|\theta_{1}(\nu|\tau)\over \theta_{1}'(0|\tau)\right|^2+{\pi\nu_{2}^2\over2\tau_{2}}\ ,
\ee
is the scalar Green function on the
torus. These Green functions are integrated over the
domain $\calT$ defined by
 \be
\label{dimar}
\calT=\{-{1\over 2}\leq \nu_1 < {1\over 2},\; 0\leq
 \nu_2 < \tau_2\}
\ee

Since $I_{nonan}$ has logarithmic branch points associated with thresholds for intermediate
on-shell massless states,  it
has singularities in $s$, $t$ and $u$ that must be extracted from the
complete expression before the analytic terms can be
determined. These thresholds arise from the region of moduli
space in which $\tau_2\to \infty$, which is the degeneration
limit of the torus.  Our procedure for separating the threshold term will be to  treat
the region with $\tau_2\ge L$ separately from the region $\tau_2\le L$, where $L\gg 1$.
In other words, we write the integral over the fundamental
domain as the sum of two terms,
\be
\label{taucut}
I(s,t,u)  = I_{\calF_L}(L; s,t,u) + I_{\calR_L}(L;s,t,u)\,,
\ee
where
\be
 I_{\calF_L}(L; s,t,u) = \int_{\calF_L} {d^2\tau\over \tau_2^2}\,
F(s,t,u;\tau)\, ,\qquad   I_{\calR_L}(L;s,t,u) =
\int_{\calR_L} {d^2\tau\over \tau_2^2}\,  F(s,t,u;\tau)\,.
\label{calfrdef}
\ee
The first term on the right-hand side is integrated over $\calF_L$, which
 is the fundamental domain cut off at $\tau_2\leq L$, and the second
 over $R_L$, which  is the semi-infinite
rectangular domain $\tau_2\ge L$, $-\half\le \tau_1 \le \half$.
Clearly, the dependence on $L$ cancels from the full
amplitude.
The first term contains the analytic part of the amplitude, together
with an $L$-dependent piece, which is also analytic in the Mandelstam invariants,
\be
 I_{\calF_L}(L; s,t,u) = I_{an}(s,t,u) + R(L; s,t,u)\,.
\label{anfl}
\ee
The $L$-dependence is contained in the function $R(L;s,t,u)$,
which has an expansion of the form
 \be
R(L;s,t,u) = \sum_{r} (d^r_1 L^{r-1} + d^r_2 L^{r-3} + \cdots+
d^r_{r/2+1}\log(L/\mu_r))\, s^r + \cdots\,,
\label{pformm}
\ee
where $\dots$ denotes terms involving $t$ and $u$.  The second term on
the right-hand side of
(\ref{taucut}) contains the nonanalytic piece of $I$, together with an
$L$-dependent piece that cancels the $L$-dependence of the first term,
\be
 I_{\calR_L}(L;s,t,u) = I_{nonan}(s,t,u) -  R(L; s,t,u)\,.
\label{nonanfl}
\ee
 The integrand of $I_{R_L}$ can be evaluated by substituting
the large-$\tau_2$ approximation and will give rise to the nonanalytic pieces,
 whereas the integral over $\calF_L$
contains purely analytic pieces.

The low energy expansion of the analytic part, $I_{\calF_L}$, in
(\ref{calfrdef}) is  obtained by  expanding the integrand $F(\tau,\bar\tau)$ in (\ref{eF}) in
powers of the scalar Feynman propagator and then integrating
over the toroidal world-sheet.  However, in order to treat the
thresholds consistently we will separate them by dividing the integral into the two pieces
given in (\ref{taucut}).  The resulting expansion of the analytic
piece is contained in
\bea
\label{anloop}
 I_{\calF_L}(L; s,t,u)&=& I_{an}(s,t,u) + R(L;s,t,u)\nn\\
& =&
   \sum_{n=0}^\infty
  \int_{\calF_L} {d^2\tau\over \tau_2^2} \int_{\calT}
\prod_{i=1}^3 {d^2\nu^{(i)}\over \tau_2} \,{1\over n!}  \calD^n\, .
\eea
The quantity $\calD$ is the linear combination,
\bea
\calD &=&\alpha' s(\calP(\nu^{(12)}|\tau) +\calP(\nu^{(34)}|\tau)-
\calP(\nu^{(13)}|\tau) -\calP(\nu^{(24)}|\tau))\nn\\
&&  +
\alpha' t (\calP(\nu^{(14)}|\tau) +\calP(\nu^{(23)}|\tau)-
\calP(\nu^{(13)}|\tau) -\calP(\nu^{(24)}|\tau))
\,,
\label{propexp}
\eea
 where $\calP(\nu^{(ij)}|\tau)$ is defined in (\ref{propmom}) and
 can be written as \cite{gv:stringloop}
\be
\calP(\nu|\tau)=   {1\over 4\pi} \sum_{(m,n)\neq(0,0)} {\tau_2\over |m\tau+n|^2} \,
    \exp\left[{2\pi i\over \tau_2} (m\nu_1\tau_2 - (m\tau_1 +n) \nu_2)\right] + C(\tau,\bar\tau)\, ,
\label{torprop} \ee
where $C(\tau,\bar\tau)$ cancels out of the
$\SL(2,Z)$-invariant combination in (\ref{propexp}).
It can also be written as
\be
\label{StringProp}
\calP(\nu|\tau)\equiv \Pinf(\nu|\tau)+\P(\nu|\tau)\,
\ee
where  $\Pinf=\lim_{\tau_2\to\infty} \calP(\nu|\tau)$ is proportional to $\tau_2$
and is given by
\be
\Pinf(\nu|\tau) =
 {\tau_{2}\over4\pi} \sum_{n\neq0} {e^{2i\pi n \, \hat\nu_{2}}\over n^2}
={\pi\tau_{2}\over2}\, \left(  \hat\nu_{2}^2-| \hat\nu_{2}|+{1\over6}\right)\,,
\label{pinfdef}
\ee
where $ \hat\nu_{2}\equiv \nu_{2}/\tau_{2}$ and  the second expression is defined in the range $-\half \le  \hat\nu_{2}\le \half$ and
is periodically repeated outside this range. The quantity $\P(\nu|\tau)$ in (\ref{StringProp})
is given by
\be
\P(\nu|\tau) =  {1\over4} \sum_{m\neq0\atop k\in\ZZ}
{1\over|m|} e^{2i\pi m (k\tau_{1}+\nu_{1})} \, e^{-2\pi
\tau_{2}|m| |k-\hat\nu_{2}|}\,.
\label{ptildedef}\ee
The  decomposition (\ref{StringProp})
will be useful for expansions at large $\tau_2$.

One way of evaluating the coefficients $J^{(0,0)}, J^{(1,0)}, J^{(0,1)},\cdots$ is to consider
the derivatives of $I_{an}$ in the small $s$, $t$ and $u$ limit, as was done in \cite{gv:stringloop}.
Thus, the coefficients in the power series expansion
\be
I_{an}(s,t,u) = \sum_{m=0}^\infty\sum_{n=0}^\infty
I_{an}^{(m,n)}\,\left(\frac{\al}{4}\right)^{m+n}\,
\frac{ s^m t^n}{m!n!}\, ,
\label{mnderiv}
\ee
are given by
\bea
\label{genderiv}
I_{an}^{(m,n)}
 &=&  \int_\calF {d^2\tau\over \tau_2^2}\,  f^{(m,n)}_{an}(\tau,\bar\tau) \,  ,
\eea
where
\bea
f^{(m,n)}_{an}(\tau,\bar\tau) &=& \left.\frac{1}{m!n!}\,
\partial^m_s\partial_t^n \,F(s,t,u;\tau)\right|_{s=t=u=0} \nn\\
&=&\int_{\calT}   \prod_{i=1}^3 {d^2\nu^{(i)}\over \tau_2} \, (4\Delta_s-4\Delta_u)^m\,
  (4\Delta_t -4\Delta_u)^n
\label{fsedf}
\eea
with $\Delta_s$, $\Delta_t$ and
$\Delta_u$ are defined in (\ref{deldefs}).

Defining the quantity $j^{(p,q)}(\tau,\bar\tau)$
 to be the integrand of $J^{(p,q)}_{\calF_L}$, so that
\be
J^{(p,q)} = \int_{\calF}
\frac {d^2\tau}{\tau_2^2}\,j^{(p,q)}(\tau,\bar\tau)\,,
\label{integ}
\ee
and comparing
 (\ref{mnderiv}) with (\ref{iform}) leads to the implicit
relation between  the integrands
$f^{(m,n)}_{an}(\tau,\bar\tau)$ and $j^{(p,q)}(\tau,\bar\tau)$,
\be
\sum_{p,q =0}^\infty
\hat\sigma_2^p\hat\sigma_3^q\,j^{(p,q)}
= \sum_{m,n=0}^\infty \,\left(\frac{\al}{4}\right)^{m+n}\,
\frac{ s^m t^n}{m!n!}\, f^{(m,n)}_{an}\,.
 \label{relatei}
\ee
We note further that  (\ref{iform}) implies
\bea
\nn
J^{(0,0)} &=& I_{an}^{(0,0)}\,,\qquad J^{(1,0)} ={1\over 4}\,  I_{an}^{(2,0)}\,,
\qquad J^{(0,1)} = -{1\over 6}\, I_{an}^{(1,2)}\,,
\qquad J^{(2,0)} ={1\over 96}\,I_{an}^{(4,0)}\,,\nn\\
J^{(1,1)}& =& - {1\over 144}\,  I_{an}^{(4,1)}\,,
\qquad J^{(3,0)} = {1\over 2880}\, I_{an}^{(5,1)}\,,\qquad  J^{(0,2)} = {1\over 5760}\, I_{an}^{(6,0)}
\label{icoeffs}
\eea

\section{The analytic terms in the ten-dimensional genus-one amplitude}
\label{sec:amplitude}

In this section we will formulate the diagrammatic expansion for determining the coefficients,
$J^{(p,q)}$,
of the higher derivative terms starting from (\ref{mnderiv}).
This will then be used to determine the coefficients of terms up to ${\alpha'}^6\,s^6\,\hR^4$.
These coefficients are obtained by
integrating the $j^{(p,q)}$'s, which are linear combinations of the
$f^{(m,n)}_{an}$'s  that have $m+n=2p+3q$, and are defined in (\ref{fsedf}).  In other
words, the coefficients are given as integrals of products of powers
of propagators between different vertex operators attached to the torus.

A term at order $\al^r$   is represented by a sum of `Feynman' diagrams with a
 total of $r$ propagators
 joining pairs of vertices.  Any propagator can join any of six
 distinct pairs of vertices separated by $\nu^{(ij)} = \nu^{(i)} -\nu^{(j)}$ ($i,j =1,2,3,4$).
A diagram is therefore labeled by a set of six numbers,
\be
\{\ell\} = \{\ell_1, \ell_2,\dots ,\ell_6\}\,,\qquad \sum_{k=1}^6 l_k =
r\, ,
\label{setofl}
\ee
which define the number of propagators that join each pair of vertices.
The labeling of the diagram is indicated in the figures in appendix~\ref{sec:ZMdiag}.
Each diagram is associated with a scalar function $D_{\{\ell\}}(\tau,\bar\tau)$
that is determined by integrating
 the positions, $\nu^{(i)}$, using the  representation (\ref{torprop})
of the propagator, which is periodic in both $\nu^{(i)}_1$ and $\nu^{(i)}_2$,
leading to conservation of the torus momentum, ${\bf p}= m+n\tau\in{\mathbb Z}+\tau {\mathbb Z}$,
 at each vertex.
The result is a
 function that depends on the topology of the diagram -- i.e. on the
 $\ell_k$'s -- and is given at order $\al^r$ by
 \begin{equation}
 \D_{\{\ell\}}(\tau,\bar\tau)={\tau_{2}^r\over (4\pi)^r }
\sum_{{\bf p}_{1},\dots,{\bf p}_{r}\in{\mathbb Z}+\tau {\mathbb Z}}\,
{\prod_{i=1}^4\delta(\sum_{j\in I_i} {\bf p}_{j})\over |{\bf p}_{1}|^2
  \cdots |{\bf p}_{r}|^2}
\ ,
\label{ddefs}
\end{equation}
where the topology of the diagram is subsumed in the values of the sets $I_i$
with $i=1,2,3,4$. The momentum conservation $\delta$-function is understood to mean
\begin{equation}
\delta({\bf p}_{1}+ {\bf p}_{2})= \delta(m_{1}+m_{2})\, \delta(n_{1}+n_{2}) \, .
\end{equation}
This condition eliminates all one-particle reducible diagrams.
In appendix~\ref{sec:ZMdiag} we will derive some detailed
properties of the
$\D_{\{\ell\}}$'s that  contribute to the expansion  up to order  $ s^8\,\hR^4$.
The net result is that $j^{(p,q)}(\tau,\bar\tau)$ is a linear combination,
\begin{equation}\label{anal}
j^{(p,q)}(\tau,\bar\tau) = \sum_{\{\ell\}}e^{(p,q)}_{\{\ell\}}\, \D_{\{\ell\}}(\tau,\bar\tau)\,,
\end{equation}
where $e^{(p,q)}_{\{\ell\}}$ is a set of constant coefficients and the sum is over all diagrams with
$\sum_{k=1}^6 l_k = r$.

In order to evaluate the integral of $j^{(p,q)}(\tau,\bar\tau)$ in
(\ref{integ}) we would like to make use of a
theorem reviewed in \cite{Terras} and restated in
appendix~\ref{sec:fund}.  This theorem states that
any function that is square integrable on $\calF$ is the sum of three
terms: (i) a function whose zero mode with respect to $\tau_1$ vanishes;
(ii) a constant; (iii)  a linear combination of incomplete theta series'.  However,
$j^{(p,q)}(\tau,\bar\tau)$ is not square integrable as it stands, since
it is easy to see that it has a
large-$\tau_2$ expansion  of the form
\be
j^{(p,q)} =
a^0_{(p,q)}\, \tau_{2}^{2p+3q} + a^1_{(p,q)}\, \tau_{2}^{2p+3q-1}
 +\cdots  + a^{4p+6q-1}_{(p,q)} \,\tau_{2}^{1-2p-3q}+ O(\exp(-\tau_{2}))\,,
\label{posneg}
\ee
where $a^{2p+3q}_{(p,q)} \equiv J^{(p,q)}$.
We can, however, proceed by subtracting the positive powers of $\tau_{2}$ in a manner
consistent with modular invariance.
This can be achieved by subtracting a suitably chosen quadratic form in nonholomorphic
Eisenstein series\footnote{This quadratic form
may not be unique but any ambiguity is irrelevant in the following.},
\be
P^{(p,q)}(\{\hat E_{r}\})= \sum_{s,s'}d^{(p,q)}_{ss'}\, \hat E_s\, \hat E_{s'}\,,
\label{peosdef}
\ee
where $d^{(p,q)}_{ss'}$ are constant coefficients and
\begin{equation}
\hat E_{s}={1\over (4\pi)^s}\,
\sum_{(m,n)\neq(0,0)}\, {\tau_{2}^s\over |m+n\tau|^{2s}}=
{2\zeta(2s)\Gamma(s)\over \pi^{s/2}}\, \tau_2^s + {2\zeta(2s-1)\Gamma(s-\half)\over \pi^{s-1/2}}\,
\tau_2^{1-s} + O(e^{-2\pi \tau_2})\,
\end{equation}
(some properties of nonholomorphic Eisenstein series are reviewed in appendix~\ref{sec:Eisenstein}).
In other words we choose the polynomial, $P^{(p,q)}$, so that it reproduces
the terms with positive powers
of $\tau_2$  in (\ref{posneg}), so that
\be
j^{(p,q)} = P^{(p,q)}(\{\hat E_{r}\}) + J^{(p,q)}+ \delta j^{(p,q)}\,,
\label{subeisenst}
\ee
where $J^{(p,q)}$ is a constant and $\delta j^{(p,q)}\to0$ as
$\tau_2\to\infty$.
Since $\delta j^{(p,q)}$ is square integrable in the
fundamental domain ${\cal F}$,  the theorem applies to it, and since we also have
\be
\int_{-\half}^\half d\tau_1\,
\delta j^{(p,q)} = {  a^{2p+3q+1}_{(p,q)} \over \tau_{2}} + \cdots +  a^{4p+6q-1}_{(p,q)}
\,\tau_{2}^{1-2p-3q}+ O(\exp(-\tau_{2})) \ne 0\,,
 \ee
the theorem implies that
\be
\int_{{\cal F}} {d^2\tau\over\tau_{2}^2}\, \delta j^{(p,q)} =0\, .
\label{theor}
\ee
Therefore, after cutting off the fundamental domain at large $\tau_{2}=L$ we have
\be
J^{(p,q)}_{\calF_L} =
\int_{{\cal F}_{L}} {d^2\tau\over\tau_{2}^2}\, j^{(p,q)} ={\pi\over3}\, J^{(p,q)}
+ \int_{{\cal F}_{L}}{d^2\tau\over\tau_{2}^2}\, P^{(p,q)}(\{\hat E_{2}\})+
O(1/L)\,
\label{jflint}
\ee
(where we have used the fact that $\int_{\calF}d^2\tau/\tau_2^2 =
\pi/3$).

The function $P^{(p,q)}$ and constants $J^{(p,q)}$
 that arise up to order order $s^6\,\hR^4$ are given in
 (\ref{beuno})-(\ref{segona}) in  appendix~\ref{sec:Expand}. The integral of
$P^{(p,q)}(\{E_r\})$ over the fundamental domain can be reduced to a
boundary integral by
using $\Delta\hat E_s =s (s-1)\,  \hat E_s$ and integrating by parts, as is reviewed
in appendix~\ref{sec:Eisenstein}.
This leads to the following results.
\begin{itemize}
\item At order $\hR^4$.  In this case $j^{(0,0)}=\D_{0}=1$, so $J^{(0,0)}=\pi/3$, as it is given by the volume of a fundamental domain for $\SL(2,Z)$.
\item At order  $\al^2\,s^2\, \hR^4$ it is easy to see that $j^{(1,0)} = \D_2=\hat E_2$
(as was found in \cite{gv:stringloop}).  Substituting $ \Delta\hat E_2 =2\,\hat E_2$ for $P^{(1,0)}$ in
the right-hand side of  (\ref{jflint}) gives a purely boundary contribution proportional to $L$
\be
J^{(1,0)}_{\calF_L}=L \,.
\label{stwor}
\ee
This $L$ dependence is canceled by the contribution from the region $\tau_{2}\geq L$,
so the result is
\be
J^{(1,0)} =0\,.
\label{onezero}
\ee
\item At order $ s^3\,\hR^4$, using (\ref{beuno}) results in
\be
 J^{(0,1)}_{\calF_L}= {\pi\over3}\, {\zeta(3)\over3}+ {\pi^3\over 567}\, L^2+O(1/L)\,,
\ee
which reproduces the result in \cite{gv:stringloop}.
The $L^2$ dependence is canceled by the contribution of the amplitude from the region $\tau_{2}\geq L$,
which was given in (3.24) of \cite{gv:stringloop} and has the form
\be
\lim_{s,t\to 0} I_{nonan\, (3);\calR_L}(s,t) = -{\pi^3\over 567}\, L^2\, \hat\sigma_3 +O(1/L)\, .
\ee
At this order there are no threshold contributions so $I_{nonan\, (3)}=0$,
and the coefficient multiplying $s^3\, \hR^4$
in the ten-dimensional effective action is
\be\label{e:c}
J^{(0,1)} =   {\pi\over3}\, {\zeta(3)\over3}\ .
\ee

\item At order $\al^4\,s^4\, \hR^4$, using (\ref{bedos}) and (\ref{asiete}) we find
\be\label{esd}\begin{split}
 J^{(2,0)}_{\calF_L}&={4\pi^4\over 42525}\, L^3+ {2\pi\over45}\,
\zeta(3)\,\log(L/\tilde \nA_4) +O(1/L)  \ ,
\end{split}\ee
where the scale $\tilde\nA_4 $ is given by
\be
\log \tilde\nA_4 = -{1\over 2}+\log(2)-{\zeta'(3)\over\zeta(3)}+{\zeta'(4)\over \zeta(4)}\ .
\label{swala}
\ee
The occurrence of the $\log L$ term originates from the presence of $\hat E^2_{2}$ in the expression for
$j^{(2,0)}$ in (\ref{bedos}), together with the use of the integral (\ref{CutIn}).
In section~\ref{sec:massthreshold} the $L^3$ and $\log(L)$ contributions will be shown to cancel with
 the contribution from $\tau_{2}\geq L$. In particular, we will see that
\be
\lim_{s,t\to 0} I_{nonan\, (4);\calR_L}(s,t)=   I_{nonan\, (4)}(s,t)
- \left({4\pi^4\over 42525}\, L^3 + \frac{2\pi}{45 }\,
\zeta(3)\, \log(L/\tilde \mu_4)\right) \hat\sigma_2^2 \,,
\label{nonamfo}
\ee
where
\be
\label{nonanfour}
I_{nonan\, (4)}(s,t) = -\frac{4\zeta(3)}{45 \pi}\,{\alpha'}^4\, s^4\log(-\al s/\nA_{4}) + perms\,.
\ee
 We will derive  the coefficient of this term (which was also derived from the unitarity argument
 in the last subsection), as well as the value of $\nA_4$, in
 section~\ref{sec:massthreshold}.
Since there is no constant $L$-independent term in (\ref{esd})
(apart from the $\log \tilde \nA_4$
associated with the $\log L$) we find
\be
J^{(2,0)}=0\, .
\label{e:d}
\ee

\item At order $\al^5\, s^5\, \hR^4$, using (\ref{betres}), we find
\be
J^{(1,1)}_{\calF_L}={\pi\over3}\, {97\over1080}\, \zeta(5)
+\frac{L^4\,{\pi }^5}{400950} + \frac{L\,{\pi }^2\,\zeta(3)}{378}+O(1/L)\ .
\label{sfive}
\ee
The power-behaved $L^4$ and $L$ contributions will again be seen to cancel the contribution from
the $\tau_{2}\geq L$ region,
\be
\lim_{s,t\to 0}  I_{nonan\, (5);\calR_L}(s,t)=
-\left(\frac{L^4\,{\pi }^5}{400950} -\frac{L\,{\pi }^2\,\zeta(3)}{378}+O(1/L)\right)\hat\sigma_2\hat\sigma_3\, ,
\ee
 but there is no $\hat\sigma_2\hat\sigma_3\, \log \,L$ term.
This is in accord with the earlier unitarity argument, which shows that there is no
threshold at the order $s^5$
so $I_{nonan\, (5)}(s,t)=0$.
The genus-one contribution to the ten-dimensional action at this order is given by the
constant term in (\ref{sfive}),
\be\label{e:e}
J^{(1,1)} =  {\pi\over 3}\, {97\over 1080}\, \zeta(5)\, .
\ee

\item At order $\al^6 s^6\, \hR^4$ there are two independent tensorial structures.
For $\sg_{2}^3\,\hR^4$,
using (\ref{becuatro}) and (\ref{aocho}), we find
\be\label{jthreezero}\begin{split}
J^{(3,0)}_{\calF_L}&={\pi\over3}\, {\zeta(3)^2\over30}
\\
&+{2\,\pi ^6\over 4729725}\,L^5 + \frac{2\,\pi^3\,\zeta(3)}{4725} \,L^2+
  \frac{11\,\pi  }{630}\,\log (L/\tilde\nA_6)\,\zeta(5) +O(1/L)\,.
\end{split}\ee
For $\sg_{3}^2\,\hR^4$, using~(\ref{segona}) and~(\ref{aocho}), we find
\be\label{jtwozero}\begin{split}
 J^{(0,2)}_{\calF_L} &={\pi\over3}\, {61\zeta(3)^2\over1080}
\\
&+{1744\,\pi ^6\over 3192564375}\,L^5 + \frac{8\,\pi^3\,\zeta(3)}{14175} \,L^2+
  \frac{\pi  }{45}\,\log (L/\tilde\nA_6)\,\zeta(5) +O(1/L)\,.
\end{split}
\ee
The scale $\tilde\nA_6$ in both of these equations is given by
\be
\log \tilde\nA_6 = - {7\over12}+ \log(2)-{\zeta'(5)\over \zeta(5)}+{\zeta'(6)\over\zeta(6)}\,.
\label{scalesix}
\ee
In this case the $\log L$ terms originate from  $\hat E_3^2$ terms in $j^{(3,0)}$ and
$j^{(0,2)}$, together with the integral in   (\ref{CutIn}).
The $L$-dependent terms in these equations will again be seen to cancel with contributions from
the $\tau_2 \ge L$ part of the integral, resulting in a net nonanalytic term of the form
$h^{(6)}(s,t,u)\, \log(-\alpha'\,s/\nA_6) + perms$, where
$h^{(6)}$ is a monomial in $s$, $t$ and $u$ of order six (defined by (\ref{nonansix}).
We see from (\ref{jthreezero}) that
the coefficient of the analytic  $\hat\sigma_{2}^3\,\hR^4$  term is
\be\label{e:f}
J^{(3,0)} = {\pi\over3}\, {\zeta(3)^2\over30}\, ,
\ee
while from $(\ref{jtwozero})$ we see that the coefficient of $\hat \sigma_{3}^2\, \hR^4$ is
\be\label{e:g}
J^{(0,2)} = {\pi\over3}\, {61\zeta(3)^2\over1080}\, .
\ee
\end{itemize}

\section{Ten-dimensional threshold terms}\label{sec:threshold}

 \begin{figure}
\centering\includegraphics[width=12cm]{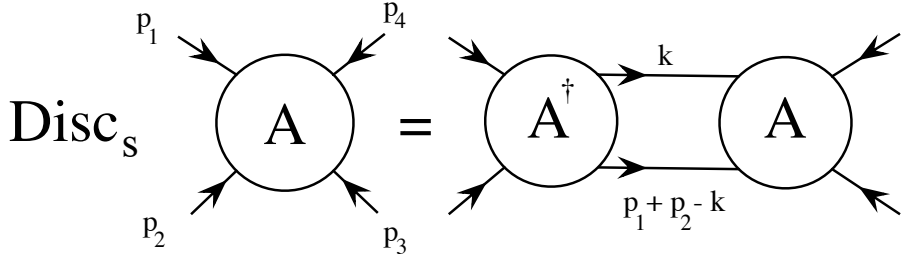}
\caption{Unitarity equation relating the two-particle $s$-channel discontinuity of the amplitude
to the square of the four-particle amplitude, integrated over the phase space of the intermediate particles
and summed over the species of these particles.}
\label{fig:Unitarity}
\end{figure}

We now turn to discuss the nonanalytic terms in the low-energy expansion of the ten-dimensional
amplitude. These are characterized by branch cuts with imaginary parts that are determined by unitarity in
a standard manner that we will review in the following subsection.
The scales of the logarithms are more difficult to evaluate and will
require a direct calculation of the genus-one integral in the region $\tau_2 \ge L$
section \ref{sec:lowthresh}.

\subsection{Two-particle unitarity}
\label{sec:twounitarity}

Unitarity identifies the discontinuity of the amplitude across the $s$-channel two-particle
cut (when $s\ge 0$, $t\le 0$)
with the product two amplitudes integrated over the phase space for the intermediate massless
two-particle states and summed over the superhelicities of these states (as illustrated in
figure \ref{fig:Unitarity}).

Our normalisations for the string S-matrix
 are such that
the tree-level and genus-one terms in the amplitude enter in the combination\footnote{In the $S$-matrix there is a power is $g_{s}$
for each external state. Here we are considering the four point amplitude, leading to an overall factor of $g_{s}^4$.}
\be
\hA=  \kappa_{(10)}^2\, g_{s}^4\,\left({1\over g_s^2}  \hA^{tree}
+ 2\pi A^{genus-1}+ O(g_s^2)\right)\, ,
\ee
where $ \kappa_{(10)}^2 =  2^6\pi^7 \, \al^4$ and  $\hA^{tree}$ has the form shown in (\ref{treesumm})
and $\hA^{genus-1}= \hA^{genus-1}_{an}+\hA^{genus-1}_{nonan}$.
At lowest order in the string coupling constant the unitarity relation
has the form
\be\begin{split}
{\rm Disc_s} \, \hA_\ze^{genus-1}(p_1,p_2,p_3,p_4) &=  -i \,{\kappa_{(10)}^2\over\alpha'} \,{\pi\over2}
\int {d^{10} k\over (2\pi)^{10}} \,\delta^{(+)}(k^2)\,\delta^{(+)}((q-k)^2)\\
&
\sum_{\{\zeta_r,\zeta_{s}\}} \hA_{\zeta_1\zeta_2\zeta_r\zeta_s}^{tree}(p_1,p_2,-k,k-q)\,
\hA_{\zeta_3\zeta_4\zeta_r\zeta_s}^{tree}(p_3,p_4,k,q-k)\,,
\label{unitarityt}
\end{split}\ee
where $\sum_{\{\zeta_r,\zeta_s\}}$ denotes the sum over all the two-particle
massless $N=8 $ supergravity states, and $\delta^{(+)}(p^2)\equiv \delta^{(10)}(p^2)\theta(p^0)$ imposes
the mass-shell condition on each intermediate massless state,
\be
k^2 = 0\,,\qquad (q-k)^2 =0\,.
\label{massshell}
\ee
Expanding both sides of (\ref{unitarityt})
 in powers of $\al$ determines the discontinuity of the genus-one
amplitude in terms of the square of the terms in the tree-level expansion.

The discontinuity of the genus-one amplitude is obtained by substituting the tree-level scattering
amplitude into (\ref{unitarityt}).  Recall
that this amplitude has  the form \cite{Green:1987sp}
\begin{equation}
\hA_\ze^{tree}(p_1, p_2, p_3, p_4)=C(s,t,u)\,
\hR_\ze^4(p_{1},p_{2},p_{3},p_{4}) \, ,
\label{treeampo}
\end{equation}
where $s,t,u$ are Mandelstam invariants, satisfying the mass-shell condition $s+t+u=0$, and
\be
C(s,t,u)= - \frac{\Gamma\left(-\frac{\al s}{4}\right) \Gamma\left(-\frac{\al t}{4}\right)
\Gamma\left(-\frac{\al u}{4}\right)}{\Gamma\left(1+\frac{\al s}{4}\right)
\Gamma\left(1+\frac{\al t}{4}\right)
\Gamma\left(1+\frac{\al u}{4}\right)}\,,
\label{Cdef}
\ee
The  unitarity relation takes a very special form in maximal supergravity (as it does in maximal
Yang--Mills), because of the self-replicating relation derived in
\cite{Bern:1998ug},
\be
\sum_{\{\zeta_r,\zeta_s\}} \hR_{\zeta_1\zeta_2\zeta_r\zeta_s}^4(p_1,p_2,k-q,-k)\,
\hR_{\zeta_3\zeta_4\zeta_r\zeta_s}^4(k,
q-k, p_3, p_4) =  s^4\,\hR_\ze^4(p_1,p_2,p_3,p_4)
\label{replice}
\ee
($q=p_1+p_2$), which simplifies the left-hand side of (\ref{unitarityt}) drastically.

The low-momentum expansion of (\ref{unitarityt}) follows by substituting the expansion of
$\hA^{tree}$ in
the right-hand side (see, for  example, \cite{gkv:twoloop}),
\bea
 \hA^{tree}&=&
{3\over \sg_{3}}\, \exp\left(-\sum_{n=1}^{\infty} {2\zeta(2n+1)\over 2n+1}
\, \sg_{2n+1}\right)\, \hR^4\nn\\
 &=&  \left( \frac{3}{\hat\sigma_3}+ 2\zeta(3)+
\zeta(5)\sg_2+{2\over 3}\zeta(3)^2\sg_3+
{1\over 2}\zeta(7)\sg_2^2+{2\over 3}\zeta(3)\zeta(5)\sg_2\sg_3\right.
\nn\\
&+&
\left. {1\over 4}\zeta(9)  \sg_2^3 +{2\over
  27}\left(2\zeta(3)^3+\zeta(9)\right)\sg_3^2
  +\cdots\right)\, \hR^4 \, .
\label{treesumm}
\eea
We see that the left-hand side of (\ref{replice}) is identified with
the lowest order term in the integrand on the right-hand side of
(\ref{unitarityt}) since $\hA^{tree} \sim \hR^4/stu$.   This means that, to lowest order in
$\al$, ${\rm Disc_s}\, \hA^{genus-1}$ is given by
\be
{\rm Disc_s}\, \hA^{genus-1}(p_1,p_2,p_3,p_4) =
 -is^2\,{\kappa_{(10)}^2\over\alpha'}\,{\pi\over2}\, \hR^4\,
\int {d^{10} k\over (2\pi)^{10}}\,\frac{\delta^{(+)}(k^2)\,\delta^{(+)}((q-k)^2)}{(p_1 - k)^2\, (p_4+k)^2\,
(p_2-k)^2\, (p_3+k)^2}\,,
\label{loword}
\ee
where we have used the expressions for the Mandelstam invariants of the tree amplitudes on either
side of the intermediate states,
\be
t' = -(p_1-k)^2\,,\qquad u'=-(p_2-k)^2\,, \qquad t^{\prime\prime} = -(p_4+k)^2\,,\qquad
u^{\prime\prime} = -(p_3+k)^2\,.
\label{mandprimes}
\ee
This reproduces the $s$-channel discontinuity of the massless box diagram, which is of order $s$.
In section~\ref{sec:masslessbox}
we will find the complete expression for the genus-one contribution, which is the
supergravity contribution, $\hA_{SUGRA}$.

The next term in the $\al$ expansion is obtained by substituting $2\zeta(3)\,{\alpha'}^3\,\hR^4$
in one of the factors of $\hA^{tree}$
on the right-hand side of the unitarity relation and the lowest-order term in
the other.  This amounts
to multiplying the  right-hand side of (\ref{replice}) by $2\zeta(3)\,{\alpha'}^3\, st'u'$ and so the expression for the
discontinuity at this order
is obtained by multiplying the integrand of (\ref{loword}) by the same factor.
The integral is proportional to $\zeta(3)\,\alpha'\,s\, S(0,0,1,1)$ that is evaluated in
appendix~\ref{sec:unitaritop}, giving
\be
{\rm Disc_s} \hA_{(4)}^{genus-1}(p_1,p_2,p_3,p_4) =-2i\pi\,\frac{-4\pi\zeta(3)}{45}\,
\left(\alpha'\,s\over4\right)^4\, \hR^4\,,
\label{fouruni}
\ee
with similar expressions for the $t$-channel and $u$-channel discontinuities.
This implies that $ \hA_{(4)}$ has the form
\be
\hA_{(4)}^{genus-1}(p_1,p_2,p_3,p_4) = -\frac{4\pi\zeta(3)}{45}\,\left({\alpha'}\over4\right)^4\,
\left(s^4\, \log(-{\alpha'\,s\over\mu_{4}})
+t^4\, \log(-{\alpha'\,t\over\mu_{4}})
+u^4\, \log(-{\alpha'\,u\over\mu_{4}})\right) \hR^4\,,
\label{fourunires}
\ee
where the scale, $\nA_4$, inside the logarithm is yet not determined --- this must await the more detailed
  analysis of the amplitude in section~\ref{sec:massthreshold}.

     There are no contributions to the discontinuity
  at order $(\alpha'\,s)^5$ since there are no terms in
  $\hA^{tree}$ of order $\alpha'\,s \, \hR^4$.  The next contribution is a
  discontinuity of order $\al^6\,s^6\, \hR^4$,
  obtained by expanding one of the
  $\hA^{tree}$
  factors on the right-hand side of (\ref{unitarityt})
  to the next non-trivial order, which means substituting the $\zeta(5)\,\hat\sigma'_2 \hR^4$
  term of
  (\ref{treesumm}) into (\ref{unitarityt}).  The resulting integral is proportional to
  $\zeta(5)\,(2\alpha'\, s\,S(0,0,3,1)+(\alpha'\,s)^3\, S(0,0,1,1))$
that is also evaluated in  appendix~\ref{sec:unitaritop}, giving
\be
{\rm Disc_s} \hA_{(6)}^{genus-1}(p_1,p_2,p_3,p_4) =-2i\pi\,{-\pi\,\zeta(5)\over 2520}
\,\left(\alpha'\over4\right)^6\,(87
\, s^6+s^4\, (t-u)^2)\,\hR^4\,,
\ee
which is attributed to a function of the form
\begin{equation}\label{nonansix}\begin{split}
\hA_{(6)}^{genus-1}(p_1,p_2,p_3,p_4)& =-{\pi\,\zeta(5)\over 2520}\,\left(\alpha'\over4\right)^6\,\Big( (87
\, s^6+s^4\, (t-u)^2)\,
\log(-{\alpha'\,s\over\mu_{6}})\\
&+(87 \, t^6+ t^4\,(s-u)^2)\, \log(-{\alpha'\,t\over\mu_{6}})
+(87 \,u^6+u^4\, (s-t)^2)\, \log(-{\alpha'\,u\over\mu_{6}})\Big)\, \hR^4\, ,
\end{split}\end{equation}
where $t$-channel and $u$-channel contributions have again be added.  The scale $\mu_6$
can again, in principle, be determined by explicit evaluation of the loop amplitude as in
section~\ref{sec:massthreshold} (although in this case we will not complete the evaluation).

At order ${\alpha'}^7\,s^7 \, \log(-\alpha'\,s)$, and beyond, contributions of a new type arise.
These come from the presence
of  higher-order terms in both factors of $\hA^{tree}$ in the unitarity equation.  They correspond
to terms in which there are stringy corrections to both propagators in the $t'$ and $t^{\prime\prime}$
channels.  For example, the total $(\alpha'\,s)^7$ contribution is proportional to the sum,
 $\zeta(3)^2\,(\alpha'\,s)^2\,(2 S(0,0,2,2) +S(1,1,1,1))$ in the notation of appendix~\ref{sec:unitaritop}.

For future reference
  it is interesting to note that
  the overall coefficient of $\log\mu_{4}$ in (\ref{e:s4cut}) is
\be
- \frac{\pi\zeta(3)}{45}\, \hat\sigma_2^2\,,
\label{mufoure}
\ee
while the overall coefficient of  $\log\mu_{6}$ in (\ref{e:s6cut}) is
\begin{equation}
-{\pi\zeta(5)\over630}\, (11 \hat\sigma_{2}^3+14\,\hat\sigma_{3}^2) \, \log\mu_{6}\,.
\end{equation}
These are the same as the coefficients of the
$r\, s^4\, \log(r^2)$, $s^6\,\log(r^2)$ and $s^6\,\log(r^2)$ terms
that arise in the compactification on a circle of radius $r$ to be
considered in section~\ref{sec:ninedim} and summarized in (\ref{fin}).
The necessity for such a cancelation is discussed following (\ref{scalel}).

\subsection{Low-momentum expansion of threshold terms}
\label{sec:lowthresh}

The constant scales inside the logarithms are not determined by
unitarity, but  by a direct calculation of the amplitude in the
large-$\tau_2$ region, which we turn to next.

 \begin{figure}
\centering\includegraphics[width=15cm]{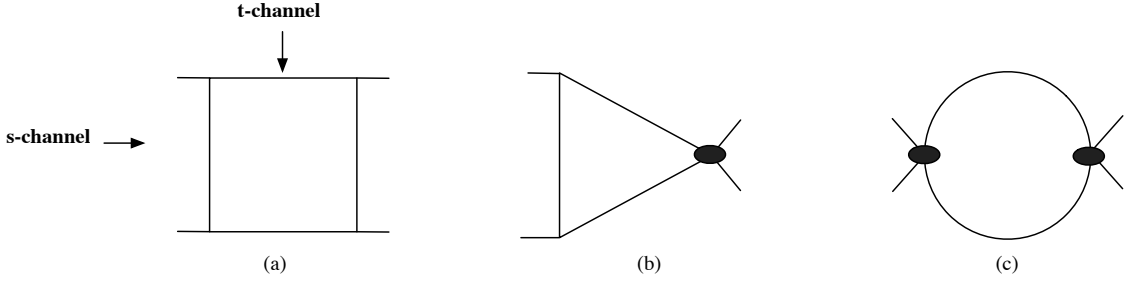}
\caption{Degeneration limits of the genus-one string amplitude. (a)  The limit in which all internal legs
are massless propagators.  This is the limit that gives the one-loop ten-dimensional
supergravity  contribution.
(b) The triangle limit, in which there are three internal massless propagators.
(c) The bubble limit in which two massless propagators are picked out.
The blobs in (a) and (b) represent higher derivative contact interactions induced
by stringy corrections.}
\label{fig:Threshold}\end{figure}

The singularities due to massless two-particle thresholds arise at genus one from the degeneration of
the torus in the limit  $\tau_{2}\to\infty$.
  As indicated in figure~\ref{fig:Threshold}(a), the zeroth order
contribution is simply the type II supergravity box diagram with massless states in all four internal lines,
where the threshold behaviour of the S-matrix has the form
$a_1\, {\alpha'}\,s\, \log (-\alpha'\,s/\nA_1)\,\hR^4$.
The constant $\nA_1$ cancels out using $s+t+u=0$.
Figure~\ref{fig:Threshold}(b) shows the discontinuity receives higher
derivative corrections on one side of the cut, due to $\hR^4$,
$s^3\,\hR^4$, and higher-order contact interactions indicated by the blob.
These correspond to threshold terms of the form ${\alpha'}^4\,s^4\, \log (-\alpha'\,s/\nA_4)$,
${\alpha'}^6\,s^6\, \log (-\alpha'\,s/\nA_6)$,
${\alpha'}^7\,s^7\, \log (-\alpha'\,s/\nA_7)$, $\dots$.
There are also corrections
with higher-order contributions on both sides of the cut, as seen in figure~\ref{fig:Threshold}(c).
These lead to higher-order threshold terms.

An important and well-known technical problem in the analysis of these threshold contributions
is that the integral representation (\ref{eLoop}) is not well defined for any values
of the Mandelstam invariants since the branch points in the $s$, $t$ and $u$ channels coincide.  This is
remedied by splitting the integration range of the vertex operator insertion points  into three regions
${\cal T}_{st}$, ${\cal T}_{us}$ and ${\cal T}_{tu}$,
that generate cuts in the $(s,t)$, the $(s,u)$ and $(t,u)$ channels, respectively.
The integrand in the ${\cal T}_{st}$ region is real for $s,t< 0$ (so that $u>0$) and may be defined
in the physical $s$-channel region ($s>0$, $t<0$) by analytic continuation.  Similarly, in the
${\cal T}_{us}$ region the integrand is defined for $s,u <0$, while in the ${\cal T}_{tu}$ region
it is defined for $t,u<0$.  Further discussion of the integral representation of the genus-one
expression may be found in \cite{D'Hoker:1994yr}.

\subsubsection{The massless supergravity amplitude}
\label{sec:masslessbox}
In the limit $\tau_2\to \infty$ the leading contribution
in the ${\cal T}_{st}$ region is given by the integral \cite{gkv:twoloop}
\begin{equation}\label{e:Threshold}
I_{{\cal T}_{st}}= \int_{L}^{\infty} {d\tau_{2}\over\tau_{2}^2}\,
\int_{{\cal T}_{st}} \prod_{i=1}^3 d\omega_{i}\,
e^{\alpha' \pi\tau_{2}\, Q(s,t)}\ ,
\end{equation}
where
${\cal T}_{st}= \{0\leq \omega_{1}\leq\omega_{2}\leq \omega_{3}\leq 1\}$, with
 $\omega_{i}=\nu_2^{(i)}/\tau_{2}$.
 The expression for  $Q(s,t)$
 \begin{equation}
 Q(s,t)=s\omega_{1}(\omega_{3}-\omega_{2})+t (\omega_{2}-\omega_{1})(1-\omega_{3})\,,
 \end{equation}
 arises by taking the asymptotic limit of the propagators, $\lim_{\tau_2\to\infty}
 \Delta_{s}\equiv \Delta^\infty_s =   \hat{\cal P}^\infty
 (\nu^{(12)})+ \hat{\cal P}^{\infty}(\nu^{(34)})$, in the
 expression~(\ref{eF}) for the one-loop amplitude (with $s$ and $t$
 negative).   We need to add the contribution to the amplitude in the regions
 ${\cal T}_{us}=\{0\leq \omega_{2}\leq\omega_{1}\leq \omega_{3}\leq 1\}$ with
 $Q(s,u)=s (\omega_{3}-\omega_{1})\omega_{2}+ u (\omega_{1}-\omega_{2})(1-\omega_{3})$
 and ${\cal T}_{tu}=\{0\leq \omega_{1}\leq\omega_{3}\leq \omega_{2}\leq 1\}$
 with $Q(t,u)=u (\omega_{2}-\omega_{1})\omega_{1}+ t (\omega_{3}-\omega_{1})(1-\omega_{2})$.
This is equivalent to evaluating the sum of the supergravity box diagrams
figure~\ref{fig:Threshold}(a) in the $(s,t)$, $(s,u)$
and $(t,u)$ channels.

It is very complicated to evaluate the detailed form of this term using the
cutoff at $\tau_2\geq L$, so we will review the method used in \cite{Green:1982sw} for
evaluating the integral using dimensional
regularization.  This will generally give a different definition of the
scale, $\nA_1$, in terms of the form $s\, \log (-\alpha'\,s/\nA_1)$.
However, in this case the scale cancels out (using $s+t+u=0$)
and so the result is identical.
In $D-2\epsilon$ dimensions, the one-loop amplitude becomes\footnote{We have made use of the
change of variables $w_{1}=\eta \xi_{1}$, $w_{2}=(1-\eta)(1-\xi_{2})+\eta\xi_{1}$
and $w_{3}=1-\eta+\eta \xi_{1}$.}
\begin{eqnarray}
I_{\calT_{st}}^{(d=D-2\epsilon)}(s,t)&=&c(D,\epsilon)\,
 \int_{0}^1 d\xi \,{ (-\alpha's \xi)^{{D\over2}-\epsilon-3}
 -(-\alpha't (1-\xi))^{{D\over2}-\epsilon-3}\over \alpha' t+\alpha' u\xi}\,,
\end{eqnarray}
where
\begin{equation}
c(D,\epsilon)=\pi^{{D\over2}-4-\epsilon}\,{ \Gamma(4-{D\over2}+\epsilon)\Gamma({D\over2}-2-\epsilon)^2\over (D-6-2\epsilon) \Gamma(D-4-2\epsilon)}\ .
\end{equation}
In order to perform the $\epsilon$ expansion we separate
the integral into  the two contributions
\be
I_{\calT_{st}}^{(d=D-2\epsilon)}(s,t)=  K_s(s,t)-K_t(s,t)\,,
\label{twoterms}
\ee
where
\begin{equation}
K_s(s,t)= c(D,\epsilon)\, \int_{0}^1 d\xi\,{ (-\alpha's \xi)^{{D\over2}-\epsilon-3}
 -(-\alpha's \xi_{*})^{{D\over2}-\epsilon-3}\over \alpha' t-\alpha'(s+t) \xi}
 \label{kdef}
\end{equation}
with $\xi_* = t/(s+t)$.
In this manner it is clear that the integral over $\xi$ does not develop a singularity in
 the limit $\epsilon\to 0$ as long as $D>4$.   In such dimensions
 $\epsilon$ singularities can only arise from the factor $c(D,\epsilon)$.
 These correspond to ultraviolet divergences that first arise as a $\epsilon$ pole
 in ten dimensions, $c(10,\epsilon)\to -\pi/(5!\,\epsilon)$ as $\epsilon\to 0$.
 For $D\le 4$ the integral may diverge at the $\xi=0$ endpoint, leading to
 infrared divergences that are seen as singularities in the
 $\epsilon\to 0$ limit.
 We now consider  the ten and nine dimensional  cases separately.

\begin{itemize}
\item In ten dimensions we have
\end{itemize}
\begin{eqnarray}
I_{\calT_{st}}^{(d=10-2\epsilon)}(s,t)
&=&-{1\over5!\,\epsilon
}\,\pi^{1-\epsilon}\,e^{-\gamma_{E}\epsilon}\left(1+{46\over15}\,\epsilon+O(\epsilon^2)\right)\,
\left({\alpha'u\over2} -\epsilon
I^{(1)}+O(\epsilon^2)\right)\\
\nn &=&  -{\alpha' u\pi\over 240\, \epsilon} + {1\over 240 }\alpha' u\,\pi
\, (\gamma_{E}+\log \pi-{46\over15})
+{\pi\over 5!}I_{\calT_{st}}^{(d=10)}(s,t)+ O(\epsilon^2)\,,
\end{eqnarray}
where $\gamma_E$ is Euler's constant.
The full result is given by adding the $(s,t)$, $(t,u)$ and $(s,u)$ contributions
and using the on-shell condition.  The pole in $\epsilon$ cancels in the sum,
leaving the ultraviolet finite result for the full amplitude in the $\epsilon\to 0$ limit,
\be
 I_{SUGRA}^{(d=10)}(s,t,u)= I_{\calT_{st}}^{(d=10)}(s,t)+
I_{\calT_{tu}}^{(d=10)}(t,u)+ I_{\calT_{us}}^{(d=10)}(u,s)\,,
\label{fullres}
\ee
where the function $I_{\calT_{st}}^{(d=10)}$ is given by
\begin{eqnarray}
  I_{\calT_{st}}^{(d=10)}(s,t)&=&\int_{0}^1 {d\xi\over \alpha' t+\alpha' u\xi}\,
  \left[(-\alpha't( 1-\xi))^2\log(-\alpha't(1-\xi))-(-\alpha's\xi)^2\log(-\alpha's\xi)\right]\, .
  \nn\\
 \label{e:IoneLog}
 \end{eqnarray}
This expression is real in the region $s<0$, $t<0$ and
has the appropriate imaginary part in other regions of the Mandelstam invariants.
For example,
in the physical region, $s>0$ and $t,u<0$, the integrand in (\ref{e:IoneLog}) is nonsingular
in the whole range of $\xi$,
so the only singularities are branch points due to the explicit $\log(-\alpha's)$
factor. The pole of the denominator at $\xi =-t/u$ is canceled by
a zero in the numerator and the integrand is finite in the range of integration, as is evident
from the form of $K_s(s,t)$ in (\ref{kdef}).
Other singularities with double discontinuities
 can be found by analytic continuation.

 Evaluating the integrals in~(\ref{e:IoneLog}) explicitly leads to
 \begin{eqnarray}
\nn I^{(d=10)}_{\calT_{st}}(s,t) &=&
 {\alpha'u\over 4}+\alpha' {st\over 2u}+ \alpha'\,{s^2 (s+3t)\, \log(-\alpha's)+t^2
 (t+3s)\,\log(-\alpha' t)\over 2u^2}
 \nn\\
 &&-\alpha'{s^2t^2\over u^3} \left(
\mathcal{L}_{2}\left(-{u\over t}\right)+\mathcal{L}_{2}\left(-{u\over s}\right)\right)\,,
 \end{eqnarray}
 where
 \begin{equation}
 \mathcal{L}_{2}(x)=\textrm{Li}_{2}(x)+\log(x)\log(1-x)= {\pi^2\over6} -\textrm{Li}_{2}(1-x)
 \end{equation}
 is a real function for all $x\geq0$ \cite{ZagierDilog}.
 Although $I_{\calT_{st}}^{(d=10)}(s,t)$ is
 a complicated expression it is easy to see that
it has the important scaling property
\begin{equation}
I_{\calT_{st}}^{(d=10)}(L s,L t)= L \, I^{(1)}(s,t)- {\alpha' u\over2} \, L\log(L)\,,
\end{equation}
which ensures that the scale of the logarithm does not contribute
after summation over the $(s,t)$, $(t,u)$ and $(s,u)$ terms.

$\bullet$
 For completeness, we also include the case of  nine dimensions, where there are no divergences when $\epsilon\to0$ and
 we can set $\epsilon=0$ directly in the expression for the amplitude,  giving
\bea
 I_{\calT_{st}}^{(d=9)}(s,t)&=&{2\over3}\, {(-\alpha's)^{3/2}+(-\alpha' t)^{3/2}\over \alpha' u}
 - 2 {(-\alpha's)^{3/2} t + (-\alpha't)^{3/2}s\over \alpha' u^2}\nn\\
&&+ {3(-{\alpha'}s)^{3/2}(-\alpha' t)^{3/2}\over(\alpha' u)^{5/2}}\, \log\left((\sqrt{-\alpha't}+ \sqrt{\alpha'u})(\sqrt{-\alpha's}- \sqrt{\alpha'u})
\over (\sqrt{-\alpha't}- \sqrt{\alpha'u})(\sqrt{-\alpha's}+ \sqrt{\alpha'u})\right) \, .
\label{ninedim}
\eea
Once again the full expression for $I_{nonan}^{(d=19)}(s,t,u)$ is
obtained by adding the contributions of $I_{\calT_{tu}}^{(d=10)}$ and
$I_{\calT_{us}}^{(d=10)}$ to $I_{\calT_{st}}^{(d=10)}$.

\subsubsection{String corrections and higher-order massless thresholds}\label{sec:massthreshold}

At higher order in the derivative expansion there are
further massless threshold effects due to the higher-order
contact terms in the blob in  figures~\ref{fig:Threshold}(b)
and \ref{fig:Threshold}(c).  These lead to factors of the form $s^r\,\log (-s/\nA_r)$, where
the scale of the logarithm does not cancel.
Luckily the threshold structure of these terms is much simpler than that of the box diagram in
figure~\ref{fig:Threshold}(a)
 since they only possess singularities in $s$ rather than overlapping singularities.
So we can go back to the cutoff procedure
outlined earlier in order to evaluate the expressions, taking care to verify that the dependence on the
cutoff $L$ indeed cancels.

The supergravity contribution considered in the last subsection was
obtained by replacing the propagators in the exponent of (\ref{eF}) by
their leading form in the large-$\tau_2$ limit.  In order to analyze
threshold terms at  higher-order in the momentum expansion we need to keep non-leading
terms at large $\tau_2$.
To do this we write
\be
\Delta_s \sim  \Delta_s^\infty + \delta_s\, ,
\label{asymdelta}
\ee
where the correction, $\delta_s$,  to the asymptotic value is given by the $k=0$
part of $\P(\nu|\tau)$ defined in (\ref{ptildedef}),
\begin{equation}
\delta_{s}=\sum_{m\neq0}{1\over 4|m|}\, \left(e^{2i\pi (m\nu^{(12)}_{1}+i |m \nu^{(12)}_{2}|)}
+e^{2i\pi (m\nu^{(34)}_{1}+i |m \nu^{(34)}_{2}|)}\right) \, ,
\end{equation}
with similar expressions for $\delta_{t,u}$.
In the $\tau_2\ge L$ region the terms with $k\ne 0$ are suppressed by powers of $e^{-L}$.
The asymptotic formula for the amplitude is therefore given by \cite{gv:stringloop},
\bea\label{e:CT}
I_{\calT_{st}; \calR_L}(s,t)&=&\int_{L}^{\infty} {d\tau_{2}\over\tau_{2}^2}\,
\int_{{\cal T}_{st}} \prod_{i=1}^3 {d^2\nu\over \tau_{2}}\,
\exp\left(\alpha' s (\Delta^\infty_{s}-\Delta^\infty_{u})
+ \alpha' t(\tilde \Delta^\infty_{t}-\tilde\Delta^\infty_{u})\right)\nn\\
&&\qquad\qquad\qquad \exp\left(\al s(\delta_s -\delta_u )+\al t(\delta_t -\delta_u)\right)\,.
\eea
The higher-order threshold contributions containing factors of $\log(-sL)$
are obtained by expanding the integrand in powers of $\delta$,
$e^{\al \delta}=1 +\al\delta +\al^2\delta^2/2 +\dots$.  The factors of $e^{\alpha'\, \Delta^\infty}$
in the integrand contribute to terms with positive powers of $L$.

In particular, the terms of order $\al^4\, s^4\log(-\al s/\nA_4)$, which will be denoted
$I^{(4)}$, is obtained from the $\delta_{s}^2$ contribution of the integrand
in~(\ref{e:CT}). Neither $\delta_t$ or $\delta_u$ contributes to this
threshold behaviour term (although
 $\delta_t$ does contribute
to the analogous $t$-channel term, ${\alpha'}^4\,t^4\, \log(-\al t/\nA_4)$).
This gives
\begin{eqnarray}
\nn  I^{(4)}_{\calT_{st}; \calR_L}(s,t)&=&{1\over 2}\sum_{m\neq0}{(-\alpha' s)^2\over (4m)^2}
\int_{L}^{\infty}{d\tau_{2}\over\tau_{2}^2}
\int_{{\cal T}_{st}} \prod_{i=1}^3  d\omega_{i}
e^{\alpha'  \pi\tau_{2}Q(s,t)-4\pi|m| \tau_{2}(\omega_{2}-\omega_{1})}\\
\nonumber&+&{1\over 2}\sum_{m\neq0}{(-\alpha' s)^2\over (4m)^2}
\int_{L}^{\infty}{d\tau_{2}\over\tau_{2}^2}
\int_{{\cal T}_{st}} \prod_{i=1}^3  d\omega_{i}
 e^{\alpha' \pi\tau_{2}  Q(s,t)-4\pi |m|\tau_{2}(1-\omega_{3})}\\
  &=&  \sum_{m\neq0} \, {(-\alpha's)^2\over (4m)^2}\, g^{(4)}(L; s,t)  \,,
\label{threshh}
\end{eqnarray}
where we note that the two terms on the right-hand side of the first equality are equal and
\begin{eqnarray}
g^{(4)}(L;s,t)&=&\int_{L}^{\infty} {d\tau_{2}\over\tau_{2}^2}
 \int_{0}^1 d\eta \, d\xi_{1}d\xi_{2}\, \eta(1-\eta)\,
e^{\alpha'\pi\tau_{2}Q(s,t)-4 \pi |m |\tau_{2}(1-\xi_{2})(1-\eta)}\,.
\end{eqnarray}
Noting further that $Q$ is given  in terms of the variables $\eta$, $\xi_1$ and $\xi_2$ by
\be
Q(s,t) = s\,\eta(1-\eta)\, \xi_1\, \xi_2 + t\,\eta(1-\eta)\,(1-\xi_1)\,(1-\xi_2)\,,
\label{qnewdef}
\ee
and performing the $\xi_2$ integral gives
\bea
 g^{(4)}(L;s,t)&=&\int_{L}^{\infty} {d\tau_{2}\over\tau_{2}^3}
 \int_{0}^1 d\eta \, d\xi_{1}\, \eta\,{
 e^{\al \pi s\tau_{2}\eta(1-\eta)\xi_{1}}-
e^{-4 \pi m \tau_{2}(1-\eta)}e^{\al\pi t\tau_2\eta(1-\eta)(1-\xi_1)}
\over \al\eta \pi(s \xi_{1}-t(1-\xi_1))+4\pi m }\,.
\label{glst}
\eea

The term of order $s^4$ in the amplitude comes from the $O(\al^3)$ correction to the tree amplitude
represented by the blob in \ref{fig:Threshold}(b).  This is obtained by expanding  $g^{(4)}$
to $O(s^2)$.  At this order it is easy to see that the term proportional to $e^{-4 \pi m \tau_{2}(1-\eta)}$
in the numerator of (\ref{glst})
is proportional to $e^{-L}$ and is negligible.  The term of the form $s^2\log(-\alpha'\,s)$
arises by expanding $e^{\al \pi s\tau_{2}\eta(1-\eta)\xi_{1}}$ to quadratic order and replacing the
denominator by $4\pi m$.
 As a result we find the $O(s^2)$ terms,
\be
g^{(4)}(L;s,t)=-{1\over 4\pi m}\,{(\alpha's\,\pi)^2\over 180}\,
\left(\log(-\alpha's \, \pi \,L/c_{e})-2/5\right)
 L^3 + O(s^3,L^{-1})\,,
\label{glres}
\ee
where $c_{e}=\exp(-\gamma_{E})$.
In obtaining this we have used the fact that
\be
\lim_{x\to 0}\int_L^\infty \frac{dt}{t^3}\, e^{-xt} = \frac{1}{2L^2}- \frac{x}{L} -
\frac{x^2}{2}\left(\log(xL) -3/2 +\gamma\right)+ o(x^2).
\label{intval}
\ee
Substituting (\ref{glres}) into (\ref{threshh}) gives the contribution
\bea\label{e:s4logsI}
I^{(4)}_{\calT_{st}; \calR_L}(s,t)&=&-
{2\pi \zeta(3)\over 45}\,\left(\frac{\alpha'}{4} \right)^4\,
s^4\log(- \alpha's\, L /\hat \mu_{4})+\cdots\,,
\eea
where the scale of the logarithm at this order is given by
\begin{equation}\label{e:muhat}
\log\hat \mu_{4}={2\over5}- \log(\pi/c_{e})\ .
\end{equation}
The ellipses indicate that a similar contribution that arises by using the $\delta_t^2$ term
in the expansion of $e^{\al t\delta_t}$, which gives  a threshold singularity in the
$t$ channel in which
figure~\ref{fig:Threshold}(b) is transformed by interchanging $s$ and $t$.
There are similarly contributions from $\calT_{us}$ and $\calT_{tu}$ regions
containing analogous terms with $s$, $t$ and $u$-channel thresholds.

Therefore the total contribution from the upper part of the fundamental domain $\tau_{2}\geq L$ to the
terms of order $s^4$ is given by the sum
\be
\begin{split}
&I^{(4)}_{\calR_L}(s,t,u) = I^{(4)}_{\calT_{st}; \calR_L}(s,t)+ I^{(4)}_{\calT_{tu}; \calR_L}(t,u)
+I^{(4)}_{\calT_{us}; \calR_L}(u,s)  \\
&= - {4\zeta(3)\over 45\pi}  \left(\frac{\alpha'}{4}\right)^4\,
s^4\,\log(-\alpha'sL\pi/c_{e})- 2/5) + (s\to t) + (s\to u)\, .
 \end{split}
 \label{totupper}
\ee
The dependence on both $L^3$ and $\log(L)$ cancel in the sum of
this contribution and the contribution from $\tau_{2}\leq L$
given in~(\ref{esd}), giving
\be\begin{split}
I^{(4)}_{nonan}(s,t,u)&=-{4\pi \zeta(3)\over 45}\, \left(\frac{\alpha'}{4}\right)^4 \left[
s^4\, \log(-\alpha' s/\nA_{4} )\right.\\
 & \left.+ t^4\, \log(-\alpha' t/\nA_{4} )+
u^4\, \log(-\alpha' u/\nA_{4} ) -6/5\right]\,,
\label{fag}
\end{split}\ee
where
\be
\log \nA_{4}= \log\hat\nA_4 - \log \tilde\nA_4 =
{9\over10} -\log\left({2\over \pi c_{e}}\right) + {\zeta'(3)\over\zeta(3)} - {\zeta'(4)\over \zeta(4)}\,,
\ee
which is the sum of the scale from the lower part of the fundamental domain,
given  in (\ref{swala}), and
the scale  in the logarithm obtained from the above computation from the large-$\tau_2$
part of the fundamental domain (\ref{e:muhat}).

Contributions to terms of order $s^5$ arise both from expanding the function $g^{(4)}(L;s,t)$ to one
further  order in $s$ and by bringing down one further power of $\delta$ from the exponential in
(\ref{e:CT}).  In this case the there are no terms proportional to $\log L$, which reflects the fact that
there is no threshold term at this order, as follows from the unitarity argument in
section~\ref{sec:twounitarity}.
In order to evaluate the thresholds at order  ${\alpha'}^{6}\,s^6 \log(-\al s/\mu_6)$,
it is necessary
to expand $g^{(4)}(L;s,t)$ to two further orders in $s$, which is fairly complicated.  In addition,
there are contributions from yet higher powers of $\delta$.  Although we have not evaluated
these terms, in principle, they will determine a scale
$\log\hat \nA_6$ (analogous to $\log\hat \nA_4$ in (\ref{e:muhat})) that combines with $\log\tilde \nA_6$
(\ref{scalesix}) to give the scale $\log \nA_6$.

At higher orders in $\alpha'$ string corrections arise on both sides of the threshold discontinuity
as we saw in section~\ref{sec:twounitarity}.  The qualitative structure of these threshold contributions
matches that of the discontinuities that we determined earlier from unitarity.
Firstly, there are contributions from the triangle diagram in
figure~\ref{fig:Threshold}(b) beyond the ones considered above.
The first of these is of order $\alpha'\,s^7\log(-\alpha'\,s)$.
In addition, a new class of contributions arises, in which both the propagators
on the left and right of the $s$-channel cut are canceled and the result
is the diagram in figure~\ref{fig:Threshold}(c), which has two blobs representing the higher
derivative vertices.

\subsection{Summary of the expansion of the ten-dimensional genus-one amplitude}
\label{sec:discussion}

To put these results in perspective we will
here summarize the low-energy expansion of the genus-one contribution
to the analytic part of the amplitude up to order $s^6\, \hR^4$.
{}At tree-level and genus one, type IIA and type IIB massless four-particle amplitudes
 are completely equivalent,
 so the results apply equally to both cases.

The analytic terms to this order are
summarized by
\begin{eqnarray}
\hA^{genus-1}_{an}(\sg_2,\sg_3)&=&{\pi\over 3}\,
\left(\sum_{p,q} J^{(p,q)}\, \sg_{2}^p \sg_{3}^q\right)\, \hR^4
 \nn\\
&=&{\pi\over 3}\, \left(1+ 0\,\sg_2+{\zeta(3)\over3}\sg_{3}
+0\,\sg_{2}^2
+{97\over 1080}\,\zeta(5)\,
\sg_{2}\sg_{3}\right.\nn\\
&&\qquad\quad\left.  + {1\over 30}\,\zeta(3)^2\sg_{2}^3+ {61\over1080} \zeta(3)^2\sg_{3}^2+
\cdots\right) \,\hR^4\,,
\label{grah}
\end{eqnarray}
(where we have indicated explicitly the vanishing $\sg_2$ and $\sg_2^2$ coefficients),
while the nonanalytic terms are contained in
\begin{equation}\begin{split}
\hA^{genus-1}_{nonan}(\sg_2,\sg_3)&=\hA^{genus-1}_{SUGRA}\\
&+ \left(\frac{\alpha'}{4}\right)^4 {4\zeta(3)\pi\over45}\,\Big(s^4\, \log(-{\alpha'\,s\over\mu_{4}})
+t^4\, \log(-{\alpha'\,t\over\mu_{4}})
+u^4\, \log(-{\alpha'\,u\over\mu_{4}}) \Big) \,\hR^4\\
 &+\left(\frac{\alpha'}{4}\right)^6 {\pi\,\zeta(5)\over2520}\,\Big( (87 \, s^6+s^4\, (t-u)^2)\,
\log(-{\alpha'\,s\over\mu_{6}})\\
&+(87 \, t^6+ t^4\,(s-u)^2)\, \log(-{\alpha'\,t\over\mu_{6}})
+(87 \,u^6+u^4\,
(s-t)^2)\, \log(-{\alpha'\,u\over\mu_{6}})
\Big)
\,\hR^4+\cdots \,,
\end{split}\label{grab}
\end{equation}
where $\hA^{genus-1}_{SUGRA}$ is the result obtained in ten-dimensional maximal
supergravity that we described earlier where we also discussed the scales $\mu_4$ and
$\mu_6$.
Notice that the terms in equations~(\ref{grah}) and
 (\ref{grab}) satisfy a `transcendentality condition' in which
$\zeta(k)$ is associated with a weight $k$, $\pi$ has weight 1 and $\log(x)$ also has
weight $1$.  The total weight of a term of order $(\al)^q$ is $q+1$.

Clearly, the separation of the amplitude into $\hA_{an}$ and $\hA_{nonan}$ depends on
the definition of the scales inside the logarithmic terms in (\ref{grab}). The sum of
the terms that depend on $\mu_4$ and $\mu_6$ is
\begin{equation}
-\frac{4\pi\zeta(3)}{45}\, \sg_2^2\,\log\mu_4\,\hR^4
-{\zeta(5)\pi\over630}\, \left(11 \sg_2^3
+14 \sg_3^2\right)\,\log\mu_6
\,\hR^4\,.
\label{grabb}
\end{equation}
 Notice that although changing the scale inside the logarithms alters the analytic part of the
 amplitude, the pattern of Riemann zeta values is different in the coefficients
 of the terms in $\hA^{genus-1}_{an}$ from those in
 $\hA^{genus-1}_{nonan}$.  In this sense, there is an objective meaning to the
  separation between the
 analytic and the nonanalytic parts as given in (\ref{grah}) and (\ref{grab}).
For example,
at order $(\alpha'\,s)^6$ one has a finite piece proportional to $\zeta(3)^2$
while the logarithmic term is multiplied by
$\zeta(5)$ in agreement with the `transcendentality' property noted earlier.  So
our choice of scale in the argument of the logarithmic terms
is not only the natural scale that arises from the
calculation of section~\ref{sec:amplitude}, but is natural if
 $\zeta(3)^2$ and $\zeta(5)$ are considered to be distinct
 coefficients.

\section{The genus-one four-particle amplitude in nine dimensions}\label{sec:ninedim}

We now turn to consider the compactification of the amplitude on a circle of radius
$r$ so that
nine-space-time dimensions are non-compact.  We will specialize to the case in which the
momenta of the scattering particles, $p_r^\mu$, have zero components in the compact direction,
so the external states are Kaluza--Klein ground states.
In nine dimensions the genus-one normal thresholds have square root
singularities instead of logarithms.  This separates the threshold terms in the
amplitude from the analytic terms in a very clear manner.  The
massive Kaluza--Klein modes also generate massive square root
thresholds that are expanded, when $\alpha' s\gg 1/r^2$, in an infinite series of terms of the
form  $(\alpha' s r^2)^n$ and therefore enter into higher order
terms in the low energy expansion.  The logarithmic
singularities characteristic of ten dimensions are recovered in
the ten-dimensional limit by a condensation of these Kaluza--Klein
thresholds \cite{grv:bigpaper}.

\subsection{General method}\label{sec:genmethod}

The expression for the integral $I$ in the compactified genus-one amplitude (\ref{eLoopa})
now has the form
\be
\label{ideff} I^{(d=9)}(s,t,u;r) = \int_{\calF} {d^2 \tau\over \tau_2^2}\,
\sqrt{\tau_2} \ \Gamma_{(1,1)}(r)\, F(s,t,u;\tau)\,  ,
\ee
 where the lattice sum factor  is given by \cite{Green:1982sw}
\begin{equation}\label{lattfac}
 \sqrt{\tau_{2}}\ \Gamma_{(1,1)}(r)
 =  \sqrt{\tau_{2}}\sum_{( m,\hat n)\in\ZZ^2}\,
 e^{-\pi\tau_{2}\left((\frac{m}{r})^2 +(\hat nr)^2\right)+2i\pi \tau_{1} {\hat m}n} \, .
\end{equation}

The loop amplitude can now be expressed in terms of
an expansion in half-integer powers of $\alpha' s$, $\alpha' t$
and $\alpha' u$.
We can separate this into two terms,
\be
I^{(d=9)}(s,t,u; r) = I_{an}^{(d=9)}(s,t,u; r)+ I_{nonan}^{(d=9)}(s,t,u; r)\,.
\label{idecom}
\ee
The analytic part has an expansion with integer powers of the Mandelstam invariants
of the form
\be
I_{an}^{(d=9)}(s,t,u; r)= \sum_{p,q=0}^\infty J^{(p,q)}(r)\,\hat \sigma_2^p \hat\sigma_3^q
\label{ian}
\ee
and is analytic in $s$, $t$ and $u$.  We will see that the coefficients can be expanded at large
$r$ as sums of powers
of $r$, together with $r \log r^2$ terms and exponentials,
\bea
J^{(p,q)}(r) &=&
 J^{(p,q)}_1\, r^{4p+6q-1} + J^{(p,q)}_2\, r^{4p+6q-3}+ \dots +J^{(p,q)}_{4p+6q}\, r^{-4p-6q+1}\nn\\
&&
+ K^{(p,q)}\, r \log(r^2\lambda_{2p+3q}) + O(e^{-r})\,,
\label{jexpan}
\eea
where the $\lambda_k$'s are constants.
In the ten-dimensional theory only the term linear in $r$ survives with coefficient $J^{(p,q)}_{2p+3q}$.
The nonanalytic part has the form
\be
I_{nonan}^{(d=9)}(s,t,u; r)=  b_1 s^{\half} + b_4 s^{\sevenh} + \cdots\, ,
\label{inonan}
\ee
where terms involving $t$ and $u$ have not been included.
The coefficients $b_i$ are independent of
$r$.  In the following we will be interested in determining the coefficients
$J^{(p,q)}_r$ in the expansion of $I^{(d=9)}_{an}$, but will not consider the coefficients
$b_n$ of the non-analytic terms in any detail (although they are
relatively easy to extract).

We will first reexpress the lattice sum by a Poisson resummation
that replaces the sum over $m$ by a sum over $\hat p$ and two relatively prime integers $(\hat m, \hat n)$,
\be
 \sqrt{\tau_{2}}\sum_{(m,\hat n)\in\ZZ^2}\,
 e^{-\pi\tau_{2}\left((\frac{m}{r})^2 +(\hat nr)^2\right)+2i\pi \tau_{1}{m\hat  n}} = r\,
\sum_{\hat p\in \ZZ\atop (\hat m,\hat n)=1}\,
 e^{-\pi\, \hat p^2\, r^2\, {| \hat m +\hat  n\tau|^2\over\tau_2}}\,.
\label{resummed} \ee
Thanks to the modular invariance of the kinematical factor $F(s,t,u;\tau)$, after separating
the zero-winding term ($\hat p =0$) one can unfold
the integral onto the semi-infinite strip \cite{Terras}, which gives
\bea
I^{(d=9)}(s,t,u;r) &=& r\,\int_{\calF} {d^2\tau\over\tau_{2}^2} \,F(s,t,u;\tau) \nn\\
&+&  r \,
\sum_{\hat p\neq0} \int_0^\infty
{d\tau_2\over \tau_{2}^2}\, e^{-\pi \hat p^2 \frac{r^2}{\tau_2}} \int_{-\half}^{\half} d\tau_1\,F(s,t,u;\tau)\,
\, . \label{unfoldhere}
\eea
The first term is the zero-winding term, which is the term that survives the $r\to \infty$ limit,
\be
 r \,\int_{\calF} {d^2\tau\over\tau_{2}^2}\,
\,F(s,t,u;\tau)= r\,I^{(d=10)}(s,t,u) \, , \label{unfoldhereII}
\ee
where the low-energy expansion of $I^{(d=10)}(s,t,u)$ was considered in sections~\ref{sec:amplitude}
 and~\ref{sec:threshold}.

Proceeding as for the ten-dimensional amplitude in
 section~\ref{sec:amplitude}, the integral $I^{(d=9)}_{\hat p\neq0}$
 can be split into an analytic and non-analytic
 parts by dividing the $\tau_2$ integral into two domains, so that
 $I^{(d=9)}=I^{(d=9)}_{(1)}+I^{(d=9)}_{(2)}$, where
\be
I^{(d=9)}_{(1)}=r\,\sum_{\hat p\neq0} \int_0^L
{d\tau_2\over \tau_2^{2}}\, e^{-\pi \hat p^2 \frac{r^2}{\tau_2}}\,
\int_{-\half}^{\half} d\tau_1\,F(s,t,u;\tau) \,
\label{partonee}
\ee
contains the analytic part of the amplitude, and
\be
I^{(d=9)}_{(2)}=  r\,
\sum_{\hat p\neq0} \int_L^\infty
{d\tau_2\over \tau_2^{2}}\, e^{-\pi\hat p^2 \frac{r^2}{\tau_2}}
 \, \int_{-\half}^{\half} d\tau_1\,F(s,t,u;\tau)\, ,
\label{parttwoe}
\ee
contains the non-analytic threshold terms.
As in the ten-dimensional case, the $L$ dependence will cancel from the final expressions.

The analytic contributions in equation~(\ref{partonee}) can be analyzed
by expanding $\int d\tau_1F(s,t,u;\tau)$ in powers of $\hat\sigma_2$ and $\hat \sigma_3$,
\be
\int_{-\half}^{+\half} d\tau_1F(s,t,u;\tau) = \sum_{p,q=0}^\infty
\hat\sigma_2^p\,\hat\sigma_3^q\, j^{(p,q)}\,.
\label{zeroexp}
\ee
As in section~\ref{sec:amplitude} the coefficients at any order in $s$, $t$, $u$,
are determined by properties of the $\D^{(0)}_{\{l\}}$ functions determined in
 appendices~\ref{sec:ZMdiag} and \ref{sec:leading}.  In those appendices we expanded these
 functions in powers of $1/\tau_2$.  In the simplest
 cases we displayed all power-behaved terms while in others
 we only displayed the positive powers and constant terms.

For the terms that are constant or negative powers of $\tau_2$ -- i.e., of the form
$a_k\, \tau_2^k$ with $k\le 0$ -- both the $\tau_2$ integral
and the $\hat p$ sum in (\ref{partonee}) are easy to evaluate.  For these terms we may simply
take $L\to \infty$ and $I^{(d=9)}_{(2)}$ in (\ref{parttwoe}) is zero.  The result is a succession
of terms of the form
\be
\hat \sigma_2^p\, \hat\sigma_3^q\,\sum_{k\leq0}\, J^{(p,q)}_k\, r^k\,,
\label{resinnine}
\ee
with $k\le 0$.
The contribution $J^{(p,q)}_{0}$ is the ten dimensional contribution
$J^{(p,q)}$ discussed in section~\ref{sec:amplitude}.

The terms in the expansion of $\int d\tau_1F(s,t,u;\tau)$
 that are positive powers of $\tau_2$ of the form $a_k\, \tau_2^k$ with $k\ge 2$
have to be treated separately since, in
the limit $\tau_2 \to \infty$,  the $\hat p$ sum in (\ref{partonee}) and (\ref{parttwoe})
 contains the factor
$e^{-\pi \hat p^2 r^2/\tau_2}\sim 1$, so it needs a Poisson resummation.
This involves first adding and
subtracting a term with $\hat p=0$.  The subtracted term,
\be
I^{(d=9)}_{(2)\hat p=0}=  r\,
 \int_L^\infty
{d\tau_2\over \tau_2^{2}} \, \int_{-\half}^{\half} d\tau_1\,F(s,t,u;\tau)\, ,
\label{peqzero}
\ee
is precisely the term studied in the last section that has thresholds of the form
$s^k\, \log(-\alpha' s L/\hat \mu_k)$ with a coefficient that ensures that it cancels
the $s^k\, \log (-\alpha'\,s/\mu_k)$ thresholds contained in the term $r\, I^{(d=10)}$ in
(\ref{unfoldhere}). This ensures that
there are no $\log (-\alpha'\,s)$ terms in the nine-dimensional expression, in accord with unitarity.
After the Poisson resummation the integer $\hat p$ is replaced
by the integer $p$ and we have
\be
I^{(d=9)}_{(2)} = \sum_{ p\in\ZZ} \int_L^\infty
{d\tau_2\over \tau_2^{3\over2}}\, e^{-\pi \, p^2 \frac{\tau_2}{r^{2}}}\,
\int_{-\half}^{\half} d\tau_1\,F(s,t,u;\tau)\,.
\ee
The $p=0$ contribution,
\be
I^{(d=9)}_{(2); p=0} = \int_L^\infty
{d\tau_2\over \tau_2^{3\over2}}\,
\int_{-\half}^{\half} d\tau_1\,F(s,t,u;\tau)\,,
\label{pzero}\ee
contains the nine-dimensional
threshold terms of the form $(-s)^{k+1/2}$, which we will not be considering in any detail in the
following.
For the $p\ne 0$ terms we can again take $L\to \infty$, in which case $I^{(d=9)}_{(2)}$ vanishes
and  $I^{(d=9)}_{(1)}$ gives a series of terms of the form (\ref{resinnine}) with $k\ge 1$.

The terms  that we have calculated of the form (\ref{resinnine}) will be summarized in
(\ref{fin}) in subsection~\ref{sec:nineD}.

However, there is a subtlety in considering the $\tau_2$ integral of terms that are linear in $\tau_2$,
i.e., terms of the form $\int d\tau_1F(s,t,u;\tau)\sim a_1\tau_2$ in (\ref{unfoldhere}).
To see this note that in this case
the $\hat p=0$ term we need to add and subtract before doing the Poisson resummation in $I^{(d=9)}_{(1)}$
  is
\be
I^{(d=9)}_{\hat p=0;a_1} = a_1r \int_0^L
{d\tau_2\over \tau_2}\,
 \, ,
\label{pnought}
\ee
which diverges.
This means that for terms linear in $\tau_2$
we cannot perform a Poisson resummation in $I^{(d=9)}_{(1)}$ but we still have to perform a
Poisson resummation in
$I^{(d=9)}_{(2)}$ since each term in the $\hat p$ sum has a factor
 $e^{-\pi\hat p^2 \frac{r^2}{\tau_2}}\sim 1$  at large $\tau_2$.  Explicitly, we have
\be
I^{(d=9)}_{(1);a_1}\equiv a_1\,r\, \sum_{\hat p\neq0} \int_0^L
{d\tau_2\over \tau_2}\, e^{-\pi \hat p^2 \frac{r^2}{\tau_2}}
\equiv a_1\, r\, f_1(L/r^2)\ ,
\label{e:fone}
\ee
where we have rescaled $\tau_2$ in order to write the expression in terms of
a function $f_{1}(L/r^2)$ which will be of later use.
After a Poisson resummation,  the $p\ne 0$ terms in $I^{(d=9)}_{(2); a_1}$ are
given by
\be
I^{(d=9)}_{(2); p\ne 0;a_1} = a_1  \sum_{ p\neq0}^\infty \int_L^\infty
{d\tau_2\over \tau_2^{1/2}}\, e^{-\pi \,  p^2 \frac{\tau_2}{r^2}} \equiv a_1
r f_2(L/r^2)\,,
\label{e:ftwo}
\ee
where we have again rescaled $\tau_2$ in order to write the expression in terms of
a function  $f_2(L/r^2)$, which will be of later use.

The dependence on $L/r^2$ in $f_1$ in equation~(\ref{e:fone}) will cancel with that of the term $f_2$ in equation~(\ref{e:ftwo}).
To see this, consider the derivatives of these terms with respect to $L$.
We have
\be
r\, {\partial f_1(L/r^2) \over \partial L} =
{r\over L}\, \sum_{\hat p\neq0} e^{-\pi \hat p^2 \frac{r^2}{L}}
\,.
\label{fone}
\ee
On the other hand Poisson resumming the integer $\hat p$ in
(\ref{e:ftwo})
\bea
r\, {\partial f_2(L/r^2) \over \partial L} &=&
- {1\over L^\half}\, \sum_{ p\neq0} e^{-\pi  p^2 \frac{L}{r^2}}\nn\\
&=& {1\over L^\half} - {r\over L} - {r\over L}
\sum_{\hat p\neq 0} e^{-\pi \hat p^2 \frac{r^2}{L}}
\,.
\label{foneII}
\eea
In order to see what this means we can use the fact that
\be
{\partial \over \partial L} \Big(I^{(d=9)}_{(1)p\ne 0;a_1}+ I^{(d=9)}_{(2)\hat p\ne 0;
    a_1}\Big)= a_1
r\, {\partial f_1(L/r^2) \over \partial L}+a_1 r\, {\partial
  f_2(L/r^2) \over \partial L}
= {a_1\over L^{1/2}} - a_1 {r\over L}\, .
\label{totalb}
\ee
Integrating over $L$, we find
\be
I^{(d=9)}_{(1)\hat p\ne 0;a_1}+I^{(d=9)}_{(2) \hat p\ne 0; a_1} =a_1 r\left(
2 \frac{L^\half}{r}  + \log\tilde\lambda + \log(r^{2}/L)\right).
\label{inttot}
\ee
where the constant $\tilde\lambda$ can be  determined by integrating between $r^2$ and $L$,
\bea\label{e:C}
\log \tilde \lambda&=& -2 +\int_{0}^1 {dt\over t} \sum_{p\neq0} e^{-\pi p^2/t} + \int_{1}^\infty{dt\over \sqrt{t}} \sum_{p\neq0} e^{-\pi p^2 t}\\
\nn&=& \gamma_E -\log(4\pi)\,.
 \eea
So the $L$-dependent terms cancel, apart from  $-a_1 r\log L$ and $2a_1 L^{1/2}$.
These two terms
cancel with the $L$-dependent parts of the $\hat p=0$
threshold terms that subtract the $\log(-\alpha'\,s)$ ten dimensional thresholds
and the $p=0$ threshold terms in~(\ref{pzero}) that add in the $(-\alpha'\,s)^{-\half}$ thresholds.

\subsection{Summary of the expansion of the nine-dimensional genus-one amplitude}\label{sec:nineD}

By using the method outlined above  and developed in detail in the
appendices, we find the following terms in the momentum expansion of the
analytic terms in the integral, $I$, that defines the four-particle amplitude in~(\ref{eLoopa})
up to order $s^8\,\hR^4$ (with partial results beyond)
\begin{equation}\label{sumario}\begin{split}
I_{an}^{(d=9)}(r;s,t) =&\, r\, I_{an}^{(d=10)}
+2\sum_{p=1}^\infty  r\int_0^\infty {d\tau_{2} \over\tau_{2}^2}
e^{-{\pi p^2 r^2\over \tau_2} }\,j^{(p,q)}\,\hat\sigma_2^p\,\hat
\sigma_3^q  \cr
=&\,  r\, I_{an}^{(d=10)}
+2\sum_{p=1}^\infty  r\int_0^\infty {d\tau_{2} \over\tau_{2}^2}
e^{-{\pi p^2 r^2\over \tau_2} } \bigg[1
+ \sg_2\ \Big( {\pi^2\over 45} \tau_2^2+{\zeta(3)\over\pi\tau_2}\Big)
\cr
&+\sg_3\, \Big(  {2 \pi^3 \tau_2^3 \over 567} + {\zeta(3)\over 3}  + {5\zeta(5)\over 4\pi^2 \tau_2^2}
+  O(e^{-\tau_2})\Big)
\cr
&+
\sg_2^2\, \Big( {8\over 315} \zeta(4)\tau_2^4 + {2\pi\zeta(3)\over 45}\, \tau_2  + {5\zeta(5)\over 12\pi \tau_2} +
 {\zeta(3)^2\over 4\pi^2 \tau_2^2}  +  {\zeta(7)\over 4\pi^3 \tau_2^3}+  O(e^{-\tau_2})\Big)
\cr
&+
\sg_2\sg_3 \, \Big( {4\pi^5\over 66825} \tau_2^5 +{\pi^2\zeta(3)\over 63}\,
\tau_2 ^2+{29\zeta(5)\over 135} +O(\tau_2^{-1})\Big)
\cr
&+
\sg_2^3\, \Big( {2\over 1001}\zeta(6) \tau_2^6 +{4\pi^3\zeta(3)\over 4725}\, \tau_2^3+
{11\pi \zeta(5)\over 630}\, \tau_2 +{\zeta(3)^3\over 30} +O(\tau_2^{-1})\Big)
\cr
&+
\sg_3^2\, \Big(  {1744 \over 675675}\zeta(6) \tau_2^6 + {16 \pi^3 \zeta(3)\over 14175} \, \tau_2^3
 +{4\pi \zeta(5) \over 180}\,  \tau_2  +{61\zeta(3)^2\over 6144}+  O(\tau_2^{-1})\Big)\bigg]\,,
\end{split}
\end{equation}
where $I_{(d=10)}^{(p,q)}$ is the ten-dimensional integral considered in section~\ref{sec:amplitude}
(the coefficient of $\hR^4$ in (\ref{grah})).
These expansions in powers of $1/\tau_2$ are valid for large $\tau_2$.  We have displayed all
powers of $\tau_2$ up to order $\hat \sigma_2^2$, but at higher orders in $s,t,u$
there are further inverse powers of $\tau_2$ beyond the order displayed, which we have not
calculated. There are also
exponentially suppressed contributions of order $e^{-\tau_2}$ at order
$\hat\sigma_3$ and above.

It is easy to integrate over $\tau_2 $ using the formula
\be
\sum_{\hat p=1}^\infty  r\int_0^\infty {d\tau_2 \over\tau_2^2}
e^{-{\pi \hat p^2 r^2\over \tau_2} } \tau_2^n = r^{2n-1} \zeta^*(2-2n)\ .
\ee
where $\zeta^*(x)= \pi^{-x/2}\, \zeta(x)\, \Gamma(x/2)$.
The contributions with $n>1$ can be obtained by using the analytic continuation $\zeta^*(2-2n)=\zeta^*(2n-1)$,
which is equivalent to performing the Poisson resummation described in
section~\ref{sec:genmethod}.
For terms linear in $\tau_2$ ($n=1$) the integral requires greater care, as
was also emphasized in section~\ref{sec:genmethod}, where
we were led to (\ref{totalb}).
The result is
\bea
I_{an}^{(d=9)}(r;s,t) &=& {\pi \over3}\bigg[ r+
r^{-1} +\sg_2\ \Big( { \zeta(3)\over 15}\, r^3 +{ \zeta(3)\over 15}\, r^{-3}\Big)
 \nn\\
&+&\sg_3\, \Big(  { \zeta(5)\over 63}\,r^5 +
{\zeta(3)\over 3}\, r +{\zeta(3) \over 3}\, r^{-1}  +  { \zeta(5) \over 63}\, r^{-5} \Big)
\nn\\
&+&
\sg_2^2\, \Big( { \zeta(7)\over 315}\,  r^7 + {2\zeta(3)\over15} \, r\,
\log(r^2\,\lambda_{4}) +
 {\zeta(5)\over 36}\, r^{-3} + { \zeta(3)^2\over 315}\, r^{-5}
+ {\zeta(7) \over 1050}\, r^{-7} \Big)
\nn\\
&+&
\sg_2 \sg_3\, \Big( {7\zeta(9) \over 2970} \, r^9 + {\zeta(3)^2 \over 21}\, r^3 +
{97\zeta(5)\over 1080}\, r
+ {29\zeta(5)\over 135}\, r^{-1} + O( r^{-3}) \Big)
\nn\\
&+&
\sg_2 ^3\, \Big( {3\zeta(11) \over 8008} \, r^{11} +{2\zeta(3)\zeta(5) \over 525} \, r^{5} +
 {11\zeta(5)\over210}\,  r\,\log(r^2\,\lambda_{6})
+{\zeta(3)^2\over 30}\, r +{\zeta(3)^2 \over 30}\, r^{-1} + O( r^{-3}) \Big)
\nn\\
&+&
\sg_3 ^2\, \Big( {109\zeta(11) \over 225225} \, r^{11}
+{8\zeta(3)\zeta(5) \over 1575} \, r^{5} + {\zeta(5)\over 15}  r\,
\log(r^2\,\lambda_{6}) +{61\zeta(3)^2\over 1080}\,r + {61\zeta(3)^2\over 6144}\,r^{-1} \,
\nn\\
&& \qquad+ O( r^{-3})\Big)+O(e^{-r})
\bigg]
\label{fin}
\eea
Note that the $r\to 1/r$ symmetry is manifest only in the first two
lines of this expression, in which there are no $e^{-r}$ terms and each power of $r$ is accompanied by
a corresponding inverse power of $r$ with identical coefficient.
At order $\hat\sigma_2^2$ and beyond, there is no such pairing of terms and here
terms that are exponentially suppressed at large $r$ play an
essential r\^ole in guaranteeing the $r\to 1/r$ symmetry.

The $(\alpha'\,s)^k\,r\, \log (r^2\,\lambda_k)$
contributions are the ones we found in (\ref{totalb}) that arise
when there is a linear dependence on $\tau_{2}$ and
are connected with the ten-dimensional threshold contributions discussed in
section~\ref{sec:massthreshold}.
The scale $\lambda_k$ in such logarithms
is determined by  combining $\log(\tilde\lambda)$ from (\ref{e:C})
and the scale  $\log(\hat \mu_{k})$ of the non-analytic contribution
discussed in section~\ref{sec:massthreshold},
\begin{equation}
\log \lambda_{k}\equiv \log (\tilde\lambda/\tilde \mu_{k})\, .
\label{scalel}
\end{equation}
It is striking that
the coefficients of the logarithm terms for $\sg_2^2$, $\sg_2^3$ and  $\sg_3^2$
agree with values based on duality
with eleven-dimensional supergravity compactified on a two-torus \cite{grv:twoloop}.

The presence of the terms with coefficients $r\, \log r^2$ is essential in ensuring a
smooth ten-dimensional limit as $r\to \infty$.  To see this, recall that in this limit an
infinite series of terms with
positive powers of $r^2 \,s$ must resum in a manner that cancels the nine-dimensional square
root thresholds \cite{grv:bigpaper}, generating the ten-dimensional logarithmic thresholds, of
the form $r\,s^k\,\log (-r^2\alpha'\,s)$.
The resummation  therefore produces  $r\,\log r^2$ terms with coefficients that are the same as the
ten-dimensional massless threshold terms (\ref{grab}) computed in section~\ref{sec:threshold}.
More precisely,
the $r\log r^2$ terms produced in the resummation must appear with the same coefficients as
$ \log \mu_4 $ and $\log \mu_6 $ in (\ref{grabb}).
From (\ref{grabb}) we see that these $r\log r^2$ contributions are precisely canceled by the
terms  $\sg_2^2\, r\,\log(r^2)$, $\sg_{3}^2\,  r\,\log(r^2)$  and
 $\sg_{2}^3\, r\,\log(r^2)$ in~(\ref{fin}), as required.

The terms in equations~(\ref{fin}) satisfy an extended transcendentality condition in which a power of
$r^{\pm(1+2m)}$ (with $m\ge 0$) contributes weight $-m$ and $\log(r^2)$ has weight $1$.
As before, $\zeta(k)$ has weight $k$ and $\pi$ has weight $1$.
The total weight of any term of order $(\al)^q$ is once again equal to $q+1$.

Although we have displayed  results up to order $(\alpha'\,s)^6\,\hR^4$,
we have also obtained partial results at all orders.
Thus, we have evaluated the coefficients of all terms of the form $\sg_2^n$
(such as $\sg_2^4$, which  corresponds
 to one of the kinematic structures
appearing at order $(\alpha'\,s)^8\,\hR^4$),
in terms of harmonic sums or multiple zeta values, using general expressions derived in the appendices.
It is notable that, at least in all cases studied here, these reduce to
the product of Riemann zeta values. This leads to the interesting
possibility that the coefficients of the momentum expansion of the
genus-one four-particle amplitude
are all rational numbers multiplying products of Riemann zeta values.
Even more interesting is the possibility (motivated by
results from eleven-dimensional supergravity on $S^1$ and on $T^2$
\cite{grv:twoloop}) that this property holds
for all genera in string perturbation theory.

We have also determined the coefficients $u_k$ of terms of the form
$u_{k}\,(\alpha'\, s)^k \zeta(2k-1)r^{2k-1}$, which contains the leading power of $r$
for a given value of $k$.  These coefficients follow from the methods described in
 appendix~\ref{sec:leading},
 and the results agree with the type IIA expressions derived by taking the one-loop  amplitude
of eleven dimensional supergravity on $T^2$ \cite{Russo:1997mk,gkv:twoloop}.
The subleading contributions of the form $v_k\, (\alpha'\, s)^k \,r^{2k-7}$
 match those obtained from the two-loop amplitude of eleven dimensional supergravity on $T^2$,
 at least up to (and including) order $(\alpha'\,s)^6\, \hR^4$ \cite{grv:twoloop}.

\acknowledgments

We would like to thank Don Zagier for many useful discussions and the derivation of the identity
in appendix~\ref{sec:Zagier}.
We thank the Galileo Galilei Institute for Theoretical Physics for the hospitality
 and the INFN for partial
support during the completion of this work
P.V. would like to thank the LPTHE of the Pierre and Marie Curie University (Paris VI) for the
hospitality  during the completion of this work, and by the RTN contrats MRTN-CT-2004-503369,
MRTN-CT-2004-005104, the ANR project ANR-06-BLAN-3\_137168 
and NORDITA for partial financial support. J.R. also acknowledges support by MCYT FPA 2004-04582-C02-01 and CIRIT GC 2005SGR-00564. 

\appendix

\section{Some mathematical background}
\subsection{Eisenstein Series}\label{sec:Eisenstein}

The non-holomorphic Eisenstein Series $E_s$ are defined by
\be
E_s=\sum_{(m,n)\neq (0,0)} {\tau_2^s\over|m+n\tau|^{2s}}\ .
\ee
We will also  use the notation $\hat E_s=E_s/(4\pi )^{s}$ and
$E^*_s= \Gamma(s) \pi^{-s}\ E_s$.
The  function $E^*_s$ can be expanded at large
$\tau_2$ as follows
\begin{eqnarray}\label{ZsExp}
E^*_{s}(\tau,\bar\tau)&=&2\zeta^*(2s)\,\tau_{2}^{s}+ 2\zeta^*(2s-1)\, \tau_{2}^{1-s}\\
\nonumber&+& 4\tau_{2}^{1\over2}\, \sum_{N\neq0} \, |N|^{s-{1\over2}}\, \sg_{1-2s}(|N|)\, K_{s-{1\over2}}(2\pi |N| \tau_{2})\, e^{2i\pi N\tau_{1}}
\end{eqnarray}
where $\sg_{k}(n)=\sum_{d|n} \, d^k$ is the $k$th divisor function of
 $n$, $K_{s}(z)$ are the  Bessel functions of the second kind, and
 \begin{equation}\label{e:zetas}
\zeta^*(s) ={\zeta(s)\Gamma(s/2)\over \pi^{s/2}}=\sum_{p\geq1} \int_{0}^\infty
{dt\over t}\, t^{s\over2}\, e^{-\pi p^2 t}\ .
\end{equation}
This function satisfies the functional equation $\zeta^*(s)=\zeta^*(1-s)$ as is easily shown using
 the Poisson resummation formula \cite{Cartier}. The function $E^*_s$
obeys an analogous relation $E^*_s=E^*_{1-s}$
and satisfy the Laplace equation
\begin{equation}\label{LaplaceZs}
\Delta_{\tau} E^*_{s}(\tau,\bar\tau) =s(s-1) E^*_{s}(\tau,\bar\tau)\ .
\end{equation}

In the main text we need to integrate the product of a pair of Eisenstein series that
appears in the integrals of the functions $P^{(p,q)}(\{E_r\})$ over $\calF_L$.
We may use (\ref{LaplaceZs}) to replace one factor of $\hat E_s$ in the integral by $\Delta \hat E_s$
and then integrate by parts to give the result
\begin{equation}\label{CutInt}
{1\over 4\zeta^*(2s)\zeta^*(2s')}\, \int_{{\cal F}_{L}}{d^{2}\tau\over\tau_{2}^2}\,
 E^*_{s}E^*_{s'}= {L^{s+s'-1}\over s+s'-1} -
 {L^{1-s-s'}\over s+s'-1} \phi(s)\phi(s')+ {L^{s-s'}\over s-s'}\phi(s')- {L^{s'-s}\over s-s'}\phi(s)+o(1)\,,
\end{equation}
where
\be
\phi(s)=\zeta^*(2s-1)/\zeta^*(2s)\,.
\label{phidef}
\ee
In section~\ref{sec:amplitude} we need to use the special cases
\be
\phi(2) =\frac{\pi\zeta(3)}{2\zeta(4)}\,,\qquad \phi(3) =\frac{3\pi \zeta(5)}{8 \zeta(6)}\,.
\ee

For $s> s'$ and in the $L\to \infty$ limit, the right-hand side of (\ref{CutInt}) contains two
terms proportional to $L^{s+s'-1}$ and $L^{s-s'}$, respectively.
When $s=s'$ the integral may be evaluated by taking the $s\to s'$ limit
of (\ref{CutInt}).
\begin{equation}\label{CutIn}
{1\over 4\zeta^*(2s)^2}\, \int_{{\cal F}_{L}}{d^{2}\tau\over\tau_{2}^2}\,
( E^*_{s})^2= {L^{2s-1}\over 2s-1} -{L^{1-2s}\over 2s-1} \phi^2(s)
 + 2\phi(s)\, \log(L/\tilde\nA_{2s})+o(1)\,,
\end{equation}
where
\be
\log(\tilde \nA_{2s})={1\over 2}\,{\phi'(s)\over \phi(s)}= {\zeta'(2s)\over\zeta(2s)} +
{\Gamma'(s-1/2)\over 2\Gamma(s-1/2)} - {\Gamma'(s)\over 2\Gamma(s)}\,.
\ee
For the cases of interest in the main text we need
\bea
\log \tilde\nA_{4}&=&{1\over2}-\log(2)+{\zeta'(3)\over\zeta(3)}-{\zeta'(4)\over \zeta(4)} \label{asiete}\\
\log \tilde\nA_{6}&=&{7\over12}- \log(2)+{\zeta'(5)\over \zeta(5)}-{\zeta'(6)\over\zeta(6)}
\label{aocho}
\eea
%

\subsection{Space of square integrable functions}\label{sec:fund}

Consider a modular function $f(\tau,\bar\tau)$ with the following zero mode expansion
\begin{equation}
f^{0}(\tau_2)\equiv \int_{-\half}^\half d\tau_1 f^{0}(\tau,\bar\tau)=\sum_{k=1}^{n}  {a_{k}\over \tau_{2}^k}  +
\sum_{N\neq0}\sum_{k=0}^{m}
 {b_{k}(|N|) \over (2\pi\tau_{2})^k} \, e^{-2\pi |N| \tau_{2}}\ .
\label{fooo}
\end{equation}
The sums over $k$ are of finite range, $n$ and $m$ are in general different. The coefficients
$a_{k}$ and $b_{k}(|N|)$ are constrained by the modular invariance
of the function $f(\tau,\bar\tau)$.  Prototypes of such functions are the  modular functions $\delta j^{(p,q)}$
arising from the derivative expansion of the string loop amplitude, where
all the positive powers in $\tau_{2}$ in the zero mode expansion have been subtracted
by polynomials in the Eisenstein series as in appendix~\ref{sec:Expand}.

In this case the integral of $f(\tau,\bar\tau)$ over a fundamental domain of $\SL(2,Z)$
converges since this function is square integrable in the fundamental domain.
We apply the following lemma given, for example, on  page 256 of \cite{Terras}:

\smallskip

\noindent{\bf Lemma}  {\sl
The space of square integrable functions on $L^2({\cal F})$ on a fundamental domain ${\cal F}=\SL(2,Z)\backslash H$
is given by the orthogonal decomposition
\be
L^2({\cal F}) = L_{0}^2({\cal F})\oplus \mathbb{C}\oplus \theta_{0}
\label{appone}
\ee
where

\be
L_{0}^2({\cal F})=\{f\in L^2({\cal F})| \int_{-1/2}^{1/2} d\tau_{1} f(\tau)=0 {for\ almost\ all\ } \tau_{2}>0\}
\label{apptwo}
\ee
and $\theta_{0}$ denotes the closed subspace of $L^2({\cal F})$
generated by the incomplete theta series,
\be
T\psi(z)= \sum_{\gamma\in\Gamma_{\infty}\backslash \Gamma} \psi(\Im m (\gamma \tau))\,
{\rm\ for\ } \tau\in H {\rm\ and\ } \Gamma_{\infty}=\{\pm \left(\begin{matrix}1&n\\
0&1\end{matrix}\right)| n\in\mathbb{Z}\}
\label{appthree}
\ee
such that
$$
\int_{0}^\infty {d\tau_{2}\over \tau_{2}^2} \psi(\tau_{2})=0=\int_{{\cal F}} {d^2\tau\over\tau_{2}^2}\,T\psi(\tau)\,,
$$
where $\psi$ is smooth with compact support on $\mathbb{R}^+$.
}

Since $f^0$ in (\ref{fooo}) does not have any component on $L_{0}^2({\cal F})$
 and does not contain any constant term  it belongs to the space $\theta_{0}$.
For any incomplete theta series in $\theta_{0}$ the integral over the fundamental domain vanishes.
Therefore for any $f\in\theta_{0}$ we
have
$$
\int_{{\cal F}}{d^2\tau\over\tau_{2}^2} \, f(\tau)=0\ .
$$

\subsection{Harmonic Sums}\label{sec:Harmonic}

 In appendix~\ref{sec:ZMdiag} we will find
different types of harmonic sums that arise upon
integrating over the vertex positions. In this subsection
  we start by discussing the simplest cases.

\subsubsection{$S(m,n)$ and $S(\a_1,\cdots,\a_n;\beta)$}

Consider the sum
\begin{equation}\label{sHnm}
 S(m,n)= \sum_{k_{1},\dots,k_{m}\neq0} \, {\delta(\sum_{1\leq i\leq m} k_{i})
 \over |k_{1}\cdots k_{m}| (|k_{1}|+\cdots+|k_{m}|)^{n}}\ ,\qquad m\geq2\ .
 \end{equation}
As shown in section~\ref{sec:Zagier}, this sum reduces to a sum over multi-zeta values (MZV's), with the
result  (for $m\geq2$)
\begin{equation}\label{e:SmnFormula}
S(m,n)= m!\,\sum_{a_{1},\dots, a_{r}\in \{1,2\}\atop a_{1}+\cdots + a_{r}=m-2} \, 2^{2(r+1)-m-n} \, \zeta(n+2,a_{1},\dots,a_{r})\ ,
\end{equation}
where $\zeta(n_{1},\dots, n_{r})$ is a multiple zeta value of weight $w=\sum_{i=1}^r s_{i}$ and depth $r$ defined by
\begin{equation}\label{eMZV}
\zeta(s_{1},\dots,s_{r})= \sum_{n_{1}> n_{2}> \cdots > n_{r}\geq1} {1\over n_{1}^{s_{1}}\cdots n_{r}^{s_{r}}}\ .
\end{equation}
In general,  $\zeta(m,n)$ does not reduce to a polynomial in zeta values.

In the following  we give  details  of $S(3,n)$, $S(4,n)$ and $S(5,1)$, specializing to the
cases needed in the main text.
\begin{itemize}
\item $S(2,n)$ is given by
\be
S(2,n)=2^{1-n}\, \zeta(n+2)\,.
\label{s2n}
\ee
\item  $S(3,n)$ is given by
 \bea S(3,n)&=& {3\over 2^{n-2}}\,\zeta(n+2,1)\nn\\
&=& {3\over 2^{n-1}}\, \left((n+2)\zeta(n+3)-\sum_{k=1}^{n}\zeta(n+2-k)\zeta(k+1)\right)\,,
\label{s3n}
\eea
which has been reduced to  zeta values using the  identity  \cite{HarmonicSum}
\begin{equation}\label{Hsum}
\zeta(n,1) =
{n\over2}\zeta(n+1)-{1\over2} \sum_{k=1}^{n-2} \zeta(n-k)\zeta(k+1)
\end{equation}
In particular, using various expressions for MZV given in the
references~\cite{HarmonicSum,borwein2,borwein,broadhurst,Petitot,EulerSum}
we have
\begin{eqnarray}
S(3,1)&=&{3\over2} \, \zeta(4)
\label{e:S31}
\\
\label{e:S32}
S(3,2)&=&6 \zeta(5)- 3\zeta(2)\zeta(3)\\
S(3,3)&=& {9\over8}\zeta(6)-{3\over4}\zeta(3)^2
\end{eqnarray}
where we have used $\zeta(2)\zeta(4)= 7\zeta(6)/4$ and $\zeta(2)^3=35\zeta(6)/8$.
\item $S(4,n)$ is given by
\begin{eqnarray}
\nonumber S(4,n)&=&{4!\over 2^n}\, \left(\zeta(n+2,2)+4\zeta(n+2,1,1)\right)\ .
\end{eqnarray}
In particular, we find
\begin{eqnarray}\label{e:S41}
S(4,1)&=& 30\zeta(5)- 12 \zeta(2)\zeta(3)\\
\label{e:S42}
S(4,2)&=&{53\over2} \zeta(6)- 18 \zeta(3)^2\\
\label{e:S43}
S(4,3)&=&-\frac{4}{5}\,\zeta(3)\,\zeta(4) - 21\,\zeta(2)\,\zeta(5) + 27\,\zeta(7)\\
S(4,4)&=&-{3\over5}\zeta(6,2)+3\,\zeta(2)\,\zeta(3)^2 - 15\,\zeta(3)\,\zeta(5) +
  \frac{463}{40}\,\zeta(8)
\end{eqnarray}
In the last expression we see the appearance, at weight 8, of the MZV  $\zeta(6,2)$ that does
not reduce to a polynomial in zeta values.
\end{itemize}
Another sum that appears is
\begin{equation}\label{e:bigsums}
S(\alpha_{1},\dots,\alpha_{n};\beta)= \sum_{m_{i}\geq1} {1\over m_{1}^{\alpha_{1}}\cdots m_{n}^{\alpha_{n}}\, (m_{1}+\cdots+m_{n})^\beta}\ .
\end{equation}
Ordering the $\alpha_{i}$ as $\alpha_{n}\geq \alpha_{n-1}\geq \cdots \geq \alpha_{1}$ and introducing the variables
$\mu_{1}=m_{1}$, and  $\mu_{r}=\mu_{r-1}+m_{r}$  for $2\leq r\leq n$, this sum can be written as

 \begin{equation}
S(\alpha_{1},\dots,\alpha_{n};\beta)= \sum_{\mu_{n}>\mu_{n-1}>\cdots>\mu_{1}\geq1}
{1\over   \mu_{n}^\beta (\mu_{n}-\mu_{n-1})^{\alpha_{n}}\cdots (\mu_{2}-\mu_{1})^{\alpha_{2}}\mu_{1}^{\alpha_{1}}}\ .
\end{equation}
A repeated  use of the identity  \cite{ZagierGanglKaneko}
\begin{equation}
{1\over m^{i} n^{j}} = \sum_{r+s=i+j\atop r,s>0}\, {\left(r-1 \atop i-1\right)\over (m+n)^r n^s}+
{\left(r-1 \atop j-1\right)\over (m+n)^r m^s}\ ,\qquad i,j>0\ ,
\end{equation}
gives, for $\alpha_{i}>0$ and $\beta>0$,
 \begin{eqnarray}
\nn S(\alpha_{n},\dots,\alpha_{1};\beta)&= &
\sum_{r_{1}+s_{1}=\alpha_{2}+r_{0}\atop r_{1},s_{1}>0}
\sum_{r_{2}+s_{2}=\alpha_{3}+r_{1}\atop r_{2},s_{2}>0}
\cdots\sum_{r_{n-1}+s_{n-1}=\alpha_{n}+r_{n-2}\atop r_{n-1},s_{n-1}>0}
\\
&\times& \prod_{i=1}^{n-1}\left[ \left(r_{i}-1 \atop \alpha_{n+1}-1\right)
+\left(r_{i}-1 \atop r_{n-1}-1\right) \right] \, \delta_{r_{0},\alpha_{1}}
\zeta(\beta+r_{n-1},s_{n-1},\dots,s_{1})
\nn
\end{eqnarray}
In the special case $\beta=0$ we find
\begin{equation}
S(\alpha_{n},\dots,\alpha_{1};0)= \prod_{i=1}^n \, \zeta(\alpha_{i})\,.
\end{equation}
The $n=2$ case gives the Witten zeta-function $S(\alpha_{2},\alpha_{1},\beta)= W(\alpha_{2},\alpha_{1},\beta)$,
with
\begin{eqnarray}
 W(\alpha_1,\alpha_2,\beta)&=&\sum_{m,n=1}^\infty {1\over m^{\alpha_2} n^{\alpha_1} (m+n)^\beta }\nn\\
&=& \sum_{r+s=\alpha_{2}+\alpha_{1}\atop r,s>0}\,\left[ \left(r-1 \atop \alpha_{2}-1\right)
+\left(r-1 \atop \alpha_{1}-1\right) \right] \, \zeta(\beta+r,s)\,,
\label{e:Wittenzeta}
\end{eqnarray}
so
\be
W(0,\alpha, \beta)=W(\alpha,0,\beta)=\zeta(\beta,\alpha)\,,\qquad
W(\alpha,\beta,0)=\zeta(\alpha)\zeta(\beta)\,.
\label{wzero}
\ee
 If $\alpha_{i}=0$ for $p$ values of $i$ we have
 \begin{eqnarray}
 S(\alpha_{n},\dots,\alpha_{p+1},0,\dots,0;\beta)&=&\zeta(\beta,\alpha_{n},\dots, \alpha_{p+1},0,\dots,0)
 \end{eqnarray}
%

\subsubsection{Evaluation of $S(m,n)$ by Don Zagier}\label{sec:Zagier}

In this appendix we give a proof due to Don Zagier 
of the general formula~(\ref{e:SmnFormula}) for the values of the sums
\begin{equation}\label{e:SmmProof}
S(m,n)=  \sum_{k_{1},\dots,k_{m}\neq0} \, {\delta(\sum_{1\leq i\leq m} k_{i})
 \over |k_{1}\cdots k_{m}| (|k_{1}|+\cdots+|k_{m}|)^{n}}\ ,\qquad (m,n\geq0)
\ .
\end{equation}
(Note that $S(m,n)=0$ if $m<2$, since then the sum is empty.)
Denoting by $r$ and $m-r$ the number of $i$  with $k_{i}>0$ and $k_{i}<0$, respectively,
and by $l$ the sum of the positive $k_{i}$, we can rewrite $S(m,n)$ as
\begin{equation}
S(m,n)= \sum_{r=0}^{m}\, {m\choose r} \, \sum_{l=1}^{\infty}\, {S_{r}(l) S_{m-r}(l)\over (2l)^n}
\end{equation}
where
\begin{equation}\label{e:Srl}
S_{r}(l)= \sum_{k_{1},\cdots k_{r}\geq1\atop k_{1}+\cdots +k_{r}=l} \, {1\over k_{1}\cdots k_{r}}
\qquad (=0 \quad \textrm{if}\quad r=0)\ .
\end{equation}
Observing that $S_{r}(l)$ is the coefficient of $x^l$ in the series expression of ${\rm Li}_{1}(x)^r$, where ${\rm Li}_{1}(x)=\sum_{k\geq1} x^k/k=-\log(1-x)$, 
we obtain 
\begin{eqnarray}
\nonumber 2^n\, S(m,n)&= &\sum_{l=1}^{\infty} \, {1\over l^n}\,  \sum_{r=0}^{m}\, {m\choose r}
{\rm coeff}_{x^ly^l}\left[ {\rm Li}_{1}(x)^r \, {\rm Li}_{1}(y)^{m-r} \right]\\
&=&\sum_{l=1}^{\infty} \, {1\over l^n}\,
{\rm coeff}_{x^ly^l}\left[ \big({\rm Li}_{1}(x)+{\rm Li}_{1}(y)\big)^{m} \right]
\end{eqnarray}
Hence the generating function $\sum_{m\geq0} S(m,n)\, X^m/m!$ is given by 
\begin{eqnarray}
\nonumber 2^n\,\sum_{m\geq0} S(m,n) \, {X^m\over m!}
&=& \sum_{l=1}^{\infty} {1\over l^{n}} \,
 {\rm coeff}_{x^ly^l} \left[\exp\big(X ({\rm Li}_{1}(x)+{\rm Li}_{1}(y))\big)
\right]\\
\nonumber &=& \sum_{l=1}^{\infty} {1\over l^{n}} \,
 {\rm coeff}_{x^ly^l}\left[ (1-x)^{-X}\, (1-y)^{-X}\right]\\
\nonumber&=&\sum_{l=1}^{\infty} \, {1\over l^{n}}\, {X+l-1\choose l}^2 \\
\nonumber&=&\sum_{l=1}^{\infty}\, {X^2\over l^{n+2}} \,\prod_{h=1}^{l-1} \left(1+{X\over h}\right)^2 \\
\nonumber&=&\sum_{l=1}^{\infty}\, {X^2\over l^{n+2}} \,\prod_{h=1}^{l-1}\, \left(1+4\sum_{a\in\{1,2\}} {(X/2)^{a}\over h^{a}}\right) \\
\label{e:Star}&=&
\sum_{r\geq0}\sum_{l>h_{1}>\cdots >h_{r}>0\atop a_{1},\dots,a_{r}\in\{1,2\}}
{4^r\, X^2 (X/2)^{a_{1}+\cdots +a_{r}}\over l^{n+2} \, h_{1}^{a_{1}}\cdots h_{r}^{a_{r}}}\ .
\end{eqnarray}
Comparing the coefficients of $X^m$ on both sides gives the desired formula
\be
S(m,n)= {m!\over 2^{m+n-2}}\, \sum_{a_{1},\dots,a_{r}\in\{1,2\}\atop a_{1}+\cdots +a_{r}=m-2}\, 
2^{2r}\, \zeta(n+2,a_{1},\dots, a_{r})\ .
\ee

\section{Properties of $\D_{\ell_{1},\dots,\ell_{6}}$}\label{sec:ZMdiag}

\begin{figure}[h]
\centering\includegraphics[]{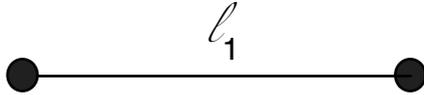}
\caption{The two-vertex diagram with $\ell_{1}$ lines connecting the two vertices
defines $\D_{\ell_{1}}$.
Each line is associated with a propagator on the torus with momentum $\p_{i}=m_{i}+n_{i}\tau$}
($i=1,\dots,l_1$).
\label{fig:vertex2}
\end{figure}

The coefficients of the terms in the analytic part of the momentum expansion
are determined in terms of the functions $\D_{\ell_{1},\dots,\ell_{6}}$ ($\sum_{k=1}^6l_k =r$)
associated with the diagrams shown in
figures~\ref{fig:vertex2}--\ref{fig:vertex4b}.
In the main part of the paper these enter in two separate manners.

1) In section~\ref{sec:ninedim} we considered compactification on a circle of radius $r$
and considered the coefficients of terms at each order up to $s^6$ that are power-behaved in $r$
for large $r$.
In this case the coefficients are determined by knowledge of the terms in $\D_{\ell_{1},\dots,\ell_{6}}$
that are power-behaved in $\tau_2$ for large $\tau_2$.  Such terms are obtained by expanding the
$\tau_1$,
\be
\D^{(0)}_{\{\ell\}}(\tau_2)=\int_{-\half}^\half \D_{\{\ell\}}(\tau,\bar\tau)
\label{dzerodef}
\ee
 in powers of $\tau_2^{-1}$.
This will also be carried out in this appendix for all the $D_{\{\ell\}}$'s that enter up to order $s^6$
(although we will not evaluate the coefficients of the negative powers of $r$ at order $s^5$ and beyond).

2)  In section~\ref{sec:amplitude}
we saw that the values of the coefficients of terms in the low-momentum
expansion in ten dimensions are determined in terms of a constant that
survives the $\tau_2\to \infty$ limit after subtracting the positive powers of $\tau_2$
with a polynomial in Eisenstein series, $P^{(p,q)}(\{\hat E_{r}\})$.
In this appendix we will here derive $P^{(p,q)}$
and the value of the constant for each $D_{\{\ell\}}$ function that enters
up to order $s^6$.
In fact, the part of $D_{\{\ell\}}$ that has positive
powers of $\tau_2$, or is constant in the large-$\tau_2$ limit, is independent of $\tau_1$ so it is
also determined by $D^{(0)}_{\{\ell\}}$.

For these reasons, our priority here is to compute the $\tau_1$ zero modes of the $D_{\{\ell\}}$'s.
For this purpose we will need to make
extensive use of the representation of the propagator
as the sum  $\Pinf(\nu)+\PP(\nu)$ given in~(\ref{StringProp}).
A few special cases were evaluated in \cite{gv:stringloop}.

\subsection{The two-vertex case}\label{sec:vertex2}

The two-vertex diagram with $\ell_{1}$ lines is associated with a modular
function of weight $\ell_{1}$, which we will denote by
\be
\D_{\ell_1}\equiv \D_{\ell_1,0,0,0,0,0}\, ,
\label{twod}
\ee
in which case we have
\begin{equation}
\D_{\ell_{1}}=\sum_{(m_i,n_i)\neq (0,0)} \, \delta(\sum_{r=1}^{l_1} \p_r)
 \prod_{i=1}^{\ell_{1}}{1\over 4\pi}{\tau_{2}\over |m_{i}+n_{i}\tau|^{2}}\ .
\end{equation}
The zero mode is given by\footnote{In this
section we use the condensed notation $\calP_{ij}=\calP(\nu^{(ij)})$,  $\Pinf_{ij}=\Pinf(\nu^{(ij)})$ and $\PP_{ij}=\PP(\nu^{(ij)})$.}
\begin{eqnarray}
\D^{(0)}_{\ell_{1}}&=& \int_{-\half}^{\half}d\tau_{1}\,\int {d\nu^{(1)}d\nu^{(2)}\over\tau_{2}^2}\,
\calP_{12}^{\ell_{1}}\\
\nn&=& \sum_{r+s=\ell_{1}} \,{\ell_{1}\choose r}
 \int_{-\half}^{\half}d\tau_{1}\, \int {d\nu^{(1)}d\nu^{(2)}\over\tau_{2}^2}\,
(\Pinf_{12})^{r}  (\PP_{12})^{s}\\
\nn&=&\sum_{r+s=\ell_{1}} \,{\ell_{1}\choose r}\,\int_{0}^1  d\hat\nu^{(1)}_{2}d\hat\nu^{(2)}_{2}\,
(\Pinf_{12})^{r}\\
\nn&\times& \sum_{m_{1},\dots,m_{s}\neq0\atop k_{1}\cdots k_{s}\in\mathbb{Z}}  {\delta(\sum_{i}m_{i})\delta(\sum_{i} m_{i}k_{i})\over 4^{s}|m_{1}\cdots m_{s}|}\,
e^{-2\pi \tau_{2}\sum_{i}|m_{i}| |k_{i}-\hat\nu_{2}^{(12)}|}\ .
\end{eqnarray}
A direct evaluation of the integrals using the identity
\begin{equation}
\int_0^1 dx_1 \int_0^{x_1} dx_2 f(x_1-x_2)= \int_0^1 dx \, (1-x)\, f(x) \, ,
\end{equation}
with $f(x)=(\Pinf(x))^n\, \exp(-2\pi \tau_{2}\sum_{i} |m_{i}||k_{i}-x|)$, and using
the periodicity of the asymptotic propagator $\Pinf(x)$, leads to
\begin{eqnarray}
\D_{\ell_{1}}^{(0)}&=&\left(\pi\tau_{2}\over12\right)^{\ell_{1}}\,{}_{2}\textrm{F}_{1}(1,-\ell_{1},3/2;3/2) +\\
\nn&+&{2\over 4^{\ell_{1}}} \sum_{k=0}^{\ell_{1}-2}\sum_{k_{1}+k_{2}+k_{3}=k} \,
{(-1)^{k_{2}}\over 6^{k_{3}}}
{\ell_{1}!(2k_{1}+k_{2})!\over (\ell_{1}-k)! k_{1}!k_{2}!k_{3}! }\,
(2\pi\tau_{2})^{k_{3}-k_{1}-1}\times\\
\nn&\times& S(\ell_{1}-k,2k_{1}+k_{2}+1)
+O(\exp(-\tau_{2}))\,,
\end{eqnarray}
where the sums $S(m,n)$ are defined in~(\ref{sHnm}) and evaluated in section~\ref{sec:Zagier}.

Now we will write down the power-behaved terms
in the large $\tau_2$ expansion for some particular cases.
 The first non-trivial case is $\D_{2}=\hat E_{2}$. The next cases are
 $\D_{3}$, $\D_{4}$ (called $B_{2}$ and $C_{4}$ in~\cite{gv:stringloop}).
We will also need $\D_{5}$ and $\D_{6}$. Their zero-mode
expansions are given by
\begin{eqnarray}\label{e:ZMD2}
\D^{(0)}_{2}&=&{1\over (4\pi)^2}\, \left[2\zeta(4)\tau_{2}^2+{ \pi\zeta(3)\over\tau_{2}}\right]\\
\D^{(0)}_{3}&=& {1\over (4\pi)^3 }\, \left[2 \zeta(6)\tau_{2}^3+ \pi^3 \zeta(3)
 +{3\pi\over 4}{\zeta(5)\over\tau_{2}^2}\right]+O(e^{-\tau_{2}})
\label{e:ZMD3}\\
\D^{(0)}_{4}&=&{1\over (4\pi)^4}\, \left[10 \zeta(8)\tau_{2}^4
+{2\pi^5\over 3}\, \zeta(3)\,\tau_{2}+ {10 \pi^3\zeta(5)\over \tau_{2}}
- {3\pi^2\zeta(3)^2\over\tau_{2}^2}+ {9\pi\over4}\,{\zeta(7)\over \tau_{2}^3}\right]
+O(e^{-\tau_{2}})
\label{e:ZMD4}\\
\label{e:ZMD5}
\D^{(0)}_{5}&=&{1\over (4\pi)^5}\,\left[
20\zeta(10)\tau_2^5  + \frac{10\,\pi^7\,\zeta(3)}{27} \,\tau_2^2
+  \frac{95\,\pi^5\,\zeta(5)}{6}\right.\\
\nn&+&\left.
  \frac{900\,\zeta(4)\,\zeta(3)^2}{\tau_2}
    + \frac{105\,\pi^3\,\zeta(7)}{4\,\tau_2^2}
    - \frac{135\,\zeta(2)\,\zeta(3)\,\zeta(5)}{\tau_2^3}
  + \frac{225\,\pi \,\zeta(9)}{16\,\tau_2^4}
\right]
+O(e^{-\tau_{2}})
\end{eqnarray}
\begin{eqnarray}
\label{e:ZMD6} \D^{(0)}_{6}&=& {1\over (4\pi)^6}\, \left[
\frac{46375\,\zeta(12)\,\tau_2^6}{691} + \frac{5\,\pi^9\,\tau_2^3\,\zeta(3)}{27} +
  2365\,\zeta(6)\,\zeta(3)^2 + \frac{140\,\pi^7\,\zeta(5)}{9}\,\tau_2\right.\\
\nn&   -&
  \frac{3\,\pi^5\,\left( 34020\,\zeta(4)\,\zeta(3) + 42120\,\zeta(2)\,\zeta(5) - 117115\,\zeta(7) \right) }
   {32\,\tau_2} -
  \frac{12150\,\zeta(4)\,\zeta(3)\,\zeta(5)}{\tau_2^2}\\
   \nn& +& \frac{45\,\pi^3\, ( 2\, \zeta(3)^3 - 14\, \zeta(3)\, \zeta(6) + 9\, \zeta(9) )   }{2\,\tau_2^3}\,
          \\
\nn &-&\left.
 \frac{4050\,\zeta(2)\,\left( {\zeta(5)}^2 + 2\,\zeta(3)\,\zeta(7) \right) }
   {8\,\tau_2^4}
+\frac{4725\,\pi \,\zeta(11)}{32\,\tau_2^5} \right]+O(e^{-\tau_{2}})
\end{eqnarray}
It is notable that all the coefficients appearing in this expansion are products of zeta values
multiplied by rational coefficients.
Thanks to this, the positive powers of $\tau_2$ in each $\D_{\{\ell\}}$  function can be matched
with a polynomial in Eisenstein series. We find
\begin{eqnarray}
\D_{2} &=& \hat E_{2}\\
\D_{3}&=& \hat E_{3} +{ \zeta(3)\over 64}+ \D_{3}^{fin}\\
\D_{4}&=&-30 \hat E_{4}+15\, \hat E_{2}^2 + \D^{fin}_{4}\\
\D_{5}&=&-375\, \hat E_{5}+175\,\hat E_{2}\hat E_{3}+{155\over12288}\,\zeta(5)+\D^{fin}_{5} \\
\D_{6}&=&-{8297625\over 691} \,\hat E_{6}+4900 \,\hat E_{3}^2+875 \hat E_{2}\hat E_{4}+
{25\over4096}\, \zeta(3)^2+\D^{fin}_{6}
\end{eqnarray}
where $\D^{fin}_{\ell_{1}}$ is such that $\lim_{\tau_{2}\to\infty} \D^{fin}_{\ell_{1}}= 0$.

 \subsection{The three-vertex case}\label{sec:vertex3}

\begin{figure}[h]
\centering\includegraphics[]{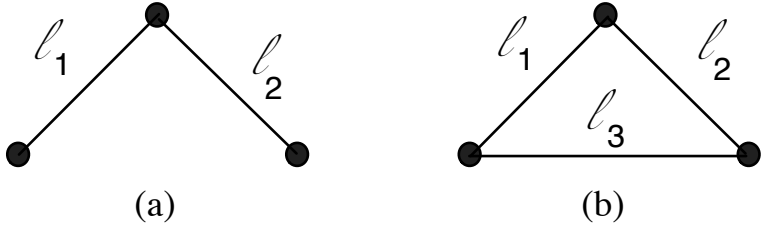}
\caption{
The two possible three-vertex diagrams, with $\ell_{1}$,
$\ell_{2}$ and $\ell_{3}$ lines joining the vertices defines $\D_{\ell_{1},\ell_{2},\ell_{3}}$.
Setting $l_3=0$ in (b) reduces the figure to the product of  two two-vertex diagrams
$\D_{\ell_{1},\ell_{2},\emptyset}=\D_{\ell_{1}}\times \D_{\ell_{2}}$ shown in (a).}
\label{fig:vertex3}\end{figure}

The three-vertex diagrams are associated with the functions
$\D_{\ell_{1},\ell_{2},\ell_{3}}\equiv
\D_{\ell_1,\ell_2,\ell_3;0,0,0}$,
which has the form
\begin{equation}
\D_{\ell_{1},\ell_{2},\ell_{3}}
=\sum_{(m^{(j)}_{i},n^{(j)}_{i})\neq(0,0)}
\prod_{1\leq r<s\leq 3}\delta\left(\sum_{k=1}^{l_r} \p^{(r)}_{k}-\sum_{k=1}^{l_s} \p^{(s)}_{k}
\right) \prod_{j=1}^3
\prod_{i=1}^{\ell_{j}}
{1\over 4\pi}{\tau_{2}\over |m^{(j)}_{i}+n^{(j)}_{i}\tau|^2}\,,
\end{equation}
where $\p_k^{(r)}$ is the momentum of the $k$th line in the $r$th leg.
Momentum conservation at each vertex implies that when one set of integers is
empty the diagram reduces to the product of two two-vertex diagrams,
 $\D_{\ell_{1},\ell_{2},\emptyset}= \D_{\ell_{1}}\times \D_{\ell_{2}}$.
Some particular cases are
\begin{eqnarray}
\nn\D_{1,1,1}&=&\hat E_{3}\, ,\\
\D_{0,2,2}&=& \hat E_{2}^2\, , \\
\nn\D_{0,2,3}&=&\D_{3}\times \hat E_{2}\, .
\end{eqnarray}

For the general case where all the $\ell_{i}$ are non zero we extract the zero mode contribution as in the previous subsection by splitting the
string propagator into its asymptotic part and the finite part at large $\tau_2$. Starting with
\begin{eqnarray}
\D^{(0)}_{\ell_{1},\ell_{2},\ell_{3}}&=& \int_{-\half}^{\half}d\tau_{1}\,\int {d\nu^{(1)}d\nu^{(2)}d\nu^{(3)}\over \tau_{2}^3}\,
\calP_{12}^{\ell_{1}} \calP_{23}^{\ell_{2}} \calP_{13}^{\ell_{3}}\,,
\label{Dexp}
\end{eqnarray}
where none of the integers $n_{i}$ is zero and splitting the propagator,
 we find the general expression
\begin{eqnarray}
\nn\D^{(0)}_{\ell_{1},\ell_{2},\ell_{3}}&=&\sum_{\ell_{i}=r_{i}+s_{i}}\,\prod_{i=1}^3 {\ell_{i}!\over r_{i}!s_{i}!}\,
\int_{0}^1  d\hat\nu^{(i)}_{2}\,
(\Pinf_{12})^{r_{1}} (\Pinf_{23})^{r_{2}} (\Pinf_{13})^{r_{3}}\times\\
&\times&
\sum_{m^{(ij)}_{i}\atop k^{(ij)}_{l}}\, \prod_{1\leq i<j\leq3}  {\delta(\sum_{i}m^{(ij)}_{i})\over |m^{(ij)}_1| \cdots  |m^{(ij)}_{s_{ij}}|}
{\delta(\sum_{i,j,l} m^{(ij)}_{l} k^{(ij)}_{l})\over 4^{s_1+s_2+s_3}}\times\\
\nn&\times& e^{-2\pi \tau_{2}\sum |m^{(ij)}_{l}| |k^{(ij)}_{l}-\hat\nu_{2}^{(ij)}|}\, .
\end{eqnarray}
It is convenient to  decompose this into different contributions:
\begin{equation}
\D^{(0)}_{\ell_{1},\ell_{2},\ell_{3}}
= \D^{\infty}_{\ell_{1},\ell_{2},\ell_{3}}+ \D^{(a)}_{\ell_{1},\ell_{2},\ell_{3}}+ \D^{(b)}_{\ell_{1},\ell_{2},\ell_{3}}
+ \D^{(c)}_{\ell_{1},\ell_{2},\ell_{3}}+ \D^{(c)}_{\ell_{2},\ell_{1},\ell_{3}}+ \D^{(c)}_{\ell_{3},\ell_{2},\ell_{1}}
\label{diffcont}
\end{equation}
where $\D^\infty$ represents the contribution where $s_{i}=0$ for all $i=1,2,3$, $\D^{(a)}$ and $\D^{(b)}$
represent the cases where $s_{i}\geq1$ for all $i=1,2,3$ and $s_{1}=0$ with $s_{2},s_{3}\geq2$, and
$\D^{(c)}$ represents the cases where $s_{1}=s_{2}=0$, $s_{3}\geq2$.
The resulting expressions involve   sums of the form
 \begin{eqnarray}
\nn S(s_{1},s_{2};s_{3};\alpha,\beta)&= &\sum_{(\underline{m^{1}}, \underline{m^{2}},
\underline{m^{3}})} {\delta(\sum_{j}m^1_{i}-\sum_{j}m^2_{j})\delta(\sum_{j}m^1_{i}-\sum_{j}m^3_{j})
 \over \prod_{j=1}^{s_{1}} |m^{1}_{j}|  \prod_{j=1}^{s_{2}} |m^{2}_{j}|
 \prod_{j=1}^{s_{3}} |m^{3}_{j}| }\\
 &\times&
 {1\over (|\underline{m^1}|+ |\underline{m^{3}}|)^{\alpha}(|\underline{m^2}|+
 |\underline{m^{3}}|)^{\beta}}  \,,
 \label{sumha}
\end{eqnarray}
 where $|\underline{m}^{i}|= \sum_{j} |m^{i}_{j}|$.  This is not one of the expressions that we analyzed in
 the earlier appendices.

{}For the term $\D^\infty_{\ell_{1},\ell_{2},\ell_{3}}$ we find
\begin{eqnarray}
\D^\infty_{\ell_{1},\ell_{2},\ell_{3}}&=&
 2\,\left(\pi\tau_{2}\over2\right)^{\ell_{123}}\,
\sum_{a_{i}+b_{i}+c_{i}=\ell_{i}} \prod_{i=1}^3 {\ell_{i}!\over a_{i}! b_{i}! c_{i}!}\,
 { (-1)^{b_{123}}\over 6^{c_{123}}}\\
 \nn &\times& {(2a_{2}+b_{2})! (2a_{3}+b_{3})!\over (2(a_{2}+a_{3})+b_{2}+b_{3}+1)!}\,
{ 1 \over(2a_{123}+b_{123}+2)}\,,
\end{eqnarray}
where we are using the compact  notation,
\be
x_{123}=x_{1}+x_{2}+x_{3}\,,\qquad  \rm{with\ \ } x=a,b,c \ \rm{or\ } n\,.
\label{compactt}
\ee
When all $s_{1}$, $s_{2}$ and $s_{3}$ are greater than 1, or $s_{1}=0$ and $s_{2}$ and $s_{3}$ greater than 2 we
have
\begin{eqnarray}
\nn\D^{(a)}_{\ell_{1},\ell_{2},\ell_{3}}&=&
{2\over 4^{\ell_{123}}} \widehat{\sum} \prod_{i=1}^3 {r_{i}!\over a_{i}!b_{i}! c_{i}!}
\, {(-1)^{b_{123}}\over 6^{c_{123}} }\,(2\pi\tau_{2})^{c_{123}-a_{123}-2}\times\\
&\times& {(2a_{3}+b_{3})! l_{1}! l_{2}!\over q_{1}! q_{2}!}\,
  S(s_{1},s_{2};s_{3};l_{1}+1,l_{2}+1)\,,
\end{eqnarray}
where the summation is over
\begin{eqnarray}
\ell_{i}&=&r_{i}+s_{i}\,,\\
r_{i}&=& a_{i}+b_{i}+c_{i}\,,\\
q_{1}+q_{2}&=&2a_{3}+b_{3}\,,\\
l_{i}&=&2a_{i}+b_{i}+q_{i}\,,
\end{eqnarray}
and
\begin{eqnarray}
\nn\D^{(b)}_{\ell_{1},\ell_{2},\ell_{3}}&=&
{2\over 4^{\ell_{123}}}
\widehat{\sum} \prod_{i=1}^3 {r_{i}!\over a_{i}!b_{i}! c_{i}!}
\, {(-1)^{b_{123}}\over 6^{c_{123}} }\,(2\pi\tau_{2})^{c_{123}-a_{123}-2}\,{(2a_{3}+b_{3})!\over q_{1}! q_{2}!}\\
 \nn&\times&
\left[\sum_{r=0}^{l_{1}}\, (-1)^{q_{2}}\,{l_{1}! (l_{2}+r)!\over r!}
 S(s_{3},s_{2};s_{1};l_{1}-r+1,l_{2}+r+1)\right.\\
&+&\left. (1\leftrightarrow 2)
 \right]
\end{eqnarray}
and when $s_{1}\geq2$ and $s_{2}=s_{3}=0$  in~(\ref{Dexp})
\begin{eqnarray}
\D^{(c)}_{\ell_{1},\ell_{2},\ell_{3}}&=&
{2\over 4^{\ell_{123}}}
\widehat{\sum_{s_{2}=0\atop s_{3}=0}} \prod_{i=1}^3 {r_{i}!\over a_{i}!b_{i}! c_{i}!}
\, {(-1)^{b_{123}}\over 6^{c_{123}} }\,(2\pi\tau_{2})^{c_{123}-a_{123}-2}\times\\
 \nn&\times&(-1)^{q_{2}}\,{(2a_{3}+b_{3})!\over q_{1}! q_{2}!(l_{2}+1)}\,
\left[(-1)^{b_{3}}\,l_{1}!
 S(s_{1},l_{1}+1)\, (2\pi\tau_{2})^{l_{2}+1}\right.\\
\nn&+&\left.(1- (-1)^{b_{3}})\,
(2a_{123}+b_{123}+1)!
 S(s_{1},2a_{123}+b_{123}+2)
 \right]\,.
\end{eqnarray}
Substituting these expressions for $D^\infty$, $D^{(a)}$, $D^{(b)}$ and $D^{(c)}$ into
(\ref{diffcont}) gives the expression for $D_{\ell_1,\ell_2\ell3}$.
This completes the general expression for the three-vertex diagrams.
We list a few explicit examples
obtained by using the above general formulas and simplifying the harmonic sums (\ref{sumha}).
Thankfully, the functions $S(s_1,s_2,s_3;\alpha,\beta)$ that we have not evaluated drop out, apart from the special cases
$\alpha=\beta=0$ and $\alpha=\beta=1$, which are simple to evaluate directly. The result for the
zero mode expansion of the three vertex functions is
\begin{eqnarray}
\D^{(0)}_{1,1,1}&=&{1\over (4\pi)^3}\,\left(2 \zeta(6)\tau_{2}^3+{3\pi\over4}\, {\zeta(5)\over\tau_{2}^2}\right)\\
\nn\D^{(0)}_{1,1,2}&=&{1\over (4\pi)^4}\,\Big({4\over3}\zeta(8)\,\tau_{2}^4
+{2\zeta(4)\zeta(3)\over \pi}\,\tau_{2}
+{5\pi\over2}{\zeta(2)\zeta(3)\over\tau_{2}}\\
&+&\left({9\over2}\zeta(2)\zeta(3)^2-10\zeta(8)\right)\,{1\over\tau_{2}^2}+
{9\pi\over16}\,{\zeta(7)\over\tau_{2}^3}\Big)+O(e^{-\tau_{2}})\\
\D^{(0)}_{1,1,3}&=&{1\over (4\pi)^5}\,\Big({42\over5}\zeta(10)\tau_{2}^5
+21\zeta(3)\zeta(6)\,\tau_{2}^2+{33\pi\over2}
\zeta(4)\zeta(5)\Big)+O({1\over\tau_{2}})\\
  \D^{(0)}_{1,2,2}&=&{1\over (4\pi)^5}\,\Big({8\over5}\zeta(10)\tau_{2}^5
 +4\pi\zeta(3)\zeta(6)\,\tau_{2}^2+24\pi\zeta(4)\zeta(5)\Big)+O({1\over\tau_{2}})
\end{eqnarray}
\begin{eqnarray}
\nn\D^{(0)}_{1,1,4}&=&{1\over (4\pi)^6}\,\Big({9940\over691}\zeta(12)\tau_{2}^6
+140\pi\zeta(3)\zeta(8)\tau_{2}^3+{525\pi\over2}\zeta(5)\zeta(6)\tau_{2}\\
&+&{1449\over2}\zeta(3)^2\zeta(6)-{450450\over691}\zeta(12)\Big)+O({1\over\tau_{2}})\\
\nn\D^{(0)}_{2,2,2}&=&{1\over (4\pi)^6}\,\Big({10615\over691}\zeta(12)\tau_{2}^6
+30\pi\zeta(3)\zeta(8)\tau_{2}^3+177\pi\zeta(5)\zeta(6)\tau_{2}\\
&+&945\zeta(3)^2\zeta(6)-{1576575\over 1382}\zeta(12)\Big)+O({1\over\tau_{2}})
\\
\nn\D^{(0)}_{1,2,3}&=&{1\over (4\pi)^6}\,\Big({4470\over691}\zeta(12)\tau_{2}^6
+30\pi\zeta(3)\zeta(8)\tau_{2}^3+{519\pi\over4}\zeta(5)\zeta(6)\tau_{2}\\
&+&{63\over2764}(33859\zeta(3)^2\zeta(6)-35750\zeta(12))\Big)+O({1\over\tau_{2}})\,.
\end{eqnarray}
The notation  $O(1/\tau_{2})$  indicates the presence of terms that are
suppressed by inverse powers of $\tau_{2}$ that we have not calculated
(and there are also exponentially suppressed terms which are not evaluated).

As before, we can associate a quadratic function of  Eisenstein series that reproduces the
positive powers of $\tau_{2}$ of these zero modes,
\begin{eqnarray}
\D_{1,1,1}&=&\hat E_{3}\\
\D_{1,1,2}&=&-{1\over 2} \hat E_{4}+{1\over 2}\hat E_{2}^2+\D_{1,1,2}^{fin}\\
32\D_{1,1,3}-24 \D_{1,2,2}&=&-{2592\over 5}\,\hat E_{5}+288\,\hat E_{2}\hat E_{3}
-{11\over1920}\zeta(5)+32\D^{fin}_{1,1,3}-24 \D^{fin}_{1,2,2}\\
\D_{1,1,4}+\D_{2,2,2}-2\D_{1,2,3}&=&-{162575\over 691}\, \hat E_{6}
+60\, \hat E_{3}^2+55\, \hat E_{2}\hat E_{4}-{7\over 245760}\zeta(3)^2\\
\nn&&+\D^{fin}_{1,1,4}+\D^{fin}_{2,2,2}-2\D^{fin}_{1,2,3}
\end{eqnarray}

\subsection{The four-vertex case}\label{sec:vertex4}

\begin{figure}[!h]
\centering\includegraphics[width=10cm]{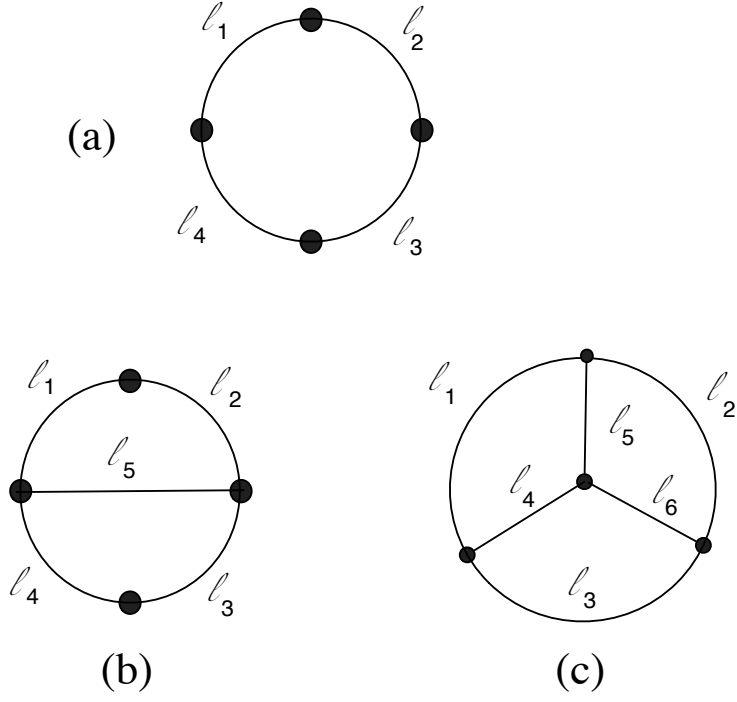}
\caption{The three possible non-degenerate four-vertex diagrams.
 (a)  defines    $\D_{\ell_{1},\ell_{2},\ell_{3},\ell_{4}}$,
 (b) defines   $\D_{\ell_{1},\ell_{2},\ell_{3},\ell_{4};\ell_{5}}$,
  (c)  defines $\D_{\ell_{1},\ell_{2},\ell_{3};\ell_{4},\ell_{5},\ell_{6}}$.}
\label{fig:vertex4a}\end{figure}

The general four-vertex diagrams shown in figure~\ref{fig:vertex4a} have
between four and six non-zero $\ell_k$'s.
We will begin by computing the general expression for the four-vertex
function of figure~\ref{fig:vertex4a}(a), in which $\ell_5=\ell_6=0$ and
 $\ell_1,\ell_2,\ell_3,\ell_4$ are non-zero and arbitrary.

\subsubsection{$\ell_5=\ell_6=0$}

In this case we define $\D_{\ell_{1},\ell_{2},\ell_{3},\ell_4}\equiv
\D_{\ell_{1},\ell_{2},\ell_{3},\ell_4;0,0}$, and  each four-vertex diagram is the modular function
\begin{equation}
\D_{\ell_{1},\ell_{2},\ell_{3},\ell_4} \equiv
\sum_{(m^{(j)}_{i},n^{(j)}_{i})\neq(0,0)}
\prod_{1\leq r<s\leq 4}\delta\left(\sum_{k=1}^{l_r} \p^{(r)} - \sum_{k=1}^{l_s}\p^{(s)}_{k}\right)
 \prod_{j=1}^4\prod_{i=1}^{\ell_{j}}
{1\over 4\pi}{\tau_{2}\over |m^{(j)}_{i}+n^{(j)}_{i}\tau|^2}
\end{equation}
In the particular case where one of the $\ell_i$ is zero, say $\ell_4=0$, this
reduces
to a product of three two-vertex functions,  $\D_{\ell_{1},\ell_{2},\ell_3,0}= \D_{\ell_{1}}\times
\D_{\ell_{2}}\times\D_{\ell_{3}}$, as in
figure~\ref{fig:vertex4b}(c), or $\D_{\ell_{1}}\times\D_{\ell_{2}}$ as in
~\ref{fig:vertex4b}(a).

For the general case where all the $\ell_{i}$ ($i=1,2,3,4$) are non zero we extract the
zero mode contribution as in the previous
subsections. We write
\begin{eqnarray}
\D^{(0)}_{\ell_{1},\ell_{2},\ell_{3},\ell_4}&=& \int_{-\half}^{\half}d\tau_{1}\,\int {d\nu^{(1)}d\nu^{(2)}d\nu^{(3)}\over \tau_{2}^3}\,
\calP_{12}^{\ell_{1}} \calP_{23}^{\ell_{2}} \calP_{34}^{\ell_{34}}  \calP_{14}^{\ell_{4}}
\label{DexpII}
\end{eqnarray}
with $\nu_4=0$.
Then we have to compute the integrals
\begin{eqnarray}
\nn\D^{(0)}_{\ell_{1},\ell_{2},\ell_{3},\ell_4}
&=&\sum_{\ell_{i}=r_{i}+s_{i}}\,\prod_{i=1}^4 {\ell_{i}!\over r_{i}!s_{i}!}\,
\int_{0}^1  d\hat\nu^{(i)}_{2}\,
(\Pinf_{12})^{r_{1}} (\Pinf_{23})^{r_{2}} (\Pinf_{34})^{r_{3}} (\Pinf_{14})^{r_{4}}\times\\
&\times&
\sum_{m^{(ij)}_{i}\atop k^{(ij)}_{l}}\, \prod_{1\leq i<j\leq4}  {\delta(\sum_{i}m^{(ij)}_{i})\over |m^{(ij)}_1| \cdots  |m^{(ij)}_{s_{ij}}|}
{\delta(\sum_{i,j,l} m^{(ij)}_{l} k^{(ij)}_{l})\over 4^{s_1+s_2+s_3+s_4}}\times\\
\nn&\times& e^{-2\pi \tau_{2}\sum |m^{(ij)}_{l}| |k^{(ij)}_{l}-\hat\nu_{2}^{(ij)}|}
\label{somme}
\end{eqnarray}
The final result can be separated into five contributions:
\begin{equation}
\D^{(0)}_{\ell_{1},\ell_{2},\ell_{3},\ell_4}
= \D^{\infty}_{\ell_{1},\ell_{2},\ell_{3},\ell_4}+  \D^{(1)}_{\ell_{1},\ell_{2},\ell_{3},\ell_4}+\D^{(a)}_{\ell_{1},\ell_{2},\ell_{3},\ell_4}+
\D^{(b)}_{\ell_{1},\ell_{2},\ell_{3},\ell_4}
+ \D^{(c)}_{\ell_{1},\ell_{2},\ell_{3},\ell_4}
\end{equation}
where $\D^\infty$ accounts for the case where $s_{i}=0$ for all $i=1,2,3,4$;
$\D^{(a)}$ contains the case where three $s_i$ vanish;
$\D^{(b)}$ contains the case where two $s_i$, namely $s_i$ and
$s_{i+1}$, vanish; $\D^{(c)}$ contains the case where $s_i$ and
$s_{i+2}$ vanish; and $\D^{(1)}$ contains all other contributions,
namely those where  $s_{i}\geq 1$ for all $i=1,2,3,4$ and those where
one of the $s_i$ is zero.

We are interested in the contributions which are not exponentially
suppressed, which means that the integration region includes points where
$\sum |m^{(ij)}_{l}| |k^{(ij)}_{l}-\hat\nu_{2}^{(ij)}|=0$.
Note that this is possible only if all $k^{(ij)}_{l}$, with
a given  $(ij)$ are equal to each other,
i.e. $k^{(ij)}_{l}\equiv k^{(ij)}$, for $(ij)=12,23,34$ and $14$. As a result,
$\delta(\sum_{i,j,l} m^{(ij)}_{l} k^{(ij)}_{l})$
becomes proportional to $\delta(\sum_{i}m^{(ij)}_{i})$ so it gives no
further restriction to the above sums in (\ref{somme}).
Therefore, we have to sum over four integers
$(k^{(12)},k^{(23)},k^{(34)},k^{(14)})$.
Since $|\hat\nu_{2}^{(ij)}|\leq 1$, the only contributions come from
the terms with $k^{(ij)}=0,1,-1$.
In order to identify such contributions, it is convenient to decompose
the integrals into four contributions:
\bea
\int_{0}^1 \int_{0}^1\int_{0}^1
d\hat\nu^{(1)}_{2}d\hat\nu^{(2)}_{2}d\hat\nu^{(3)}_{2}
&=& \int_{0}^1d\hat\nu^{(1)}_{2} \int_{0}^{\hat\nu^{(1)}_{2}}
d\hat\nu^{(2)}_{2}
\int_{0}^{ \hat\nu^{(2)}_{2}} d\hat\nu^{(3)}_{2}  +
\int_{0}^1d\hat\nu^{(1)}_{2} \int_{0}^{\hat\nu^{(1)}_{2}}
d\hat\nu^{(2)}_{2}
\int_{ \hat\nu^{(2)}_{2}}^1 d\hat\nu^{(3)}_{2}
\non\\
&+&
\int_{0}^1d\hat\nu^{(2)}_{2} \int_{0}^{\hat\nu^{(2)}_{2}}
d\hat\nu^{(1)}_{2}
\int_0^{ \hat\nu^{(2)}_{2}} d\hat\nu^{(3)}_{2} +
\int_{0}^1d\hat\nu^{(2)}_{2} \int_{0}^{\hat\nu^{(2)}_{2}}
d\hat\nu^{(1)}_{2}
\int_{ \hat\nu^{(2)}_{2}}^1 d\hat\nu^{(3)}_{2}
 \non\\
&\equiv & J_1+J_2+J_3+J_4
\eea
Now it is easy to recognize the possible contributions of
$(k^{(12)},k^{(23)},k^{(34)},k^{(14)})$.
For example, for $J_1$, the values of $k^{(ij)}$ which will give rise to
contributions to the zero mode that are not exponentially suppressed are
$(0,0,0,0),(1,1,1,1),(0,0,1,1),(1,0,0,1),(0,1,0,1)$.
The next step is to consider each separate contribution and perform
the integral.
Computing  the integrals in all cases, we find an explicit
 expression for the general four-vertex function.
We will here omit the explicit expression since it is very cumbersome.
The expression involves, in addition to the harmonic sums of the
previous subsection $S(m,n)$ and $S(s_1,s_2,s_3;\alpha,\beta )$, new
harmonic sums that are given by
 \begin{eqnarray}
\nn H_1(s_{1},s_{2},s_{3},s_4;\alpha,\beta,\gamma )&=
&\sum_{(\underline{m^{1}}, \underline{m^{2}},\underline{m^{3}},
\underline{m^{4}})} {\delta(\sum_{j}m^1_{i}-\sum_{j}m^2_{j})\,
\delta(\sum_{j}m^3_{i}-\sum_{j}m^4_{j})
 \over \prod_{a=1}^4\left( \prod_{j=1}^{s_{a}} |m^{a}_{j}|\right) }\\
 &\times&
 {1\over (|\underline{m^1}|+
   |\underline{m^{4}}|)^{\alpha}(|\underline{m^2}|+
   |\underline{m^{4}}|)^{\beta}
(|\underline{m^3}|+ |\underline{m^{4}}|)^{\gamma }}
 \end{eqnarray}
 \begin{eqnarray}
\nn H_2(s_{1},s_{2},s_{3},s_4;\alpha,\beta,\gamma )&=
&\sum_{(\underline{m^{1}}, \underline{m^{2}},\underline{m^{3}},
\underline{m^{4}})} {\delta(\sum_{j}m^1_{i}-\sum_{j}m^2_{j})\,
\delta(\sum_{j}m^3_{i}-\sum_{j}m^4_{j})
 \over \prod_{a=1}^4\left( \prod_{j=1}^{s_{a}} |m^{a}_{j}|\right) }\\
 &\times&
 {1\over (|\underline{m^2}|+
   |\underline{m^{3}}|)^{\alpha}(|\underline{m^1}|+
   |\underline{m^{4}}|)^{\beta}
(|\underline{m^3}|+ |\underline{m^{4}}|)^{\gamma }}
 \end{eqnarray}
 \begin{eqnarray}
\nn H_3(s_{1},s_{2},s_{3},s_4;\alpha,\beta,\gamma )&=
&\sum_{(\underline{m^{1}}, \underline{m^{2}},\underline{m^{3}},
\underline{m^{4}})} {\delta(\sum_{j}m^1_{i}-\sum_{j}m^2_{j})\,
\delta(\sum_{j}m^3_{i}-\sum_{j}m^4_{j})
 \over \prod_{a=1}^4\left( \prod_{j=1}^{s_{a}} |m^{a}_{j}|\right) }\\
 &\times&
 {1\over (|\underline{m^2}|+
   |\underline{m^{4}}|)^{\alpha}(|\underline{m^1}|+
   |\underline{m^{2}}|)^{\beta}
(|\underline{m^3}|+ |\underline{m^{4}}|)^{\gamma }}
 \end{eqnarray}
 where $|\underline{m}^{1}|= \sum_{j} |m^{12}_{j}|$,
 $|\underline{m}^{2}|= \sum_{j} |m^{23}_{j}|$, $|\underline{m}^{3}|=
 \sum_{j} |m^{34}_{j}|$, $|\underline{m}^{4}|= \sum_{j} |m^{14}_{j}|$.
 In addition, the result involves particular cases of these harmonic
 sums:
\bea
H_0(s_1,s_2;\alpha,\beta,\gamma )& \equiv &H_3(0,0,s_1,s_2;\alpha,\beta,
\gamma )=
H_3(0,s_1,0,s_2;\gamma,\alpha,\beta )\non\\
S_0 (s_1,s_2;\alpha,\beta )&\equiv & S(0,s_2,s_1;\alpha ,\beta )
\eea
Running a {\tt Mathematica} program with the complete expression, we find
the following large-$\tau_{2}$ expansions for the zero modes,
\bea
\D^{(0)}_{1,1,1,1} &=& {1\over (4\pi)^4}\Big(2\zeta(8)\tau_2^4 +
{5\pi \zeta(7) \over 8 \tau_2 ^3}\Big) \\
\D^{(0)}_{1,1,1,2} &=& {1\over (4\pi)^5}\Big( {6\zeta(10)\over
  5}\tau_2^5 + 2\pi \zeta(3)\zeta(6)\tau_2^2
-{1\over 2}\pi \zeta(4)\zeta(5)\non\\
&+&{21\pi\zeta(2)\zeta(7)\over
  8\tau_2^2}-{3\zeta(2)\zeta(3)\zeta(5)\over \tau_2^3} +
{43\pi\zeta(9)\over 64\tau_2^4}\Big)+O(e^{-\tau_{2}})\\
\D^{(0)}_{1,1,2,2}&=& {1\over (4\pi)^6}\Big( {612\over691}\zeta(12)\tau_2^6
 + {8\pi\over3} \zeta(3)\zeta(8)\tau_2^3
-\pi \zeta(5)\zeta(6)\tau_{2}+{7\over12}\zeta(3)^2\zeta(6)\Big)+O({1\over\tau_{2}})\\
\D^{(0)}_{1,2,1,2}&=&{1\over (4\pi)^6}\Big( {612\over691}\zeta(12)\tau_2^6
 + {8\pi\over3} \zeta(3)\zeta(8)\tau_2^3
-\pi \zeta(5)\zeta(6)\tau_{2}+21\zeta(3)^2\zeta(6)\Big)+O({1\over\tau_{2}})\\
\nn\D^{(0)}_{1,1,1,3}&=&{1\over (4\pi)^6}\Big( {5625\over691}\zeta(12)\tau_2^6
 + 20\pi \zeta(3)\zeta(8)\tau_2^3
-{15\pi\over4} \zeta(5)\zeta(6)\tau_{2}+{63\over4}\zeta(3)^2\zeta(6)\Big)+O({1\over\tau_{2}})\,,\\
\eea
From these expressions we can determine the expression quadratic in $\hat E_r$'s that reproduces the terms
with positive powers of $\tau_2$, so that,
\begin{eqnarray}
\D_{1,1,1,1}&=& \hat E_{4}\\
\D_{1,1,1,2}&=& -{8\over5}\hat E_{5}+\hat E_{2}\hat E_{3}+\D^{fin}_{1,1,1,2}\\
\D_{1,1,2,2}&=&-{3658\over 2073} \hat E_{6}- {1\over 3}\hat E_{3}^2+{4\over3}\hat E_{2}\hat E_{4}+ {\zeta(3)^2 \over184320}+\D^{fin}_{1,1,2,2}\\
\D_{1,2,1,2}&=& -{3658\over 2073} \hat E_{6}- {1\over 3}\hat E_{3}^2+{4\over3}\hat E_{2}\hat E_{4}+ {\zeta(3)^2 \over184320}+\D^{fin}_{1,2,1,2}\\
\D_{1,1,1,3}&=&-{10415\over 691}\hat E_{6}-{5\over4}\hat E_{3}^2+10\,\hat E_{2}\hat E_{4}+{\zeta(3)^2\over 245760}+\D^{fin}_{1,1,1,3}
\end{eqnarray}
We saw in section~\ref{sec:amplitude} and in appendix~\ref{sec:fund} that
the constant part in the large $\tau_2$ expansion is needed in order to determine the
coefficient of $s^k\, \hR^4$ in the ten-dimensional theory.

In addition to the preceding results, which are needed in order to determine the coefficients of the terms
up to order $s^6\, \hR^4$ that are summarized in the main text, we have also evaluated certain terms
at higher order.
For the vertex functions appearing
in the calculation of the genus-one coefficient of  $s^7\,\hR^4$ we find
\begin{eqnarray}
\left.\D^{(0)}_{1,1,1,4}\right|_{cste}&=&-{167\over123863040}\, \zeta(7); \quad
\left.\D^{(0)}_{1,1,2,3}\right|_{cste}={727\over 82575360}\, \zeta(7);\\
\nn\left.\D^{(0)}_{1,2,1,3}\right|_{cste}&=&{727\over 82575360}\, \zeta(7);\quad
\left.\D^{(0)}_{1,2,2,2}\right|_{cste}={733\over 61931520}\, \zeta(7)
\end{eqnarray}
For  $s^8\, \hR^4$, we find
\be
\left.\D^{(0)}_{1,1,1,5}\right|_{cste}={223\over 49545216 }\zeta(3)\zeta(5)\ ,\quad
\left.\D^{(0)}_{2,2,2,2}\right|_{cste}=-{173\over 15482880}\zeta(3)\zeta(5)
\ee
\be
\left.\D^{(0)}_{1,2,1,4}\right|_{cste}=\left.\D^{(0)}_{1,1,2,4}\right|_{cste}=-{199\over 61931520  }\zeta(3)\zeta(5)\ ,\quad
\left.\D^{(0)}_{1,1,3,3}\right|_{cste}=-{69\over 9175040}\zeta(3)\zeta(5),\
\ee
\be
\left.\D^{(0)}_{1,3,1,3}\right|_{cste}=-{449\over55050240 }\zeta(3)\zeta(5)\ ,\quad
\left.\D^{(0)}_{1,2,2,3}\right|_{cste}=\left.\D^{(0)}_{2,1,2,3}\right|_{cste}=-{89\over 10321920
}\zeta(3)\zeta(5) .
\ee
These coefficients are strikingly simple --a rational number times
$\zeta(3)\zeta(5)$ -- even though
they arise after  summing a huge number of terms.

\subsubsection{$\ell_5$ and/or $\ell_6\ne 0$}

\begin{figure}[!h]
\centering\includegraphics[width=10cm]{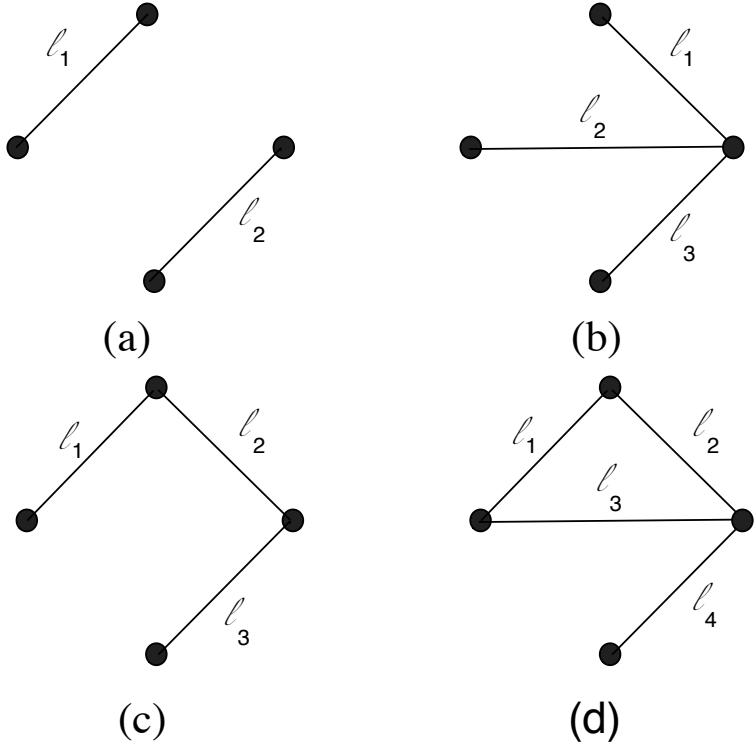}
\caption{This figure shows all the degenerate four-vertex diagrams that reduce to products
of two-vertex and three-vertex diagrams.}
\label{fig:vertex4b}\end{figure}

Let us now consider particular four-vertex diagrams where  $\ell_5$ and/or
$\ell_6$ are different from zero. In some cases, they reduce to
products
of lower-point vertex diagrams, as in
figure~\ref{fig:vertex4b}(d), given by $\D_{\ell_{1}\ell_{2}\ell_{3}}\times\D_{\ell_{4}}$.

More generally, the diagrams are those of figure~\ref{fig:vertex4a}(b)
and figure~\ref{fig:vertex4a}(c).
Although we have not evaluated the $\D$ functions in these cases for arbitrary nonzero values of
$\ell_r$, we have computed the positive powers of $\tau_2$ and the constant part in two
special
cases that are needed for evaluating terms of order $s^k\,\hR^4$
up to $k=6$. One of these is the diagram in figure~\ref{fig:vertex4a}(c) with all $l_r=1$,
for which the zero mode is
\begin{eqnarray}
\D_{1,1,1;1,1,1}^{(0)}&=&{1\over (4\pi)^6}\,\left({138\over691}\,\zeta
(12)\,\tau_{2}^6+6\pi\,\zeta(5)\zeta(6)\tau_{2}+O(1/\tau_{2})\right)\,,
\end{eqnarray}
which leads to the expression
\begin{eqnarray}
\D_{1,1,1;1,1,1}&=&- {2791\over691}\, \hat E_{6}+2\, \hat E_{3}^2+ \D_{1,1,1;1,1,1}^{fin}\,.
\end{eqnarray}
The other special case that we have evaluated is that of figure~\ref{fig:vertex4a}(b) with $\ell_r=1$,
for which the zero mode is
\begin{eqnarray}
\D_{1,1,1,1;1}^{(0)}&=&{1\over (4\pi)^5}\,{4\over5}\zeta(10)\tau_{2}^5+{\zeta(5)\over
30720}+O(1/\tau_{2})\,,
\end{eqnarray}
which leads to
\begin{eqnarray}
\D_{1,1,1,1;1}&=&{2\over 5}\hat E_{5} + {\zeta(5)\over30720}+\D_{1,1,1,1;1}^{fin}\,.
\end{eqnarray}

However, we will see in the next section that in order to determine $j^{(0,2)}$
we also need to evaluate $\D_{1,1,1,1;2}$ and $\D_{1,1,1,2;1}$.  These will be obtained by a slightly
different procedure in appendix~\ref{sec:leading}.

\section{Diagrammatic expansion of the coefficients}\label{sec:Expand}

We will here present the expansions (\ref{anal}) of the $j^{(p,q)}$'s as linear combinations of
$\D_{\{\ell\}}$'s for values of $p$ and $q$
up to order $2p+3q=8$. We will also substitute the expressions derived in the last section
for $\D_{\{\ell\}}$ in terms of Eisenstein series that are used in evaluating the
coefficients in the ten-dimensional case and the large-$\tau_2$ power series expansions
that are needed for obtaining the $r$-dependent coefficients in the nine-dimensional case.

We will present the expressions for each $j^{(p,q)}$ function in the form
\bea
j^{(p,q)} &=& \sum_{\{\ell\}}e_{\{\ell\}}^{(p,q)}\, \D_{\{\ell\}}  \nn\\
&=&  P^{(p,q)}(\{\hat E_{r}\}) + J^{(p,q)} + \delta j^{(p,q)}\,,
\label{jpqfun}
\eea
where $\delta j^{(p,q)}$ decreases by at least a power of $\tau_2$ at large $\tau_2$.
The first equality shows the diagrammatic decomposition and
the second equality shows how the positive powers of $\tau_2$ are incorporated in a
quadratic form in Eisenstein series and will be appropriate for the ten-dimensional calculations.
Furthermore, substituting the expansions of $\D^{(0)}_{\{\ell\}}$ derived in the previous appendix
into the first line of (\ref{jpqfun}) gives the expressions contained in the nine-dimensional
case in (\ref{sumario}).

The list of functions of interest to us is the following:
\begin{eqnarray}j^{(0,1)}&=&{4^3\over3!}\,\left( 8
\D_{1,1,1}+2 \D_{3}\right) \nn \\
&=&{4^3\over3!}\,\left( 10
\hat E_{3}+{\zeta(3)\over 32}\right) + \delta j^{(0,1)}\,,
\label{beuno}
\end{eqnarray}
where the $\D_{3}$ contribution\footnote{In \cite{gv:stringloop}
  $\D_3$ was denoted $B_{2}$.} is represented by the two-vertex diagram of figure~\ref{fig:vertex2}
of appendix~\ref{sec:vertex2} with $\ell_1=2$,
and $\delta j^{(0,1)}$.

\begin{eqnarray}
\nn j^{(2,0)}&=&{4^4\over4!}\,
\left(9 \D_{2}^2+ 6 \D_{1,1,1,1}+ \D_{4}\right)\\
&=&{4^4\over4!}\,
\left(24 \hat E_{2}^2- 24\hat E_{4}\right)+\delta j^{(2,0)}\,,
\label{bedos}
\end{eqnarray}
where $\D_4$ is the $\ell_1=4$ contribution to figure~\ref{fig:vertex2}.

\begin{eqnarray}
\nn j^{(1,1)}&=&{5\over6}\,{4^5\over 5!}\,
\left(2\D_5+96 \D_{2}\D_{1,1,1}+28 \D_3 \D_{2}
+32\D_{1,1,3}-24 \D_{1,2,2}
+24\D_{1,1,1,2} - 48 \D_{1,1,1,1;1} \right)\\
&=&{5\over6}\,{4^5\over 5!}\,
\left(-{9864\over5}\hat E_{5}+1080\hat E_{3}\hat E_{2}+{97\over7680}\zeta(5)\right)+\delta j^{(1,1)} \,,
\label{betres}
\end{eqnarray}
where $\D_{\ell_1,\ell_2\ell_3}$,  $\D_{\ell_1,\ell_2\ell_3,\ell_4}$ and
$\D_{\ell_1,\ell_2\ell_3,\ell_4;\ell_5}$ are defined by
figure~\ref{fig:vertex3}(b), figure~\ref{fig:vertex4a}(a) and
figure~\ref{fig:vertex4a}(b), respectively.

\begin{eqnarray}
\nn  j^{(3,0)}&=&{4^6\over6!}\,
\left(
-5 \D_{3}^2+{45\over2}\, \D_{2}\,\D_{4}+45 \D_{1,1,1}^3
+{1\over2} \D_{6}+60 \D_{1,1,1,3}-90 \D_{1,1,2,2} + 45 \D_{1,2,1,2}\right)\\
&=&{4^6\over6!}\,
\left(-{9501120\over 691} \hat E_{6}+3960 \hat E_{3}^2+2880 \, \hat E_{2}\hat E_{4}
+{3\over 512}\,\zeta(3)^2\right)+\delta j^{(3,0)}\, .
\label{becuatro}
\end{eqnarray}

 \be\begin{split}
 j^{(0,2)}&={4^6\over 6!}\,
\big({2\over3}\D_{6}+{100\over3} \D_{3}^2 -10 \D_{2} \D_{4}-20 \D_{2}^3+160\D_{1,1,1}\D_{3}\\
&+40 (\D_{2,2,2}+ \D_{1,1,4}-2 \D_{1,2,3})
-20( 3\D_{1,2,1,2}-12 \D_{1,1,2,2}+4 \D_{1,1,1,3})\\
&+80 \D_{1,1,1;1,1,1}+240(
 \D_{1,1,1,1;2}-2 \D_{1,1,1,2;1})\big)\cr
 &={4^6\over 6!}\,
\left(- {12345120\over691} \, \hat E_{6}+5040\, \hat E_{3}^2+3840 \, \hat E_{2}\hat E_{4}
+ {61\over 6144}\,\zeta(3)^2\right)+\delta j^{(0,2)}\,,
\end{split}
\label{segona}
\ee
where $\D_{\ell_1,\ell_2\ell_3,\ell_4;\ell_5,\ell_6}$ is defined by  figure~\ref{fig:vertex4a}(c)

All of the $\D$ functions that arise in these expressions were evaluated in appendix~\ref{sec:ZMdiag}
apart from $\D_{1,1,1,1;2}$ and $\D_{1,1,1,2;1}$, which will be determined by a different procedure in
appendix~\ref{sec:leading}.

We also note the sum of diagrams that arise at the next two orders, even though we will not
evaluate the $\D$ functions that arise in these cases in this paper.
The coefficient multiplying $\sg_2^2\,\sg_3\,\hR^4$  is given by
\begin{eqnarray}
\nn j^{(2,1)} &=&{4^7\over7!}\, {2\cdot 4^5\over 3^3\cdot 5}\,
\bigg(
24 \D_{1, 1, 5} - 60\D_{1, 2, 4} + 40\D_{1, 3, 3} +
 90 \D_{1, 1, 1, 4} - 120 \D_{1, 1, 2, 3} +
 180\D_{1, 2, 1, 3}
\\
&-& 90 \D_{1, 2, 2, 2} + 30 \D_2(16 \D_{1, 1, 3} - 12 \D_{1, 2, 2} +
   5 \D_2 \D_3) + 120 \D_{1,1,1}\D_4 +
 25 \D_3\D_4 + 39\D_2\D_5 + \D_7
\non\\
&-&
 480 \D_{1, 1, 1, 3; 1} +
   360(\D_{1, 1, 2, 2; 1} - \D_{1, 2, 1, 2; 1} +
     \D_{1, 2, 2, 1; 1} ) \bigg)\,.
\end{eqnarray}
The coefficient multiplying $\sg_2^4\hR^4$  is given by
\be\begin{split}
j^{(4,0)}&={4^8\over8!}\, {128\over 135}\,
\bigg(
\D_8+84 \D_2 \D_6 -56 \D_3 \D_5 +105 \D_4^2+1260 \D_2^2\D_4
-560 \D_2 \D_3^2 \\
&+
336 \D_{1, 1, 1, 5} - 1680 \D_{1, 1, 2, 4} +
 1120 \D_{1, 1, 3, 3} + 840 \D_{1, 2, 1, 4} -
 3360 \D_{1, 2, 2, 3} \\
&+
560 \D_{1, 3, 1, 3} +
 1680 \D_{2, 1, 2, 3} + 630 \D_{2, 2, 2, 2}\bigg)\, .
 \end{split}\ee
The coefficient  multiplying $\sg_{2}\sg_{3}^2\,\hR^4$ is
\be\label{lastd}\begin{split}
j^{(1,2)}&={4^8\over8!}\,{128\over 135}\,
\bigg(\D_8 +700\,\D_2\,\D_3^2 - 420\,\D_2^2\,\D_4 - 35\,\D_4^2 +
  420\,\D_{1,1,1}\,\D_5 + 154\,\D_3\,\D_5 \\
  & +  14\,\D_2\,\D_6  + 1680\,\D_3\,\D_{1,1,3} + 1260\,\D_2\,\D_{1,1,4}
 +  70\,\D_{1,1,6}\cr
 & - 1260\,\D_3\,\D_{1,2,2} - 2520\,\D_2\,\D_{1,2,3}\\
  & -
  210\,\D_{1,2,5} + 70\,\D_{1,3,4} + 1260\,\D_2\,\D_{2,2,2} + 210\,\D_{2,2,4}\\
  &-
  140\,\D_{2,3,3} - 84\,\D_{1,1,1,5} + 1470\,\D_{1,1,2,4} - 1120\,\D_{1,1,3,3}\\
  &-
  210\,\D_{1,2,1,4} + 2940\,\D_{1,2,2,3} - 140\,\D_{1,3,1,3} -
  1680\,\D_{2,1,2,3}\\
  & -  630\,\D_{2,2,2,2} +  420\,\D_{1,1,1,1;4} -
  1680\,\D_{1,1,1,2;3}
- 1260\,\D_{1,2,2,1;2}\\
  & +
  3360\,\D_{1,1,1,3;2} - 2100\,\D_{1,1,1,4;1}
   - 1260\,\D_{1,1,2,2;2} + 840\,\D_{1,1,2,3;1}\\
  & +
  2520\,\D_{1,2,1,2;2} - 4200\,\D_{1,2,1,3;1}
    + 1260\,\D_{1,2,2,2;1}+ 840\,\D_{1,2,3,1;1}\\
  & +  1680\,\D_{1,1,3;1,1,1} + 1260\,\D_{2,1,2;1,1,1} - 2520\,\D_{2,2,1;1,1,1}\bigg)
\end{split}\ee
All the $\D$'s in the last two equations have been evaluated apart from
$\D_{\ell_1,\ell_2,\ell_3;\ell_4,\ell_5,\ell_6}$ and
$\D_{\ell_1,\ell_2,\ell_3;\ell_4,\ell_5,\ell_6}$ with
all $\ell_k$'s $\ne 0$.  These coefficients that have not been evaluated arise
from the `Mercedes' diagram, figure~\ref{fig:vertex4a}(c).

\section{The zero mode of $\D_{\ell_{1},\dots,\ell_{6}}$  by another method}\label{sec:leading}
The calculations of appendix~\ref{sec:ZMdiag} become very
difficult at relatively low orders, as is evident from the fact that
we did not evaluate certain $\D$ functions that were needed in order to determine $j^{(0,2)}$ in
 (\ref{sec:ZMdiag}).  Here  we will here present
an alternative method for evaluating these functions, even though this also has technical difficulties.

The different diagrams that arise in the low energy expansion can be computed by using
the representation (\ref{torprop}) of the propagator. We proceed by considering
the following  generalization of the $\D$ functions,
\be
C_{s_1,s_2,\dots,s_{N+1}}=\sum_{(m_i,n_i)\neq (0,0)} \prod_{i=1}^{N+1}
 {\tau_2^{s_i}\over |m_i+n_i\tau|^{2s_i}  } \delta(\sum_k
  m_k)\delta(\sum_k n_k)\,,
\label{dero}
\ee
where $s_i$ are integers and the sum over $k$ involves a subset of the $\{ m_i,n_i\}$,
according to the topology of the diagram, which needs to be specified in order to define
this expression completely.  Certain special cases of these functions
correspond to certain $D$ functions.  For example,
\be
\D_3 = \frac {1}{(4\pi)^3} C_{1,1,1}\,, \qquad
\D_{1,1,1,1;1}= \frac{1}{(4\pi)^5}C_{2,2,1}\,
\label{cddef}
\ee
(although information on the topology of the diagram needs to be specified in completely
defining the function $C_{2,2,1}$).

\subsection{Leading and subleading powers of $\tau_2$}

We will start by determining
the first two terms in the expansion of (\ref{dero}) at large $\tau_2$.
To illustrate the method, we start with a particular example,
\be
C_{s_1,s_2,s_3}=\sum_{(m_1,n_1)\neq (0,0)\atop (m_2,n_2)\neq(0,0)} {\tau_2^{s_1+s_2+s_3}\over
|m_1+n_1\tau|^{2s_1} |m_2+n_2\tau|^{2s_2} |m_1+m_2+(n_1+n_2)\tau|^{2s_3}}\,.
\ee
The zero mode expansion has the general form
\bea
C^{(0)}_{s_1,s_2,s_3} &=& a_{s_1s_2s_3}\ \tau_2^{s_1+s_2+s_3}+
b_{s_1s_2s_3}^1 \ \tau_2^{1+s_1-s_2-s_3}+
b_{s_1s_2s_3}^2 \ \tau_2^{1+s_2-s_1-s_3}
\nn\\
&+&
b_{s_1s_2s_3}^3 \ \tau_2^{1+s_3-s_1-s_2}+
\cdots+c_{s_1s_2s_3} \ \tau_2^{1-s_1-s_2-s_3}+O\big( \exp(-c\tau_2 )\big)\, .
\eea
So we begin by computing $a_{s_1s_2s_3}$ and $b_{s_1s_2s_3}^i$.

The leading term comes by setting $n_1=n_2=0$ in the sums.
This gives
\bea
a_{s_1s_2s_3} &=& \sum_{m_1,m_2}{}' {1\over
  |m_1|^{2s_1}|m_2|^{2s_2}|m_1+m_2|^{2s_3}}
\non\\
 &=& 2 W(2s_1,2s_2,2s_3)+ 2 W(2s_2,2s_3,2s_1)+2 W(2s_3,2s_1,2s_2)\,,
\eea
where  $\sum_{m_1,m_2}'$ excludes $m_1m_2(m_1+m_2)=0$ and $W(a,b,c)$ is the Witten zeta-function
of (\ref{e:Wittenzeta}) whose values are tabulated using the methods of the Appendix~\ref{sec:Harmonic} . For example,
one gets $a_{1,1,1}=2\zeta(6)$.

The term with coefficient $b^i_{s_1s_2s_3}$ comes from $n_i=0$
(where we define $n_3=n_1+n_2$). Consider in particular the contribution with $n_1=0$.
This is given by
\be
V_{s_1s_2s_3}^1 \equiv \sum_{m_1,n_2\neq 0}\sum_{m_2}
{\tau_2^{s_1+s_2+s_3}\over
|m_1|^{2s_1} |m_2+n_2\tau|^{2s_2} |m_1+m_2+n_2\tau|^{2s_3}}\, .
\ee
Applying the Poisson resummation formula for the summation over $m_2$ gives
\be
V_{s_1s_2s_3}^1 = \sum_{m_1,n_2\neq 0}\sum_{w}
{\tau_2^{s_1+s_2+s_3}\over
|m_1|^{2s_1}} \int_{-\infty}^{+\infty} d\mu {e^{2\pi i w(\mu -n_2\tau_1) }\over
 |\mu+i n_2\tau_2|^{2s_2} |m_1+\mu+i n_2\tau_2|^{2s_3}}\, .
\ee
Now the only dependence on $\tau_1$ is in the factor
$e^{-2\pi i w n_2\tau_1}$. The zero mode is independent of
$\tau_1$, so it arises  from the $w=0$ term in the sum.
Introducing a new integration variable
$\nu=\mu/(n_2\tau_2)$ leads to
\be
V_{s_1s_2s_3}^1 \bigg|_{\rm pert} = \sum_{m_1,n_2\neq 0}
{\tau_2^{1+s_1-s_2-s_3}\over
|m_1|^{2s_1} |n_2|^{2s_2+2s_3-1}} \,
\int_{-\infty}^{+\infty} d\nu {1 \over
 (\nu^2+1)^{s_2} ((\nu+ {m_1\over n_2\tau_2})^2 +1)^{s_3}}\, .
\label{verti}
\ee
We now consider the limit of large $\tau_2$.
When the sum over $m_1$ of $1/|m_1|^{2s_1}$ is convergent
(i.e. $s_1>\ha $)
the leading term of the expansion of the integrand in powers of $1/\tau_2$ is finite.
For the leading term the integral reduces to
$$
\int_{-\infty}^{+\infty} d\nu {1 \over
 (\nu^2+1)^{s_2+s_3}} = {\sqrt{\pi}\Gamma(s_2+s_3-\ha)\over \Gamma(s_2+s_3)}\, ,$$
so
\be
V_{s_1s_2s_3}^1 \bigg|_{\rm pert} = \tau_2^{1+s_1-s_2-s_3}
2\zeta(2s_1) 2\zeta(2s_2+2s_3-1){\sqrt{\pi}\Gamma(s_2+s_3-\ha)\over
\Gamma(s_2+s_3)}\, .
\ee
Therefore
\be
b^1_{s_1s_2s_3}={4\sqrt{\pi}\Gamma(s_2+s_3-\ha)\over \Gamma(s_2+s_3)}\zeta(2s_1)\zeta(2s_2+2s_3-1)\ .
\ee
As a check, we now compare with the expansions that were calculated earlier.
For $\D^{(0)}_{3}$, we find
\be
(4\pi)^3\, \D_3^{(0)}= C^{(0)}_{1,1,1} =2\zeta(6) \tau_2^3+ \pi^3 \zeta(3) + O(\tau_2^{-2})\, ,
\label{twel}
\ee
in agreement with the expansion given in section~\ref{sec:vertex2}.
We have multiplied $b_{1,1,1}^1$ by 3 since there are three identical subleading contributions
$ b_{1,1,1}^i$, $i=1,2,3$.
Similarly, we reproduce the leading and subleading terms of the
diagrams $\D_{1,1,1,2}=C_{1,1,3}/(4\pi)^5$, $\D_{1,1,1,1;1}=C_{2,2,1}/(4\pi)^5$.

We now turn to consider the more general modular function given by (\ref{dero}).
The expansion now has the form
\be
C_{s_1,s_2,\dots,s_{N+1}}=a_{s_1,\dots,s_{N+1}}\
\tau_2^{s_1+\cdots+s_{N+1}}
+ b^{ij}_{s_1,\dots,s_{N+1}}\
\tau_2^{1+s_1+\cdots-s_i-s_j+\cdots+s_{N+1}}+\cdots\,,
\ee
where
\be
a_{s_1,\dots,s_{N+1}} = \sum_{m_1,\dots,m_N}{}'{1\over
  |m_1|^{2s_1}\cdots|m_N|^{2s_N}|m_1+\cdots+m_N|^{2s_{N+1} } }\, .
\label{deth}
\ee
The second term arises by setting to zero all $n_k$ except
$n_i$ and $n_j$ (with the understanding that $n_{N+1}\equiv \sum_k n_k$).
The  corresponding contribution $V^{i,j}$ is the following one:
\be
V_{s_1,\dots,s_{N+1}}^{N,N+1} \equiv \sum_{m_i\neq 0, i<N\atop n_N\neq 0}
\sum_{m_N}
{\tau_2^{s_1+\cdots+s_{N+1}}\over
|m_1|^{2s_1}\cdots|m_{N-1}|^{2s_{N\! -\! 1}} |m_N+n_N\tau|^{2s_N}
|m_1+\cdots+m_N+n_N\tau|^{2s_{N+1}}}
\ee
Performing a Poisson resummation in $m_N$ we arrive at
\be
V_{s_1,\dots,s_{N+1}}^{N,N+1} \bigg|_{\rm pert} = \sum_{m_1,,\dots,,m_{N-1}\neq 0}
\sum_{n_N\neq 0}
{\tau_2^{1+s_1+\cdots+s_{N-1}-s_N-s_{N+1}}\ J\over
|m_1|^{2s_1}\cdots|m_{N-1}|^{2s_{N-1}}|n_N|^{2s_N+2s_{N+1}-1}}\ ,
\ee
$$
J=\int_{-\infty}^{+\infty} d\nu {1 \over
 (\nu^2+1)^{s_N} \big( (\nu+ {1\over n_2\tau_2}(m_1+\cdots+m_{N-1}) )^2 +1\big)^{s_{N+1}}}\ .
$$
Now we would like to extract the leading term in the expansion in powers of
$1/\tau_2$ of $J$.
When $s_1,\dots, s_{N-1}>1/2$, the sums over $m_i$ of
$1/|m_i|^{2s_i}$ are
convergent and  the leading term is simply obtained by setting
$\tau_2=\infty $ inside the integral.
This gives
$$
J={\sqrt{\pi}\, \Gamma(s_N+s_{N+1}-\ha)\over \Gamma(s_N+s_{N+1})}\, ,
$$
so that
\be
b^{N,N+1}_{s_1,\dots,s_{N+1}}={2^N \sqrt{\pi}\, \Gamma(s_N+s_{N+1}-\ha)\over
  \Gamma(s_N+s_{N+1}) }
\zeta(2s_1)\cdots\zeta(2s_{N-1}) \zeta(2s_N+2s_{N+1}-1)\ .
\label{bbr}
\ee
Applying (\ref{deth}) and (\ref{bbr}), we reproduce the first two terms in the expansion of
$\D_4^{(0)}$ given in
section~\ref{sec:vertex2}.
\be
(4\pi)^4\D_4^{(0)}= C^{(0)}_{1,1,1,1} =  10\zeta(8) \tau_2^4+
{2\pi^5 \over 3}\, \zeta(3)\tau_2+\cdots
\ee
We have included a combinatorial factor $4!/2!2!$  multiplying the  subleading term,
 which counts the number of identical subleading terms obtained by setting different $n_i$
 to zero.
Similarly, we reproduce the first two terms in the expansion of $\D_5=C_{1,1,1,1,1}/(4\pi)^5$
and $\D_{1,1,3}=C_{1,1,1,2}/(4\pi)^5$ (where the precise definition of these $C$ functions
requires a specification of the topology of the diagram in addition to the values of the integers
$s_i$).

The method outlined in this subsection is particularly useful for diagrams that correspond
to $\D$ functions of the general
form  $\D_{\ell_1,\ell_2,\ell_3;\ell_4,\ell_5,\ell_6}^{(0)}$ where
five or six of the $\ell_i$ are different from zero,
for which we did not derive the general formula in section~\ref{sec:ZMdiag}.
As an application we compute the first two terms in the
expansion of $\D_{1,1,1;1,1,1}^{(0)}$, defined as the zero $\tau_1$ mode of
\bea
\D_{1,1,1;1,1,1}&=&
{1\over(4\pi)^6} \sum_{m_i,n_i}{}'
{\tau_2^6\over |m_1+n_1\tau|^2 |m_2+n_2\tau|^2 |m_3+n_3\tau|^2
|m_1+m_2+(n_1+n_2)\tau|^2}
\nn\\
&\times &
{1\over  |m_1+m_3+(n_1+n_3)\tau|^2|m_1+m_2+m_3+(n_1+n_2+n_3)\tau|^2  }\, .
\label{arfe}
\eea
The expansion of the zero $\tau_1$ mode  has the form
\be
\D_{1,1,1;1,1,1}^{(0)}={1\over(4\pi)^6} \Big( a\tau_2^6 + b\tau_2 + O(\tau_2^{-1}) \Big)\, .
\label{vabe}
\ee
The leading term arises by setting all $n_i=0$,  giving the sum
\be
a=\sum_{m_1,m_2,m_3}{}' \ {1\over m_1^2 m_2^2 m_3^2 (m_1-m_2)^2(m_1-m_3)^2(m_1-m_2-m_3)^2}\,.
\ee
The symbol $\sum'$ indicates that the sum does not contain those
$m_i$ where the denominator vanishes.
Computing this sum gives
\be
a={138\over 691} \zeta(12)\, .
\label{abe}
\ee
The subleading term arises from four identical sums, in which:
a)  $n_1=n_2=0$; b) $n_1=n_3=0$; c) $n_1=n_3, \ n_2=0$;
d) $n_1=n_2, \ n_3=0$.
Following the above procedure for case a)
we perform a Poisson resummation on $m_3$, leading to the result
\be
b = 4 \sum_{m_1,m_2,n_3}{}' \ {1\over m_1^2m_2^2(m_1-m_2)^2} {1\over (n_3)^5}
\int_{-\infty}^{+\infty} d\nu {1\over (\nu^2+1)^3} ={6\pi }\ \zeta(6)\zeta(5)\ .
\label{bebe}
\ee

\subsection{Systematic large-$\tau_2$ expansion}

We return to the general modular function (\ref{dero}).
The power-behaved terms of the zero mode expansion
are $\tau_2^{s_1+s_2+\cdots+s_{N+1}}$, $\dots$, $\tau_2^{1-s_1-s_2-\cdots-s_{N+1}}$.
In order to obtain  the coefficients of these terms one proceeds as follows.
The different contributions can be organized in terms of the different subsets $\{n_k\}$ where all
$n_k$  are zero
(the most suppressed contribution is the one where none of the $n_i$ vanishes).
These subsets can easily be visualized
by considering a graphical representation of the modular function (see
figure~\ref{fig:vertex4a}) and remove propagator lines
in all possible ways leaving diagrams containing closed loops only. For example, in
figure~\ref{fig:vertex4a}(a) one could cut all
propagators in $\ell_4$ setting $n_1=\cdots=n_{\ell_4}=0$,  as long as  $\ell_{1,3}\geq 2$ and
$\ell_2$ is either zero or  $\ell_2\geq 2$,
so that only closed loops remain in the diagram.
For a given subset $\{n_k\}$ of vanishing $n_k$,
one performs Poisson resummation in all remaining $m_i$ variables with $i\neq k$.
The resulting integral is then computed explicitly.
The next step is to perform the remaining summations explicitly.
This step becomes more complicated in diagrams with a large number of propagator lines.
In this step one can drop contributions which are exponentially suppressed at large $\tau_2$.

As an example, we come back to (\ref{verti}) with $s_1=s_2=s_3=1$. Using
\be
\int_{-\infty}^{+\infty} d\nu {1 \over
 (\nu^2+1) ((\nu+ a)^2 +1)} = {2\pi \over 4+ a^2}\ ,
\ee
we find
\be
V_{111}^1 =2\pi
\sum_{m_1,n_2\neq 0} {1\over m_1^{2} |n_2|} {\tau_2^2\over
4n_2^2\tau_2^2+m_1^2} \, .
\label{vrt}
\ee
Now use
\be
\sum_{m_1\neq 0}
{\tau_2^2\over m_1^2(4n_2^2\tau_2^2+m_1^2)} ={\pi^2\over 12n_2^2}+{1\over
16n_2^4\tau_2^2}-{\pi\over 8n_2^3\tau_2}\coth(2n_2\pi\tau_2) \ .
\ee
Noting that
$n_2^{-3}\coth(2n_2\pi\tau_2)\to |n_2|^{-3}$, modulo exponentially
suppressed terms, as $\tau_2\to \infty$,
it is straightforward to perform the remaining sum
over $n_2$, giving zeta values.
 Adding the multiplicity factor 3 and the leading $\tau_2^3$ term, we obtain
\be
(4\pi)^3\, \D_3^{(0)}= C_{1,1,1}^{(0)}=2\zeta(6) \tau_2^3+ \pi^3 \zeta(3)
-{\pi^6\over 60\tau_2}+{3\pi \zeta(5)\over 4\tau_2^2}+V_{111}'\ ,
\label{ssel}
\ee
where $V_{111}'$ represents the contribution where none of the $n_i$ vanishes. This is computed as follows.
By performing Poisson resummation in $m_1$ and $m_2$ we find
\be
V_{111}'=\sum_{n_1,n_2}{}'  \int_{-\infty}^{+\infty} d\mu_1 d\mu_2
{\tau_2^3\over |\mu_1 +in_1\tau_2|^2|\mu_2 +in_2\tau_2|^2|\mu_1+\mu_2 +
i(n_1+n_2)\tau_2|^2 }+O(\exp(-2\pi \tau_2))\ .
\ee
Introducing new integration variables $\mu_{1}=\nu_{1}|n_{1}|\tau_2$
and $\mu_{2}=\nu_{2}|n_{2}|\tau_2$
 and computing the integrals, we find
\be
V_{111}'={\pi^2\over \tau_2}\ S(3,1) = {3\pi^2\over 2\tau_2}\zeta(4)
=\frac{\pi^6}{60\tau_2}\, ,
\ee
where $S(m,n)$ is defined in (\ref{sHnm}) and we have used (\ref{e:S31}).
Thus $V'_{111}$ cancels the similar
contribution in (\ref{ssel}). The final result reproduces (\ref{e:ZMD3})
obtained by using the asymptotic form of the propagator.

We now calculate the first terms in the expansion of
$\D_{1,1,1,1;2}$ and $\D_{1,1,1,2;1}$, which  are given by
\be
\D_{1,1,1,1;2}= \sum_{m_1,m_2,m_3\atop n_1,n_2,n_3}\! {\tau_2^6\over
|m_1+n_1\tau|^4 |m_2+n_2\tau|^4|m_3+n_3\tau|^2|m_1+m_2+m_3+(n_1+n_2+n_3)\tau|^2}\,,
\label{dones}
\ee
and
\bea
\D_{1,1,1,2;1}&=& \sum_{m_1,m_2,m_3 \atop n_1,n_2,n_3}\! {\tau_2^6\over |m_1+n_1\tau|^4
|m_2+n_2\tau|^2|m_3+n_3\tau|^2}
\nn\\
&\times &
{1\over |m_1+m_2+(n_1+n_2)\tau|^2
|m_2+m_3+(n_2+n_3)\tau|^2}\,.
\label{donetwo}
\eea
The leading term is of order $\tau_2^6$ and
arises from the contribution with $n_1=n_2=n_3=0$. There are
various contributions according to which $n_i$ are zero. Our
aim is to compute the expansion up to the term constant in $\tau_2$.
Keeping only these contributions, we find
\bea
(4\pi)^6\
\D_{1,1,1,1;2}&=&  {5047\zeta(12)\over 691}\,
\tau_2^6+V_{12}+4V_{13}+2V_1+O(\tau_2^{-1})
 \non\\
(4\pi)^6\ \D_{1,1,1,2;1}&=&  {802\zeta(12)\over 691}\, \,
\tau_2^6+\tilde V_{12}+2\tilde V_{13}+\tilde V_{23}+\tilde V_1+O(\tau_2^{-1})
\eea
Here
$V_{ij}$ indicates the contribution with $n_i=n_j=0$ and $V_i$ the contribution with $n_i=0$.
To compute the leading term of order $\tau_2^6$ we have used (for $m_1\neq 0$)
\bea
\sum_{m_2}{}'
{1\over (m_1+m_2)^2 m_2^2} &=& {4\zeta(2)\over m_1^2} -{6\over
m_1^4}
\non\\
\sum_{m_2}{}' {1\over (m_1+m_2)^4 m_2^2} &=& {2\zeta(4)\over m_1^2} +{8\zeta(2)\over m_1^4} -{15\over m_1^6}
\non\\
\sum_{m_2}{}' {1\over (m_1+m_2)^4 m_2^4} &=& {4\zeta(4)\over m_1^4} +{40\zeta(2)\over m_1^6} -{70\over m_1^8}
\label{sumitas}
\eea

Next, for each $V_{ij},\ \tilde V_{ij}$ we perform
Poisson resummation in the $m_k$ with $k\neq i,j$ and integrate over the
resulting continuous variable $\nu_k$. We find
\bea
V_{12} &=& 2\pi \sum_{m_1,m_2,n_3}{\tau_2^5\over m_1^4 m_2^4 |n_3|}{1\over 4n_3^2\tau_2^2+(m_1+m_2)^2}
\non\\
V_{13} &= & {\pi\over 2} \sum_{m_1,m_3,n_2}{\tau_2^3\over m_1^4 m_3^2 |n_2|^3}{12n_2^2\tau_2^2+(m_1-m_3)^2
\over (4n_2^2\tau_2^2+(m_1-m_3)^2)^2}
\non\\
\tilde V_{12} &=&   2\pi\sum_{m_1,m_2,n_3}{\tau_2^5\over m_1^4 m_2^2(m_1-m_2)^2 |n_3|}{1\over 4n_3^2\tau_2^2+m_2^2}
\non\\
\tilde V_{13} &=&  2\pi\sum_{m_1,m_3,n_2}{\tau_2^5\over m_1^4 m_3^2|n_2|}
{m_1^2+m_3^2-m_1m_3+12n_2^2\tau_2^2\over (4n_2^2\tau_2^2+m_1^2)(4n_2^2\tau_2^2+m_3^2)(4n_2^2\tau_2^2+(m_1-m_3)^2)}
\non\\
\tilde V_{23} &=& {\pi\over 2} \sum_{m_2,m_3,n_1}{\tau_2^3\over (m_2-m_3)^2 m_2^2 m_3^2 |n_1|^3}
{12n_1^2\tau_2^2+m_2^2
\over (4n_1^2\tau_2^2+m_2^2)^2}
\eea
Using the identities (\ref{sumitas}) we  find
\bea
V_{12} &=& {14\pi\over 3}\zeta(8)\zeta(3)\tau_2^3-{7\pi\over 2}\zeta(5)\tau_2+{\pi^6\over 180}\zeta(6)+O(\tau_2^{-1})
\non\\
V_{13} &= & {21\pi\over 4}\zeta(5)\tau_2-{\pi^6\over 90}\zeta(6)+O(\tau_2^{-1})
\non\\
\tilde V_{12} &=&   {4\pi\over 3}\zeta(8)\zeta(3)\tau_2^3-{5\pi\over 4}\zeta(5)\tau_2+{\pi^6\over 360}\zeta(6)+O(\tau_2^{-1})
\non\\
\tilde V_{13} &=&  {21\pi\over 4}\zeta(5)\tau_2-{\pi^6\over 90}\zeta(6)+O(\tau_2^{-1})
\non\\
\tilde V_{23} &=& {3\pi\over 2}\zeta(5)\tau_2+O(\tau_2^{-1})
\eea
To compute the contribution $V_1$ we Poisson resum the integers $m_2$ and $m_3$. This gives
\be
V_1=\sum_{m_1,n_2,n_3}{\tau_2^6 \over m_1^4}\int_{-\infty}^{+\infty} d\mu_2d\mu_3
{1 \over |\mu_2+in_2\tau_2|^4 \mu_3+in_3\tau_2|^2 |m_1+\mu_2-\mu_3+i(n_2-n_3)\tau_2|^2}
\ee
where we have dropped exponentially suppressed terms. After introducing $\nu_{2,3}$ integration variables by
$\mu_{2,3}=\nu_{2,3}|n_{2,3}|\tau_2 $ we get
\be
V_1=\sum_{m_1,n_2,n_3}{1 \over m_1^4|n_2|^3|n_3|}\int_{-\infty}^{+\infty} d\nu_2d\nu_3
{1 \over (\nu_2^2+1)^2 (\nu_3^2+1) (({m_1\over\tau_2} +\nu_2|n_2|-\nu_3|n_3|)^2+(n_2-n_3)^2)}
\ee
Since we are interested in the leading term $O(\tau_2^0)$ we can set $m_1=0$ in the integrand.
The sum over $m_1$ then gives $2\zeta(4)$. Computing the integrals,
one finds sums which can be reduced to Witten zeta functions. The final result is
\be
V_1={7\pi^6\over 360} -{\pi^6\over 180}\zeta(3)^2
\ee
Similarly, we find $\td V_1=V_1$.

To summarize, we have found
\bea
\D_{1,1,1,1;2}&=&  {5047\zeta(12)\over  (4\pi)^6\, 691 }
\, \tau_2^6
+{14\pi\zeta(8)\zeta(3)\over
(4\pi)^6\,3}\tau_2^3+{35\pi\zeta(6)\zeta(5)\over (4\pi)^6\,2}\tau_2 -{2\zeta(6)\zeta(3)^2\over (4\pi)^6\, 21}+O(\tau_2^{-1})
 \non\\
 &=&-{27965\over2073}\, \hat E_{6}+ {35\over6} \, \hat E_{3}^2+{7\over3}\,\hat E_{2}\hat E_{4}
-{2\zeta(6)\zeta(3)^2\over (4\pi)^6\, 21}+ \D_{1,1,1,1;2}^{fin}
\non
\\
\non\D_{1,1,1,2;1}&=&   {802\zeta(12)\over (4\pi)^6\,691}\,  \tau_2^6+
{4\pi\zeta(8)\zeta(3)\over (4\pi)^6\,3}\tau_2^3+{43\pi\zeta(6)\zeta(5)\over
(4\pi)^6\,4}\tau_2-{\zeta(6)\zeta(3)^2\over (4\pi)^6\, 21}+O(\tau_2^{-1}) \\
&=&-{3435\over691}\, \hat E_{6}+ {43\over12} \, \hat E_{3}^2+{2\over3}\,
\hat E_{2}\hat E_{4}-{\zeta(6)\zeta(3)^2\over (4\pi)^6\, 21}+
\D_{1,1,1,2;1}^{fin}
\label{ddoce}
\eea
The results (\ref{ddoce}), together with  (\ref{vabe}), (\ref{abe}), (\ref{bebe}) are
used in (\ref{segona}) -- combined with the other vertex functions found
earlier -- to find the corresponding $\sg_3^2$ terms in (\ref{sumario}).

\section{Phase-space integrals for two-particle unitarity}
\label{sec:unitaritop}

We will here evaluate the phase-space integrals that arise in the unitarity analysis of
section~\ref{sec:twounitarity}.
This will involve a number of basic integrals that result from expanding the tree amplitudes in
(\ref{unitarityt}) in powers of $\hat\sigma'_2=(\alpha'/ 4)^2\,(s^2+{t'}^2+{u'}^2)$,
 $\hat\sigma_3'=(\alpha'/4)^3\,3st'u'$,
$\hat \sigma_2^{\prime\prime}=(\alpha'/4)^2\, (s^2+{t^{\prime\prime}}^2+{u^{\prime\prime}}^2)$ and
$\hat \sigma_3^{\prime\prime}=(\alpha'/4)^3\,3st^{\prime\prime}u^{\prime\prime}$, where
$t'$ $u'$, $t^{\prime\prime}$, $u^{\prime\prime}$ are defined in terms of the internal and external
momenta in (\ref{mandprimes}).
This leads to integrals of the general form
\be
S(a,b,c,d) =\int d^{10} k\,\delta^{(+)}(k^2)\,\delta^{(+)}((p_1+p_2-k)^2)
\, (t')^{a-1}\, (u')^{b-1}\, (t^{\prime\prime})^{c-1}
(u^{\prime\prime})^{d-1}\,,
\label{prototype}
\ee
where $a,b,c,d$ are integers.

The $\delta^{(+)}$ functions impose the mass-shell conditions.
If we choose the centre of mass frame in which $q^\mu = p^\mu_1+ p^\mu_2=
(s^\half, \vec 0_9)$ (where $\vec 0_k$ is the $k$-dimensional zero vector) we can write
\be
\delta^{(+)}((q-k)^2)= \frac{1}{2s}\delta (k^0-{1\over2}\, s^\half)\,,\qquad
\delta^{(+)}(k^2) = \frac{1}{s}\,  \delta(k^0 - |\vec k|)\,.
\label{deltacon}
\ee
so that $k^\mu$ has the form
\be
k^\mu = (k^0,{\vec  k}) = \frac {s^\half}{2}\, (1, {\vec n}_9)\,,
\label{kform}
\ee
where ${\vec n}_9$ is the unit nine-vector.
The integral (\ref{prototype})
may be evaluated by choosing the momenta, which satisfy the on-shell masslessness condition,
to take the following form,
\bea
p_1^\mu&=& \frac {s^\half}{2}\, (1,\,1\,, \vec 0_8) \,\qquad p^\mu_2 =\frac {s^\half}{2}\, (1,\,-1\,, \vec 0_8)
\nn\\
p_3^\mu &=& \frac {s^\half}{2}\, (-1,\, \cos \rho,\,\sin\rho,\, \vec 0_7) \,,\qquad
p_4^\mu = \frac {s^\half}{2}\,
(-1,\,- \cos \rho,\,- \sin\rho,\, \vec 0_7)\nn\\
k^\mu &=& \frac {s^\half}{2}\, (1,\, \cos \theta,\,\sin\theta\cos\phi,
\,\sin\theta\sin\phi\, {\vec n}_7)  \,,
\label{momcoord}
\eea
where ${\vec n}_7$ is the unit seven-vector and the scattering angle, $\rho$, is given by
\be
\cos\rho = \frac{t-u}{s}\,.
\label{rhodef}
\ee
Changing variables from $k^i$ ($i=1,\dots,9$)
 to $\theta$, $\phi$, ${\vec n}_7$, the measure of integration becomes
  \be
 d^{10}k\, \delta^{(+)}((q-k)^2)\, \delta^{(+)}(k^2) =
 \frac{s^{3}}{2^{6}}\, (\sin\theta)^7\, (\sin\phi)^6\, d\theta\, d\phi\,
 d^7{\vec n}_7\, \delta({\vec n}^2_7-1)\, .
 \label{measure}
 \ee
In this parametrization we have
\bea
t' &=& 2p_1\cdot k = -\frac{s}{2}\,(1-\cos\theta)\, ,\quad t^{\prime\prime} =
-2p_4\cdot k = -\frac{s}{2}\, (1+\cos\theta\cos\rho +
\sin\theta\cos\phi\sin\rho)\, ,\nn\\
u'&=& 2p_2\cdot k = -\frac{s}{2}\, (1+\cos\theta)\,,
\quad u^{\prime\prime}= -2p_3\cdot k = -\frac{s}{2}\, (1-\cos\theta\cos\rho
-\sin\theta\cos\phi\sin\rho)\,.
\label{paramstu}
\eea
The integral over the seven dimensional  unit vector $\vec n_{7}$ gives an overall factor of
 the volume of the six-dimensional
sphere vol$(S^6)=16\pi^3/15$.

Substituting the term of order $\zeta(3)\,\hR^4$ introduces a factor of $st^{\prime\prime}
u^{\prime\prime}$
into one of the
tree amplitudes on the right-hand side of (\ref{unitarityt}), with the lowest order term in the other,
 leading to an integral of the form (\ref{prototype}) with $a=b=0$ and $c=d=1$,
\begin{equation}
S(0,0,1,1)= 2\zeta(4)\,s\,,
\label{e:s4cut}\end{equation}
which determines the coefficient of the threshold terms of order $\al^4\,s^4\,\log(-s/\mu_4)$.

Similarly, substituting the expansion of one tree-level amplitude at order
$\zeta(5)\,\hat\sigma_2\, \hR^4$ and the lowest order term in the other in~(\ref{unitarityt})
leads to an integral of the form (\ref{prototype}) with $a=b=0$ and $c=d=2$,
\be
S(0,0,3,1) ={\zeta(4)\over56}\,s\,(31 \, s^2+ (t-u)^2)\,,
\label{e:s6cut}
\ee
which determines the coefficient of the threshold terms of order
$\al^6\,s^6\,\log(-s/\mu_6)$.



\begin{thebibliography}{99}

\bibitem{Green:1981yb}
  M.~B.~Green and J.~H.~Schwarz,
 { \sl Supersymmetrical String Theories,}
  Phys.\ Lett.\  B {\bf 109} (1982) 444.

\bibitem{Green:1987sp}
  M.~B.~Green, J.~H.~Schwarz and E.~Witten,
  {\sl Superstring Theory. Vol.~1: Introduction,}
{\it  Cambridge, Uk: Univ. Pr. ( 1987) 469 P. ( Cambridge Monographs On Mathematical Physics)}


 \bibitem{Green:1987mn}
  M.~B.~Green, J.~H.~Schwarz and E.~Witten,
 { \sl Superstring Theory. Vol. 2: Loop Amplitudes, Anomalies And Phenomenology,}
{\it  Cambridge, Uk: Univ. Pr. ( 1987) 596 P. ( Cambridge Monographs On Mathematical Physics)}



\bibitem{gv:stringloop}  M.B. Green and  P. Vanhove,
{\sl The low energy expansion of the one-loop type II superstring amplitude},
Phys. Rev. {\bf D 61}, 104011 (2000) [arXiv:hep-th/9910056].

\bibitem{grv:twoloop} M. Green, J. Russo and P. Vanhove,  {\sl Modular properties of the
four-graviton type II superstring amplitude}, in preparation.

\bibitem{Terras} A. Terras, "Harmonic Analysis on Symmetric Spaces and
  Applications I", Springer-Verlag 1985


\bibitem{grv:bigpaper} M. Green, J. Russo and P. Vanhove,
{\sl Non-renormalisation Conditions in Type II String Theory and Maximal Supergravity},
JHEP {\bf 0702} (2007) 099
  [arXiv:hep-th/0610299].


  \bibitem{gkv:twoloop}
  M.~B.~Green, H.~h.~Kwon and P.~Vanhove,
 {\sl Two loops in eleven dimensions,}
  Phys.\ Rev.\ D {\bf 61} (2000) 104010
  [arXiv:hep-th/9910055].


\bibitem{Bern:1998ug}
  Z.~Bern, L.~J.~Dixon, D.~C.~Dunbar, M.~Perelstein and J.~S.~Rozowsky,
  {\sl On the relationship between Yang-Mills theory and gravity and its
  implication for ultraviolet divergences,}
  Nucl.\ Phys.\  B {\bf 530} (1998) 401
  [arXiv:hep-th/9802162].

\bibitem{D'Hoker:1994yr}
  E.~D'Hoker and D.~H.~Phong,
 { \sl The Box graph in superstring theory,}
  Nucl.\ Phys.\  B {\bf 440} (1995) 24
  [arXiv:hep-th/9410152].

\bibitem{Green:1982sw}
  M.~B.~Green, J.~H.~Schwarz and L.~Brink,
{\sl N=4 Yang-Mills And N=8 Supergravity As Limits Of String Theories,}
  Nucl.\ Phys.\ B {\bf 198} (1982) 474.

\bibitem{ZagierDilog}
D. Zagier, {\sl The Dilogarithm Function} in
  {\sl Frontiers in number theory, physics, and geometry 2:
  On Conformal Field Theory, Discrete Groups, and Renormalisation
  Proceedings, Meeting, Les Houches,
  France, March 9-21, 2003,}
  P.~Cartier, B.~Julia, P.~Moussa and P.~Vanhove (Ed).






 \bibitem{Russo:1997mk}
  J.~G.~Russo and A.~A.~Tseytlin,
  {\sl One-loop four-graviton amplitude in eleven-dimensional supergravity,}
  Nucl.\ Phys.\ B {\bf 508} (1997) 245
  [arXiv:hep-th/9707134].



\bibitem{Cartier} P.~Cartier {\sl An introduction to zeta functions}
in {\sl From Number Theory to Physics} (Les Houches, 1989),
 Michel Waldschmidt, Pierre Moussa, Jean-Marc Luck, Claude Itzykson, eds.,
 Springer-Verlag, Berlin, 1992, pp. 1-63


\bibitem{HarmonicSum} Eric W. Weisstein et al. "Harmonic Number." From MathWorld--A Wolfram Web Resource. http://mathworld.wolfram.com/HarmonicNumber.html

\bibitem{borwein2}
  J. M. Borwein,
D. M. Bradley,
D. J. Broadhurst, {\sl Evaluations of k-fold Euler/Zagier sums:
a compendium of results for arbitrary k },
the electronic journal of combinatorics 4 (no.2) (1997),1

\bibitem{borwein}
J.M. Borwein and R. Girgensohn,
{\sl Evaluation of Triple Euler Sums,}
The electronic journal of combinatorics  3 (1996) R23.

\bibitem{broadhurst} D. Broadhurst,  {\sl Private communication}

\bibitem{Petitot} Michel Petitot, {\sl Base de Grobner des Multiple Zeta Values en Maple (poids 12)} {\tt http://www2.lifl.fr/\~{}petitot/}

\bibitem{EulerSum} Eric W. Weisstein. "Euler Sum." From MathWorld--A Wolfram Web Resource. http://mathworld.wolfram.com/EulerSum.html

  \bibitem{ZagierGanglKaneko} H. Gangl, M. Kaneko, D. Zagier {\sl
  Double zeta values and modular forms},
  in ``Automorphic forms and Zeta functions'', Proceedings of the conference in memory of Tsuneo Arakawa, World Scientific, 71--106, (2006)



\end{thebibliography}
\end{document}